\documentclass[10pt]{article}
\usepackage[top=2cm,left=2cm,right=2cm]{geometry}
\usepackage[numbers,square]{natbib}

\usepackage[italian,english]{babel}
\usepackage{epsfig}
\usepackage{hyperref}
\usepackage{amssymb}
\usepackage{amsfonts}
\usepackage{epsf}
\usepackage{rotating}
\usepackage{graphicx}
\usepackage{amsmath}
\usepackage{fancyhdr}
\usepackage{lineno}
\usepackage{subfigure}
\usepackage{babel}
\usepackage{graphics}
\usepackage{geometry}
\usepackage{pstricks}
\usepackage{color}

\makeatletter
\def\cleardoublepage{\clearpage\if@twoside
\ifodd\c@page
\else\hbox{}\thispagestyle{empty}\newpage
\if@twocolumn\hbox{}\newpage\fi\fi\fi}
\makeatother

\let\a=\alpha   \let\b=\beta   \let\g=\gamma   \let\d=\delta
\let\e=\epsilon    \let\h=\eta     
    \let\k=\kappa  \let\l=\lambda  
                 \let\r=\rho
\let\s=\sigma        \let\f=\phi
\let\c=\chi         \let\w=\omega
\def\sq{\,\raise.5pt\hbox{$\nbox{.09}{.09}$}}
\def\sqb{\,\raise.5pt\hbox{$\overline{\nbox{.09}{.09}}$}}

\def\e{\epsilon}
\def\a{\alpha}
\def\b{\beta}

\def\l{\lambda}

\def\g{\gamma}

\def\r{\rho}

\def\ds#1{#1\kern-1ex\hbox{/}}
\def\dsh{h\kern-1.2ex /}

\newcommand{\bea}{\begin{eqnarray}}
\newcommand{\eea}{\end{eqnarray}}

\def\nn{\nonumber}

\def\beq{\begin{equation}}
\def\eeq{\end{equation}}

\def\ba{\begin{eqnarray}}
\def\ea{\end{eqnarray}}

\newcommand{\beqa}{\begin{eqnarray}}
\newcommand{\eeqa}{\end{eqnarray}}

\newcommand{\si}{\sigma}

\newcommand{\ksl}{k \! \! \!  /}

\newcommand{\pd}{\partial}

\newcommand{\mD}{\mathcal{D}}

\newcommand{\bh}{\bar{\h}}

\def\th{\theta}

\newcommand{\bes}{\begin{subequations}}
\newcommand{\ees}{\end{subequations}}

\begin{document}
\begin{center}
\vspace{4.cm}

{\bf \large
Gravity and the Neutral Currents: \\ Effective Interactions from the Trace Anomaly}

\vspace{1cm}

{\bf Claudio Corian\`{o}, Luigi Delle Rose and Mirko Serino}

\vspace{1cm}

{\it Departimento di Fisica, Universit\`{a} del Salento \\
and  INFN-Lecce, Via Arnesano 73100, Lecce, Italy\footnote{claudio.coriano@unisalento.it, luigi.dellerose@le.infn.it, mirko.serino@le.infn.it}
}\\
\vspace{.5cm}
\begin{abstract}
We present a complete study of the one graviton-two neutral gauge bosons vertex at 1-loop level in the electroweak theory. This vertex provides the leading contribution to the interaction between the Standard Model and gravity, mediated by the trace anomaly, at first order in the inverse Planck mass and at second order in the electroweak expansion. At the same time, these corrections are significant for precision studies of models with low scale gravity at the LHC. We show, in analogy with previous results in the QED and QCD cases, that the anomalous interaction between gravity and the gauge current of the Standard Model, due to the trace anomaly, is mediated, in each gauge invariant sector, by effective massless scalar degrees of freedom. We derive the Ward and Slavnov-Taylor identities characterizing the vertex. Our analysis includes
the contributions from the improvements of the scalar sector, induced by a conformally coupled Higgs sector in curved space.
\end{abstract}
\end{center}
\newpage

\section{Introduction}
This work is the fourth in a sequence of investigations \cite{Armillis:2009pq,Armillis:2010qk,Armillis:2010pa}, motivated by 
the original analysis of \cite{Giannotti:2008cv}, aimed at studying the precise structure of the anomalous effective action which describes the anomalous breaking of scale invariance in the Standard Model (SM). Here we expand and fill in the details of a previous study \cite{Coriano:2011ti}. 

This breaking is induced by the trace anomaly \cite{Duff:1977ay,Duff:1993wm} and can be extracted from the exact computation of a set of diagrams involving, to leading order in the gravitational constant and in the gauge couplings, the graviton-gauge-gauge vertex. The work is a natural extension and an application of remarkable classical studies \cite{{Freedman:1974gs},Callan:1970ze,Adler:1976zt,Collins:1976yq} of the energy momentum tensor and of the corresponding trace anomaly in gauge theories. 

In the case of a gravitational background characterized by a small deviation with respect to the flat spacetime metric, this vertex is described by the correlation function containing one insertion of the energy momentum tensor (EMT) (denoted as $T$) on the correlation function of two gauge currents (denoted as $V, V'$). If we allow only conformally coupled scalars and
operators only up to dimension-4 in the Langrangian \cite{Freedman:1974gs} \cite{Callan:1970ze}, the EMT is uniquely defined by gravity and takes the form of a symmetric and (on-shell) conserved expression. In the massless limit, which in our case is equivalent to dealing with
an unbroken theory (i.e. before electroweak symmetry breaking) the EMT is classically (on-shell) traceless.

As remarked in \cite{Giannotti:2008cv} and in our previous studies in the context of QED \cite{Armillis:2009pq} and QCD \cite{Armillis:2010qk}, the study of this correlator is interesting in several ways and allows to address some important issues concerning anomaly-mediated interactions between the SM and gravity. At the same time, this program is part of an attempt to characterize rigorously in quantum field theory the effective action which describes the interaction between matter and gravity beyond tree level, showing some interesting features, such as the appearence of effective massless scalar degrees of freedom as mediators of the breaking of scale invariance \cite{Giannotti:2008cv}, in close analogy with what found in the case of chiral gauge theories \cite{Armillis:2009sm,Armillis:2009im,Armillis:2008bg,Coriano:2008pg}. Beside these theoretical motivations, these corrections find direct application in collider studies of low scale gravity, a point that we will address in a related work.

In a theory such as the SM, the breaking of scale invariance is related both to the trace anomaly and to the spontaneous breaking of the gauge symmetry by the Higgs mechanism  \cite{Coriano:2011ti}, and both contributions may become significant in some specific scenarios. For example, the enduring discussion over the cosmological implications of the quantum breaking of scale invariance has spanned decades \cite{Dolgov:1993vg, Corradini:2007gd}, since the work of Starobinsky \cite{Starobinsky:1980te}, with his attempt to solve the problem of the cosmological "graceful exit" that predated inflationary studies. At the same time, the treatment of the trace anomaly using more refined
approaches such as the world-line formulation, has allowed for new ways to investigate the corresponding effective action
\cite{Bastianelli:2002qw}.

The computation of the effective action which underlies this interaction is, in principle,  rather challenging not only for the large number of diagrams involved, but also because of the need of a consistent way to define these interactions. The ambiguity
present in the definition of the fermion contributions, for instance, requires particular care, due to the presence
of axial-vector and vector currents in an external gravitational background. These have been analyzed building on the results of  \cite{Armillis:2010pa}, which provides the ground for the extensions contained in the present study. The current analysis is far more involved than any previous study of ours, due to the appearance of a larger set of diagrams in the perturbative expansion. Their definition requires a suitable set of Ward and Slavnov-Taylor identities (STI's) which need to be identified from scratch and that we are going to discuss in fair detail. These are essential in order to establish the correctness of the computation and of the chosen regularization scheme, which is dimensional regularization with on-shell renormalization.

When we move from an exact gauge theory to a theory with spontaneous breaking of the gauge symmetry such as the SM, the contributions coming from the trace anomaly and from mass corrections are harder to disentangle, since the massless limit is not an option. However, even under these conditions, there are two possible ways of organizing the contributions to the 1-loop effective action which may turn out handy. The first expansion, obviously, is the usual $1/m$ expansion, where $m$ is a large electroweak mass, valid below the electroweak scale. The second has been first discussed in a previous work \cite{Armillis:2009sm} and is characterized by the isolation of the anomalous massless pole contribution from the remaining subleading $O(m^2/s)$ corrections. These can be extracted from a complete computation.

The goal of this work is to discuss the role of the interactions mediated by the conformal anomaly  using as a realistic example the Lagrangian of the SM, by focusing our investigation on the neutral currents sector. A similar analysis will be presented for the charged current sector in a forthcoming separate work.  These contributions play a role, in general, also in scenarios of TeV gravity and as such are part of the radiative corrections to graviton-mediated processes at typical LHC energies. 

\subsection{Organization of this work}
Our work is organized as follows. In section \ref{EMT} we will provide the basic definition of the energy momentum tensor in a curved spacetime, followed by a direct computation of all of its components according to the Lagrangian of the SM
(section \ref{tmunusection}). We then move to briefly summarize some important issues which concern the structure of the effective action, highlighting its perturbative properties, first among them the appearance of massless (scalar) effective degrees of freedom (anomaly poles) in the QED and QCD cases. In
sections \ref{mastersection} and \ref{BRSTsection} we derive the fundamental Ward and Slavnov-Taylor identites which define the structure of the $TVV'$ vertex, expanded in terms of its $TAA$, $TAZ$ and $TZZ$ contributions, where $T$ couples to the graviton and $A$ and $Z$ are the photon and the neutral massive gauge boson, respectively. Complete results for all the amplitudes are given in section \ref{resultsection}, expressed in terms of a small set of form factors. As we are going to show, the contribution to the anomaly comes from a single form factor in each amplitude, multiplying a unique tensor structure. These form
factors are characterized by the appearance of a massless pole with a residue that can be related to the beta function of the theory and which is the signature of the anomaly \cite{Armillis:2009im}. We have extensively elaborated in previous works on the significance of such contributions in the ultraviolet region (UV) \cite{Coriano:2011ti}.

In the presence of spontaneous symmetry breaking the perturbative expansion of these form factors can be still arranged in the form of a $1/s$ contribution, with $s$ being the invariant mass of the graviton line, plus mass corrections of the form $v^2/s$, with $v$ being the electroweak vev.   The computation shows that the trace part of the amplitude is then clearly dominated at large energy (i.e for $s \gg v^2$) by the pole contribution, as we will discuss in section \ref{discussionsection}. Our conclusions and perspectives are given in section \ref{conclusions}. Several technical points omitted from the main sections have been included in the appendices to facilitate the reading of those more involved derivations.

\section{ The EMT of the Standard Model: definitions and conventions}
\label{EMT}
The expression of a symmetric and conserved EMT for the SM, as for any field theory
Lagrangian, may be obtained, more conveniently, by coupling the corresponding Lagrangian to the gravitational field, described
by the metric $g_{\mu\nu}$ of the curved background
\beq S = S_G + S_{SM} + S_{I}= -\frac{1}{\kappa^2}\int d^4 x \sqrt{-g}\, R + \int d^4 x
\sqrt{-g}\mathcal{L}_{SM} + \frac{1}{6} \int d^4 x \sqrt{-g}\, R \, \mathcal H^\dag \mathcal H      \, ,
\eeq
where $\kappa^2=16 \pi G_N$, with $G_N$ being the four dimensional Newton's constant and $\mathcal H$ is the Higgs doublet.
We recall that Einstein's equations take the form
\beq\label{Einsteinfunct.1}\,
\frac{\d}{\d g^{\mu\nu}(x)}S_G =- \frac{\d}{\d g^{\mu\nu}(x)}[ S_{SM} + S_I ]
\eeq
and the EMT in our conventions is defined as
\beq  T_{\mu\nu}(x)  = \frac{2}{\sqrt{-g(x)}}\frac{\d [S_{SM} + S_I ]}{\d g^{\mu\nu}(x)},
\eeq
or, in terms of the SM Lagrangian, as
\beq \label{TEI spaziocurvo}
\frac{1}{2} \sqrt{-g} T_{\mu\nu}{\equiv} \frac{\pd(\sqrt{-g}\mathcal{L})}
{\pd g^{\mu\nu}} - \frac{\pd}{\pd x^\s}\frac{\pd(\sqrt{-g}\mathcal{L})}{\pd(\pd_\s g^{\mu\nu})}\, ,
\eeq
which is classically covariantly conserved   ($g^{\mu\rho}T_{\mu\nu; \rho}=0$).  In flat spacetime, the covariant derivative is replaced by the ordinary derivative, giving the ordinary conservation equation ($ \pd_\mu T^{\mu\nu} = 0$).

We use the convention $\eta_{\mu\nu}=(1,-1,-1,-1)$ for the metric in flat spacetime, parameterizing its deviations from the flat case as
\beq\label{QMM} g_{\mu\nu}(x) = \h_{\mu\nu} + \kappa \, h_{\mu\nu}(x)\,,\eeq
with the symmetric rank-2 tensor $h_{\mu\nu}(x)$ accounting for its fluctuations.

 In this limit, the coupling of the Lagrangian to gravity is given by the term
\beq\label{Lgrav} \mathcal{L}_{grav}(x) = -\frac{\kappa}{2}T^{\mu\nu}(x)h_{\mu\nu}(x). \eeq
The corrections to the effective action describing the coupling of the SM to gravity that we will consider in our work
are those involving one external graviton and two gauge currents. These correspond to the leading contributions to the anomalous breaking of scale invariance of the effective action in a combined expansion in powers of $\kappa$ and of the electroweak coupling ($g_2$) (i.e. of $O(\kappa \, g_2^2)$).

Coming to the fermion contributions to the EMT, we recall that the fermions are coupled to gravity using the spin connection $\Omega$ induced by the
curved metric $g_{\mu\nu}$. This allows to define a spinor derivative $\mathcal{D}$ which transforms covariantly under local Lorentz transformations. If we denote with $\underline{a},\underline{b}$ the Lorentz indices of a local free-falling frame, and denote with
$\s^{\underline{a}\underline{b}}$ the generators of the Lorentz group in the spinorial representation, the spin connection takes the form
\beq
 \Omega_\mu(x) = \frac{1}{2}\s^{\underline{a}\underline{b}}V_{\underline{a}}^{\,\nu}(x)V_{\underline{b}\nu;\mu}(x)\, ,
\eeq
where we have introduced the vielbein $V_{\underline{a}}^\mu(x)$. The covariant derivative of a spinor in a given representation
$(R)$ of the gauge symmetry group, expressed in curved $(\mathcal{D}_{\mu})$ coordinates is then given by
\beq \mathcal{D}_{\mu} = \frac{\pd}{\pd x^\mu} + \Omega_\mu   + A_\mu,\eeq
where $A_\mu\equiv A_\mu^a\, T^{a  (R)}$ are the gauge fields and $T^{a (R)}$ the group generators,
giving a Lagrangian of the form
\beqa \mathcal{L}& = & \sqrt{-g} \bigg\{\frac{i}{2}\bigg[\bar\psi\g^\mu(\mathcal{D}_\mu\psi)
 - (\mathcal{D}_\mu\bar\psi)\g^\mu\psi \bigg] - m\bar\psi\psi\bigg\}.       \eeqa

\section{Contributions to $T_{\mu\nu}$}
\label{tmunusection}
In this section we proceed with a complete evaluation of the EMT for the SM Lagrangian coupled to gravity. We will do so for the entire quantum Lagrangian of the SM, which includes also the contributions from the ghosts and the gauge-fixing terms. Details on our conventions for this section have been collected in appendix (\ref{conventions}). \\
The full EMT is given by a minimal tensor $T^{Min}_{\mu\nu}$ (without improvement) and a term of improvement, $T^I_{\mu\nu}$, generated by the conformal coupling of the scalars
\bea
T_{\mu\nu} = T^{Min}_{\mu\nu} + T^I_{\mu\nu} \,,
\eea
where the minimal tensor is decomposed into
\bea
T^{Min}_{\mu\nu} = T^{f.s.}_{\mu\nu} + T^{ferm.}_{\mu\nu} + T^{Higgs}_{\mu\nu} + T^{Yukawa}_{\mu\nu} + T^{g.fix.}_{\mu\nu} + T^{ghost}_{\mu\nu}.
\eea
\subsection{The gauge and fermion contributions}
The contribution from the gauge kinetic terms derived from the field strengths of the SM is
\beqa
T^{f.s.}_{\mu\nu}
& = &
\eta_{\mu\nu}\frac{1}{4}\left[F^a_{\r\s}F^{a\,\r\s} + Z_{\r\s}Z^{\,\r\s}
       + F^A_{\r\s}F^{A\,\r\s} + 2 W^+_{\r\s}W^{-\,\r\s}\right] \nn \\
&-&   F^a_{\mu\r}F^{a\,\r}_\nu - F^A_{\mu\r}F^{A\,\r}_\nu
       - Z_{\mu\r}{Z_\nu}^{\rho} - W^+_{\mu\r}W^{-\,\r}_\nu - W^+_{\nu\r}W^{-\,\r}_\mu, \,
\eeqa
where $F^a_{\mu\nu}$, $F^A_{\mu\nu}$, $Z_{\mu\nu}$ and $W^{\pm}_{\mu\nu}$ are respectively the field strengths of the gluon, photon, $Z$ and $W^{\pm}$ fields defined in appendix (\ref{conventions}). The fermion contribution is rather lengthy and we give it here for a single fermion generation
\beqa \label{TEIfermioni}
T^{ferm.}_{\mu\nu}
&=& - \h_{\mu\nu}\mathcal L_{ferm.} + \frac{i}{4}\bigg\{\bar\psi_{\nu_e}\g_\mu\stackrel{\rightarrow}{\pd}_\nu\psi_{\nu_e}
+ \bar \psi_e\g_\mu\stackrel{\rightarrow}{\pd}_\nu \psi_e  + \bar \psi_u\g_\mu\stackrel{\rightarrow}{\pd}_\nu \psi_u
+ \bar \psi_d\g_\mu\stackrel{\rightarrow}{\pd}_\nu \psi_d\nn\\
&-&  i\bigg[\frac{e}{\sqrt{2}\sin\th_W}\bigg(\bar\psi_{\nu_e} \g_\mu \frac{1-\g^5}{2}\psi_e\,W^+_\nu
+ \bar \psi_e\g_\mu\frac{1-\g^5}{2}\psi_{\nu_e}\,W^-_\nu\bigg)\nn\\
&+&  \frac{e}{\sin2\th_W}\bar\psi_{\nu_e} \g_\mu\frac{1-\g^5}{2}\psi_{\nu_e} Z_\nu
- \frac{e}{\sin2\th_W}\bar \psi_e\g_\mu\bigg(\frac{1-\g^5}{2} - 2\sin^2\th_W\bigg)\psi_e\,Z_\nu \nn\\
&+&  \frac{e}{\sqrt{2}\sin\th_W}\bigg(\bar \psi_u\g_\mu\frac{1-\g^5}{2}\psi_d\,W^+_\nu + \bar \psi_d\g_\mu
\frac{1-\g^5}{2}\psi_u\,W^-_\nu \bigg)\nn\\
&+&  \frac{e}{\sin2\th_W}\bar \psi_u\g_\mu\bigg(\frac{1-\g^5}{2} - 2\sin^2\th_W\frac{2}{3}\bigg)\psi_u\,Z_\nu
- \frac{e}{\sin2\th_W}\bar \psi_d\g_\mu\bigg(\frac{1-\g^5}{2} - 2\sin^2\th_W\frac{1}{3}\bigg)\psi_d\,Z_\nu\nn\\
&+&  e A_\nu\bigg(- \bar \psi_e\g_\mu \psi_e + \frac{2}{3}\,\bar \psi_u\g_\mu \psi_u - \frac{1}{3}\,\bar \psi_d\g_\mu \psi_d
\bigg) + g_s G^a_\nu\bigg(\bar \psi_u\g_\mu t^a \psi_u + \bar \psi_d\g_\mu t^a \psi_d \bigg)\bigg]\nn\\
&-&  \bar\psi_{\nu_e}\g_\mu\stackrel{\leftarrow}{\pd}_\nu\psi_{\nu_e} -\bar \psi_e\g_\mu\stackrel{\leftarrow}{\pd}_\nu \psi_e
  - \bar \psi_u\g_\mu\stackrel{\leftarrow}{\pd}_\nu \psi_u - \bar \psi_d\g_\mu\stackrel{\leftarrow}{\pd}_\nu \psi_d\nn\\
&-&  i\bigg[\frac{e}{\sqrt{2}\sin\th_W}\bigg(\bar\psi_{\nu_e} \g_\mu \frac{1-\g^5}{2}\psi_e\,W^+_\nu
    + \bar \psi_e\g_\mu\frac{1-\g^5}{2}\psi_{\nu_e}\,W^-_\nu\bigg)\nn\\
&+&  \frac{e}{\sin2\th_W}\bar\psi_{\nu_e} \g_\mu\frac{1-\g^5}{2}\psi_{\nu_e} Z_\nu
- \frac{e}{\sin2\th_W}\bar \psi_e\g_\mu\bigg(\frac{1-\g^5}{2} - 2\sin^2\th_W\bigg)\psi_e\,Z_\nu \nn\\
&+&  \frac{e}{\sqrt{2}\sin\th_W}\bigg(\bar \psi_u\g_\mu\frac{1-\g^5}{2}\psi_d\,W^+_\nu + \bar \psi_d\g_\mu\frac{1-\g^5}{2}\psi_u\,W^-_\nu \bigg)\nn\\
&+& \frac{e}{\sin2\th_W}\bar \psi_u\g_\mu\bigg(\frac{1-\g^5}{2} - 2\sin^2\th_W\frac{2}{3}\bigg)\psi_u\,Z_\nu
- \frac{e}{\sin2\th_W}\bar \psi_d\g_\mu\bigg(\frac{1-\g^5}{2} - 2\sin^2\th_W\frac{1}{3}\bigg)\psi_d\,Z_\nu\nn\\
&+&  e A_\nu\bigg(- \bar \psi_e\g_\mu \psi_e + \frac{2}{3}\,\bar \psi_u\g_\mu \psi_u - \frac{1}{3}\,\bar \psi_d\g_\mu \psi_d\bigg)
+ g_s G^a_\nu\bigg(\bar \psi_u\g_\mu t^a \psi_u + \bar \psi_d\g_\mu t^a \psi_d \bigg)\bigg] + (\mu \leftrightarrow \nu)\bigg\}   \,, \nn \\
\eeqa
where $\psi_{\nu_e}$, $\psi_e$, $\psi_u$ and $\psi_d$ are the Dirac spinors describing respectively the electron neutrino, the electron, the up and the down quarks while $\mathcal L_{ferm.}$ is given in appendix (\ref{conventions}).

\subsection{The Higgs contribution}
Coming to the contribution to the EMT from the Higgs sector, we recall that the scalar Lagrangian for the Higgs fields ($\mathcal H$) is given by
\beq \mathcal{L_{\mathcal H}} = (D^\mu \mathcal H)^\dagger(D_\mu \mathcal H)
+ \mu_{\mathcal H}^2 \mathcal H^\dagger \mathcal H - \lambda(\mathcal H^\dagger \mathcal H)^2\quad
\mu_{\mathcal H}^2,\l >0   \, ,\eeq
with the covariant derivative defined as
\beq D_\mu = \pd_\mu - i g W^a_\mu T^a - i g' B_\mu Y ,  \eeq
where, in this case, $T^a=\sigma^a/2$ are the generators of $SU(2)_L$, $Y$ is the hypercharge and the coupling constants $g$ and $g'$ are defined by
$e = g \, \sin \th_W = g' \cos \th_W$. As usual we parameterize
the vacuum $\mathcal H_0$ in the scalar sector in terms of the electroweak vev $v$ as
\beq \label{VEVHiggs}
\mathcal H_0 =
\left(\begin{array}{c} 0 \\ \frac{v}{\sqrt{2}} \end{array}\right)
\eeq
and we expand the Higgs doublet in terms of the physical Higgs boson $H$ and the two Goldstone bosons $\phi^{+}$, $\phi$ as
\bea
\mathcal H = \left(\begin{array}{c} -i \phi^{+} \\ \frac{1}{\sqrt{2}}(v + H + i \phi) \end{array}\right),
\eea
then the masses of the Higgs ($m_H$) and of the $W$ and $Z$ gauge bosons are given by
\beq m_H =  \sqrt{2}\,\mu_{\mathcal H}\, ,\quad M_W  =  \frac{1}{2}g v\, , \quad M_Z = \frac{1}{2}\sqrt{g^2 +
g'^2}\,v.\,  \eeq
We obtain for the energy-momentum tensor of the Higgs contribution the following expression
\beqa \label{TEIHiggs}
T^{Higgs}_{\mu\nu}
& = & -\h_{\mu\nu}\mathcal L_{Higgs} + \pd_\mu H\pd_\nu H + \pd_\mu \f\pd_\nu \f
      + \pd_\mu \f^+\pd_\nu \f^- + \pd_\nu \f^+\pd_\mu \f^- \nn\\
&+&  M_Z^2 Z_\mu Z_\nu + M_W^2(W^+_\mu W^-_\nu + W^+_\nu W^-_\mu)\nn\\
&+&  M_W\left(W^-_\mu\pd_\nu \f^+ + W^-_\nu\pd_\mu \f^+ + W^+_\mu\pd_\nu \f^- + W^+_\nu\pd_\mu \f^- \right)
+ M_Z(\pd_\mu \f Z_\nu + \pd_\nu \f Z_\mu)\nn\\
&+&  \frac{e M_W}{\sin\th_W}H\left( W^+_\mu W^-_\nu + W^+_\nu W^-_\mu \right)
    + \frac{e M_Z}{\sin 2\th_W}H \left(Z_\mu Z_\nu\right)\nn \\
&-&  \frac{e}{2\sin\th_W}\left[W^+_\mu\left(\f^-\stackrel{\leftrightarrow}{\pd}_\nu(H + i \f)\right)
     - W^+_\mu\left(\f^-\stackrel{\leftrightarrow}{\pd}_\nu(H + i \f)\right)\right]\nn \\
&-&  \frac{e}{2\sin\th_W}\left[W^-_\mu\left(\f^+\stackrel{\leftrightarrow}{\pd}_\nu(H - i \f)\right)
+ W^-_\nu\left(\f^+\stackrel{\leftrightarrow}{\pd}_\mu(H - i \f)\right)\right]\nn \\
&+&  i e \left( A_\mu + \cot 2\th_W Z_\mu \right)\left(\f^- \stackrel{\leftrightarrow}{\pd}_\nu \f^+\right)
+ i e \left( A_\nu + \cot 2\th_W Z_\nu \right)\left(\f^- \stackrel{\leftrightarrow}{\pd}_\mu \f^+\right)\nn \\
&-&  \frac{e}{\sin 2\th_W} \left[Z_\mu\left( \f\stackrel{\leftrightarrow}{\pd_\nu}H \right)
+ Z_\nu\left( \f\stackrel{\leftrightarrow}{\pd_\mu}H \right)\right]\nn \\
&-&  i e M_Z\sin\th_W\left[ Z_\mu\left(W^+_\nu \f^- - W^-_\nu \f^+ \right)
+ Z_\nu\left(W^+_\mu \f^- - W^-_\mu \f^+ \right)\right]\nn \\
&-&  i e M_W\left[ A_\mu\left(W^-_\nu - W^+_\nu \f^-\right) + A_\nu\left(W^-_\mu - W^+_\mu \f^-\right)\right]\nn \\
&+&  \frac{e^2}{4\sin^2\th_W}H^2\left[\left(W^+_\mu W^-_\nu + W^+_\nu W^-_\mu + 2 Z_\mu Z_\nu\right)\right]\nn \\
&-&  \frac{i e^2}{2\cos\th_W} H \left[Z_\mu\left(W^+_\nu \f^- - W^-_\nu \f^+\right)
+ Z_\nu\left(W^+_\mu \f^- - W^-_\mu \f^+\right)\right]\nn \\
&+& \frac{e^2}{4\sin^2\th_W}\f^2\left(W^+_\mu W^-_\nu + W^+_\nu W^-_\mu + 2 Z_\mu Z_\nu\right)
+  \frac{e^2}{\sin\th^2_W}\f^+ \f^- \left(W^+_\mu W^-_\nu + W^+_\nu W^-_\mu\right)\nn \\
&-&  \frac{i e^2}{2\sin\th_W} H \left[A_\mu\left(W^-_\nu \f^+ - W^+_\nu \f^-\right)
+ A_\nu\left(W^-_\mu \f^+ - W^+_\mu \f^- \right)\right]\nn \\
&+&   \frac{e^2}{2\cos\th_W} \f \left[Z_\mu\left(W^+_\nu \f^- + W^-_\nu \f^+\right)
+ Z_\nu\left(W^+_\mu \f^- + W^-_\mu \f^+\right)\right] \nn \\
&-&   \frac{e^2}{2\sin\th_W} \f \left[ A_\mu\left(W^-_\nu \f^+ + W^+_\nu \f^-\right)
+ A_\nu\left(W^-_\mu \f^+ + W^+_\mu \f^- \right)\right]\nn \\
&+&  e^2\cot^2 2\th_W \f^+ \f^- Z_\mu Z_\nu + e^2 \f^+ \f^- A_\mu A_\nu
+   2 e^2\cot 2\th_W \f^+ \f^- \left(A_\mu Z_\nu + A_\nu Z_\mu \right)\, .
\eeqa
In the Higgs Lagrangian $\mathcal L_{Higgs}$ and in the third line of the previous equation we have bilinear mixing terms involving the massive gauge bosons and their Goldstone. These terms will be canceled in the $R_{\xi}$ gauge by the EMT coming from the gauge-fixing contribution.

\subsection{Contributions from the Yukawa couplings}
The expression of the contributions coming from the Yukawa couplings are derived from the Lagrangian
\beq \mathcal{L}_{Yukawa} = \mathcal{L}^l_{Yukawa} + \mathcal{L}^q_{Yukawa}\, ,\eeq
where the lepton part is given by
\beq \label{termineYukawa}\mathcal{L}^l_{Yukawa} = -\lambda_e \bar L \, \mathcal H \, \psi_e^R - \lambda_e \, \bar\psi_e^R \, \mathcal H^\dag \, L,\,
\eeq
while the quarks give
\beq
\mathcal{L}^q_{Yukawa} =
- \lambda_d \, \bar Q \, \mathcal H \, \psi_d^R
- \lambda_d \, \bar \psi_d^R \, \mathcal H^\dag \, Q
- \lambda_u \, Q_i \, \e^{ij}\mathcal H^\ast_j \, \psi_u^R
- \lambda_u \, \bar \psi_u^R (\e^{ij}\mathcal H^\ast_j)^\dag \, Q_i \, .\eeq
In the previous expressions the coefficients $\lambda_e$, $\lambda_u$ and $\lambda_d$ are the Yukawa couplings, $L = (\psi_{\nu_e} \,\, \psi_e)_L$ and $Q = (\psi_u \, \, \psi_d)_L$ are the lepton and quark $SU(2)$ doublet while the suffix $R$ on the spinors identifies their right components.
The contribution from this sector to the total EMT is then given by
\beqa \label{TEIYukawa}
T_{\mu\nu}^{Yukawa}
& = & -\h_{\mu\nu}\mathcal{L}_{Yukawa} \nn \\
& = & \h_{\mu\nu}\bigg\{ m_e\bar \psi_e \psi_e + m_u \bar \psi_u \psi_u + m_d \bar \psi_d \psi_d
      + i\,\frac{e}{\sqrt{2}\sin\th_W}\bigg[\frac{m_e}{M_W} \left(\f^-\bar \psi_e P_L \psi_{\nu_e} - \f^+\bar \psi_{\nu_e} P_R \psi_e\right)\nn\\
&+&     \frac{m_d}{M_W} \left(\f^-\bar \psi_d P_L \psi_u - \f^+\bar\psi_u P_R \psi_d \right)
      + \frac{m_u}{M_W} \left(\f^+\bar\psi_u P_L \psi_d - \f^-\bar\psi_d P_R \psi_u\right)\bigg]\nn\\
&+&     i\,\frac{e}{2\sin\th_W}\bigg[\frac{m_e}{M_W}\f\left(\bar \psi_e P_R \psi_e - \bar\psi_e P_L \psi_e \right)
      + \frac{m_d}{M_W}\f\left(\bar \psi_d P_R \psi_d - \bar \psi_d P_L \psi_d \right) \nn \\
&+& \frac{m_u}{M_W}\f\left(\bar \psi_u P_L \psi_u - \bar\psi_u P_R \psi_u \right)\bigg]
+     \frac{e \, H}{2\sin\th_W \, M_W} \bigg[m_e \bar \psi_e \psi_e + m_d \bar \psi_d \psi_d
      + m_u \bar \psi_u \psi_u\bigg] \bigg\}\, .
\eeqa
In the expression above we have used standard conventions for the chiral projectors $P_{R\,,L} = ({1 \pm \gamma^5})/{2}$. For simplicity we consider only one generation of fermions.
\subsection{Contributions from the gauge-fixing terms}
The contribution of the gauge-fixing Lagrangian can be computed is a similar way. We will work in the
$R_\xi$ gauge where we choose for simplicity the same gauge-fixing parameter $\xi$ for all the gauge sectors. In this case we obtain (see also appendix (\ref{conventions}))
\beqa\label{TEIgaugefixing}
T^{ g.fix. }_{\mu\nu}
& = & \frac{1}{\xi}\bigg\{G^a_\nu\pd_\mu(\pd^\si G^a_\si) + G^a_\mu\pd_\nu(\pd^\si G^a_\si)\nn\\
&+&   A_\nu\pd_\mu(\pd^\si A_\si) + A_\mu\pd_\nu(\pd^\si A_\si)
  +    Z_\nu\pd_\mu(\pd^\si Z_\si) + Z_\mu\pd_\nu(\pd^\si Z_\si)\nn \\
&+&  \frac{1}{2}\bigg[W^+_\mu\pd_\nu(\pd^\si W^-_\si) + W^+_\nu\pd_\mu(\pd^\si W^-_\si) +
       W^-_\mu\pd_\nu(\pd^\si W^+_\si) + W^-_\nu\pd_\mu(\pd^\si W^+_\si)\bigg] \bigg\}\nn \\
&-&  \h_{\mu\nu}\bigg\{-\frac{1}{2\xi}(\pd^\si A_\si)^2 - \frac{1}{2\xi}(\pd^\si Z_\si)^2
  -   \frac{1}{\xi}(\pd^\si W^+_\si)(\pd^\r W^-_\r) - \frac{1}{2\xi}(\pd^\s G^a_\s)^2\nn \\
&+&  \frac{1}{\xi}\pd^\r(A_\r\pd^\si A_\si) + \frac{1}{\xi}\pd^\r(Z_\r\pd^\si Z_\si)
  +   \frac{1}{\xi}\pd^\r\left[W^+_\r\pd^\si W^-_\si + W^-_\r\pd^\si W^+_\si\right]\nn\\
&+&  \frac{1}{\xi}\pd^\r(G^a_\r\pd^\s G^a_\s)\bigg\}
  +   \h_{\mu\nu}\frac{\xi}{2}M^2_Z \f\f + \h_{\mu\nu}\xi M^2_W \f^+ \f^- \nn \\
&-&   M_Z(Z_\mu \pd_\nu\phi + Z_\nu \pd_\mu\phi)
    - M_W(W^+_\mu\pd_\nu\phi^- + W^+_\nu\pd_\mu\phi^- + W^-_\mu\pd_\nu\phi^+ + W^-_\nu\pd_\mu\phi^+ )\, .
\eeqa
\subsection{The ghost contributions}
Finally, from the ghost Lagrangian one obtains the ghost contribution to the EMT, which is given by
\beqa\label{TEIGhost}
T^{ghost}_{\mu\nu}
& = &  -\h_{\mu\nu}\mathcal{L}_{ghost} +   \pd_\mu\bar{c}^a \left(\pd_\nu\d^{ac} + g_s f^{abc}G^b_\nu\right)c^c
  +    \pd_\nu\bar{c}^a \left(\pd_\mu\d^{ac} + \alpha_s f^{abc}G^b_\mu\right)c^c\nn \\
&+&   \pd_\mu\bar{\h}^Z\pd_\nu\h^Z + \pd_\nu\bar{\h}^Z\pd_\mu\h^Z
  +    \pd_\mu\bar{\h}^A\pd_\nu\h^A + + \pd_\nu\bar{\h}^A\pd_\mu\h^A\nn \\
&+&   \pd_\mu\bar{\h}^+\pd_\nu\h^- + \pd_\nu\bar{\h}^+\pd_\mu\h^-
  +    \pd_\mu\bar{\h}^-\pd_\nu\h^+ + \pd_\nu\bar{\h}^-\pd_\mu\h^+\nn \\
&+&   i g \bigg\{\pd_\mu\bar{\h}^+\left[W_\nu^+ (\cos\th_W \h^Z + \sin\th_W\h^A)
  -    (\cos\th_W Z_\nu + \sin\th_W A_\nu)\h^+ \right] \nn \\
&+&   \pd_\nu\bar{\h}^+\left[W_\mu^+ (\cos\th_W \h^Z + \sin\th_W\h^A)
  -    (\cos\th_W Z_\mu + \sin\th_W A_\mu)\h^+ \right]\nn \\
&+&   \pd_\mu\bar{\h}^-\bigg[\h^- (\cos\th_W Z_\nu + \sin\th_W A_\nu)
  -    (\cos\th_W\h_Z + \sin\th_W \h_A)W_\nu^-\bigg]\nn \\
&+&   \pd_\nu\bar{\h}^-\bigg[\h^- (\cos\th_W Z_\mu + \sin\th_W A_\mu)
  -    (\cos\th_W\h_Z + \sin\th_W \h_A)W_\mu^-\bigg]\nn \\
&+&   \pd_\mu(cos\th_W\bar{\h}^Z + \sin\th_W\bar{\h}^A )\left[W^+_\nu\h^-
  -    W^-_\nu\h^+\right]\nn \\
&+&   \pd_\nu(cos\th_W\bar{\h}^Z + \sin\th_W\bar{\h}^A )\left[W^+_\mu\h^-
  -    W^-_\mu\h^+\right]\bigg\}\, ,
\eeqa
where $c^a$, $\eta^A$, $\eta^Z$ and $\eta^{\pm}$ are respectively the ghost of the gluon, photon, $Z$ and $W^{\pm}$ bosons while $\mathcal L_{ghost}$ is the SM Lagrangian for the ghost fields defined in appendix (\ref{conventions}).
\subsection{The EMT from the terms of improvement}
The terms of improvement contribute with an EMT of the form
\bea
T^I_{\mu\nu} = - \frac{1}{3} \bigg[ \partial_{\mu} \partial_{\nu} - \eta_{\mu\nu} \, \Box \bigg] \mathcal H^\dag \mathcal H = - \frac{1}{3} \bigg[ \partial_{\mu} \partial_{\nu} - \eta_{\mu\nu} \, \Box \bigg] \bigg( \frac{H^2}{2} + \frac{\phi^2}{2} + \phi^{+}\phi^{-} + v \, H \bigg).
\eea
\section{The integrated anomaly and the nonlocal action}

Before dealing with the actual computation of the various vertices of the neutral currents sector involving one insertion of
the EMT, we briefly review the issue of the extraction of the anomaly poles from these correlators, in order to render our treatment self-contained. We proceed from the QED case and then move to QCD.

We recall that the expression of the trace anomaly \cite{Duff:1993wm}

\beq
T_\mu^\mu= -\frac{1}{8} \left[ 2 b \,C^2 + 2 b' \left( E - \frac{2}{3}\square R\right) + 2 c\, F^2\right]
\label{anomalyeq}
\eeq
brings in the problem of defining an appropriate action whose EMT satisfies Eq. \ref{anomalyeq}. Such an action,
obtained by integration of the anomaly (anomaly-induced action) can be searched for by trial and error and is, in general, nonlocal. The solution was given by Riegert long ago \cite{Riegert:1984kt} in the form
\beqa
&& \hspace{-.6cm}S_{anom}[g,A] = \frac {1}{8}\int d^4x\sqrt{-g}\int d^4x'\sqrt{-g'} \left(E - \frac{2}{3} \square R\right)_x
 G_4(x,x')\left[ 2b\,C^2
 + b' \left(E - \frac{2}{3} \square R\right) + 2c\, F_{\mu\nu}F^{\mu\nu}\right]_{x'}
\label{Riegertactions}
\eeqa
where $b$, $b'$ and $c$ are parameters.  For the case of a single fermion in an abelian gauge theory they are given by $b = 1/320 \, \pi^2$,  $b' = - 11/5760 \, \pi^2$,
and $c= -e^2/24 \, \pi^2$. $C^2$ is the square of the Weyl tensor and $E$ is the Euler density given by
\beqa
C^2 &=& C_{\lambda\mu\nu\rho}C^{\lambda\mu\nu\rho} = R_{\lambda\mu\nu\rho}R^{\lambda\mu\nu\rho}
-2 R_{\mu\nu}R^{\mu\nu}  + \frac{R^2}{3}  \\
E &=& ^*\hskip-.1cm R_{\lambda\mu\nu\rho}\,^*\hskip-.1cm R^{\lambda\mu\nu\rho} =
R_{\lambda\mu\nu\rho}R^{\lambda\mu\nu\rho} - 4R_{\mu\nu}R^{\mu\nu}+ R^2.
\eeqa
 The notation $G_4(x,x')$ denotes the Green's function of the
differential operator defined by
\beq
\Delta_4 \equiv  \nabla_\mu\left(\nabla^\mu\nabla^\nu + 2 R^{\mu\nu} - \frac{2}{3} R g^{\mu\nu}\right)
\nabla_\nu = \square^2 + 2 R^{\mu\nu}\nabla_\mu\nabla_\nu +\frac{1}{3} (\nabla^\mu R)
\nabla_\mu - \frac{2}{3} \square R.\,
\label{operator4}
\eeq
As shown in \cite{Mottola:2006ew,Giannotti:2008cv} 
Performing repeated variations of the "anomaly induced" action \ref{Riegertactions} with respect to the background metric $g_{\mu\nu}$ and to the 
$A_{\alpha}$ gauge field, here taken as a background, one can reproduce the anomalous contribution of correlators 
with multiple insertions of the EMT or of gauge currents. Notice that an anomaly-induced action does not reproduce the homogeneous contributions to the anomalous trace Ward identity, which require an independent computation in order to be identified. Obviously, as the rank of the correlator increases the perturbative study of these correlation functions becomes more and more involved. Notice also that such an action does not account for all those terms which are responsible for the explicit breaking of scale invariance. In the case of the Standard Model such terms are obviously present in the spontaneosly broken phase of the theory and provide important corrections to the anomalous correlators. 

An important issue concerns the reformulation of this action in such a way that its interactions become local. This important point 
has been analyzed in \cite{Mottola:2006ew}. The authors introduce two scalar fields $\varphi$ and $\psi$ which satisfy fourth order differential equations
\bes\bea
&& \Delta_4\, \varphi = \frac{1}{2} \left(E - \frac{2}{3} \square R\right)\,,
\label{auxvarphi}\\
&& \Delta_4\, \psi = \frac{1}{2}C_{\lambda\mu\nu\rho}C^{\lambda\mu\nu\rho} 
+ \frac{c}{2b} F_{\mu\nu}F^{\mu\nu} \,,
\label{auxvarpsi}
\eea
\label{auxeom}
\ees
\hspace{-.35cm}
which allow to express the nonlocal action in the local form 
\bea
S_{anom} = b' S^{(E)}_{anom} + b S^{(F)}_{anom} + \frac{c}{2} \int\,d^4x\,\sqrt{-g}\ 
F_{\mu\nu}F^{\mu\nu} \varphi\,,
\label{allanom}
\eea
where
\bea
&& S^{(E)}_{anom} \equiv \frac{1}{2} \int\,d^4x\,\sqrt{-g}\ \left\{
-\left(\square \varphi\right)^2 + 2\left(R^{\mu\nu} - \frac{R}{3} g^{\mu\nu}\right)(\nabla_{\mu} \varphi)
(\nabla_{\nu} \varphi) + \left(E - \frac{2}{3} \square R\right) \varphi\right\}\,;\nonumber\\
&& S^{(F)}_{anom} \equiv \,\int\,d^4x\,\sqrt{-g}\ \left\{ -\left(\square \varphi\right)
\left(\square \psi\right) + 2\left(R^{\mu\nu} - \frac{R}{3}g^{\mu\nu}\right)(\nabla_{\mu} \varphi)
(\nabla_{\nu} \psi)\right.\nonumber\\
&& \qquad\qquad\qquad + \left.\frac{1}{2} C_{\lambda\mu\nu\rho}C^{\lambda\mu\nu\rho}\varphi +
\frac{1}{2} \left(E - \frac{2}{3} \square R\right) \psi \right\}\,.
\label{SEF}
\eea
\vspace{-.4cm}

\noindent
The equations of motion for $\psi$ and $\varphi$  (\ref{auxvarpsi}) can be obtained by varying \ref{allanom} with respect to these fields. Notice that in momentum space, these equations, being quartic, show the presence of a double pole in the corresponding energy momentum tensor. This can be defined, as usual, by varying \ref{allanom} with respect to the background metric. The reduction of this double pole to a single pole has been discussed in the same work, using a perturbative formulation of the local action around the flat metric background. In particular the field $\varphi$ has to be assumed of being of first order in the metric fluctuation $h_{\mu\nu}$. With this assumption, the quartic pole is reduced to a single pole and the action takes the simpler form

\bea
S_{anom}[g,A]  \rightarrow  -\frac{c}{6}\int d^4x\sqrt{-g}\int d^4x'\sqrt{-g'}\, R_x
\, \square^{-1}_{x,x'}\, [F_{\alpha\beta}F^{\alpha\beta}]_{x'}.
\label{SSimple}
 \eea
Notice that this action is valid to first order in metric variations around flat space. Its local expression is given by
\bea
S_{anom} [g,A;\varphi,\psi'] =  \int\,d^4x\,\sqrt{-g} 
\left[ -\psi'\square\,\varphi - \frac{R}{3}\, \psi'  + \frac{c}{2} F_{\alpha\beta}F^{\alpha\beta} \varphi\right].
\label{effact}
\eea
The equations of motion of the auxiliary fields are also now of second order and take the form
\bes\bea
&&\psi' \equiv  b \square\, \psi\,, \label{tpsidef}\\
&&\square\,\psi' =  \frac{c}{2}\, F_{\alpha\beta}F^{\alpha\beta} \label{tpsieom}\,,\\
&&\square\, \varphi = -\frac{R}{3}\,.
\label{phieom}\eea  
\ees
$R$, in the equations above, is the linearized version of the Ricci scalar
 \beq
 R_\equiv \partial^x_\mu\, \partial^x_\nu \, h^{\mu\nu} - \square \,  h, \qquad h=\eta_{\mu\nu} \, h^{\mu\nu}.
 \eeq
 A similar approach can be followed in the case of the chiral anomaly \cite{Armillis:2008bg,Coriano:2008pg,Giannotti:2008cv}.

A perturbative test of the pole structure identified in the anomaly induced action is obtained by a direct computation of the correlator $TAA$, with the insertion of the EMT on the photon 2-point function  $(AA)$ at nonzero momentum transfer. This test has been performed in QED \cite{Giannotti:2008cv,Armillis:2009pq} and generalized to QCD in the 2-gluon case
\cite{Armillis:2010qk}. The advantage of a complete computation of the correlator, respect to the variational solution found by inspection, is that it gives the possibility of extracting also the mass corrections to the pole behaviour \cite{Armillis:2009pq}.
In fact, anomaly-induced actions analogous to Eq. (\ref{Riegertactions}) are not available for spontaneosuly broken gauge theories coupled to gravity. The origin of the pole contribution in the effective action can be attributed to a special region of the triangle diagram 
- which is responsible for the generation of the trace anomaly at perturbative level - in momentum space. This region is identified by a computation of the spectral denstity $\rho(s)$ of this diagram which turns out to be proportional to a delta function $(\delta(s))$, with 
$s$ denoting the virtuality of the graviton line (see the discussion in \cite{Giannotti:2008cv}). A similar behaviour of the spectral density is found for the anomaly loop \cite{Dolgov:1971ri}. 

\begin{figure}[t]
\centering
\includegraphics[scale=0.8]{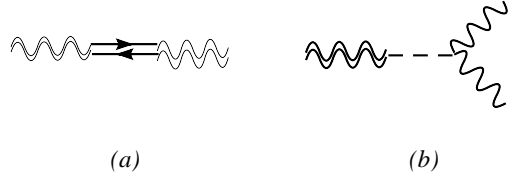}\hspace{1cm}
\label{poles}
\caption{The pole contribution as a collinear exchange of a fermion/antifermion pair (a) and the diagrammatic representation of the anomaly pole (b).}
\label{poleb}
\end{figure}

The kinematical region which is responsible for such behaviour is briefly illustrated in Fig. \ref{poleb}. For instance, in the QED case this singular spectral density is generated when we set on-shell the two fermion lines of the anomaly loop, cut in the $s$-channel.  In this configuration the virtual graviton decays into two on-shell fermions which move collinearly before reaching the final state, where they decay into two photons (Fig. \ref{poleb} (a)). The exchange of a simple pole (Fig. \ref{poleb} (b)) accounts for the contribution coming from this kinematical region, and should be viewed as a dynamical effect. The similarity between the gravitational and the chiral case is indeed rather striking, since the decay of an axial vector current into two vector currents, which is the source of the axial anomaly, can be equally described by a diagram similar to (Fig. \ref{poleb} (b)), with the role of the scalar exchange taken by an interpolating field with the quantum numbers of the pion, and the graviton replaced by an axial-vector current.

\section{The master equation of the Ward identities}
\label{mastersection}
In this section we proceed with the derivation of the Ward identities describing the conservation of the EMT starting from the case of a simple model, containing a scalar, a gauge field and a single fermion in a curved spacetime and then moving to the case of the full SM Lagrangian. In both cases we start with the derivation of two master equations from which the Ward identities satisfied by a specific correlator can be extracted by functional differentiations.

We denote with $S[V_{\underline{a}}^\mu,\f,\psi,A_\mu]$ the action of the model. Its expression depends on the vielbein, the fermion field $\psi$, the complex scalar field $\phi$ and the abelian gauge field $A_\mu$. We can use this action and the vielbein to derive a useful form of the EMT
\beq \Theta^{\mu\nu} = - \frac{1}{V}\frac{\d S}{\d V^{{\underline{a}}}_\mu}V^{{\underline{a}}\nu},\eeq
in terms of the determinant of the vielbein $V(x){\equiv} \left|V^{\underline{a}}_\mu(x)\right|$.
Notice that this expression of the EMT is non-symmetric. The symmetric expression can be easily defined by the relation
 \beq T^{\mu\nu} = \frac{1}{2}(\Theta^{\mu\nu} + \Theta^{\nu\mu})\,
 \label{tmunu} \eeq
that will be used below.\\
We introduce the generating functional of the model, given by
\beqa\label{Z} Z[V,J,J^\dag,J^\mu,\c,\bar\c]
&=& \int \mD\f\mD\f^\dag\mD\psi\mD\bar{\psi}\mD A_\mu\,\mbox{exp}\bigg\{i S[V,\f,\psi,A_\mu]  + i\int d^4x\, \bigg[J^\dag(x)\f(x) + \f^\dag(x)J(x)\nn\\
&+& \bar{\c(x)}\psi(x) + \bar{\psi}(x)\c(x) + J^\mu(x)A_\mu(x)\bigg]\bigg\}\, ,\eeqa
where we have denoted with $J(x)$, $J^\mu(x)$ and $\chi(x)$ the sources for the scalar, the gauge field and the spinor field respectively.
We will exploit the invariance of $Z$ under diffeomorphisms for the derivation of the corresponding Ward identities.
For this purpose we introduce a condensed notation to denote the functional integration measure of all the fields
\beqa
\label{SourcesToymodel}
\mD \Phi {\equiv} \mD\f\mD\f^\dag\mD\psi\mD\bar{\psi}\mD A_\mu\,
\eeqa
and redefine the action with the external sources included
\beqa
\label{ActionToyModel}
\tilde{S}&            =          & S + i\int d^4x\,\left( J^\mu A_\mu + J^\dag(x)\f(x) + \bar{\c}(x)\psi(x) + \textrm{h.c.}\right).\,
\eeqa
Notice that we have absorbed a factor $\sqrt{-g}$ in the definition of the sources, which clearly affects their transformation under changes of
coordinates (see also appendix (\ref{ward})).

The condition of diffeomorphism invariance of the generating functional $Z$ gives
\beq\label{covW}
Z[V,J,J^\dag,\c,\bar{\c},J^\mu] = Z[V',J',J^{'\,\dag},\c',\bar{\c}',J^{'\,\mu}]\, ,\eeq
where we have allowed an arbitrary change of coordinates $x^{'\,\mu} = F^\mu(x)$ on the spacetime manifold, which can be
parameterized locally as $x^{'\,\mu} = x^\mu + \e^\mu(x)$.
 The measure of integration is invariant under such changes $( \mD\Phi' = \mD\Phi)$ and we obtain to first order in $\e^\mu(x)$
\beqa\label{preWard}
\int \mD\Phi \, e^{i \tilde{S}}
& = & \int \mD\Phi \quad e^{i \tilde{S}}\,\bigg( 1  + i\int d^4xd^4y\, \bigg\{- V\Theta^\mu_{\,\,\,\underline{a}}\bigg[-\d^{(4)}(x-y)\pd_\nu V^{\underline{a}}_\mu(x)
- [\pd_\mu\d^{(4)}(x-y)]V^{\underline{a}}_\nu\bigg]\nn\\
&& - \pd_\nu[\d^{(4)}(x-y)J^\dag(x)]\f(x) - \f^\dag(x)\pd_\nu[\d^{(4)}(x-y)J(x)]\nn\\
&& - \pd_\nu[\d^{(4)}(x-y)\bar{\c}(x)]\psi(x) - \bar{\psi}(x)\pd_\nu[\d^{(4)}(x-y)\c(x)] \nn\\
&& - [\pd_\nu[J^\mu(x)\d^{(4)}(x-y)] + [\pd_\r\d^{(4)}(x-y)]\d^\mu_\nu J^\r(x)]A_\mu(x)\bigg\}\e^\nu(y)\bigg).
\eeqa
This expression needs some further manipulations in order to be brought into a convenient form for the perturbative test. Using some results of appendix \ref{ward} we rewrite it in an equivalent form and then perform the flat space-time limit to obtain
\beqa\label{Ward}
 &&  \int\mD\Phi\, e^{i\tilde{S}}\, \bigg[\pd_\a T^{\a\b}(y) - J^\dag(y)\pd^\b\f(y)
            - \pd^\b\f^\dag(y)J(y)\nn  -  J^\a(y)\pd^\b A_\a(y) + \pd_\a[J^\a(y) A^\b(y)]\nn\\
  && -  \pd^\b\bar{\psi}(y)\c(y) - \bar{\c}(y)\pd^\b\psi(y)  -      \frac{1}{2}\pd_\a\bigg(\frac{\d S}{\d\psi(y)}\s^{\a\b}\psi(y)
            - \bar{\psi}(y)\s^{\a\b}\frac{\d S}{\d\bar{\psi}(y)}\bigg)\bigg] = 0\,.
\eeqa
A more general derivation is required in the case in which we have a theory which is SM-like, where we have more fields to consider. The master formula that one obtains is slightly more involved, but its structure is similar. Before specializing the derivation to the neutral sector of the SM we discuss the Ward identity for the amputated Green functions obtained from this functional integral.
\subsection{The master equation for connected and 1PI graphs}
We can extend the above analysis by deriving a different form of the master equation in terms of the generating functional of the connected graphs ($W$) or, equivalently, directly
in terms of the effective action ($\Gamma$), which collects all the 1-particle irreducible (1PI) graphs.  The Ward identities for the various correlators are then obtained starting from these master expressions via functional differentiation. For this purpose we extend the generating functional given in (\ref{Z}) by coupling the model to a weak external gravitational field $h_{\alpha\beta}$
\beqa
Z[J,J^\dag,J^{\mu},\c,\bar{\c},h_{\a\b}] =  \int \mD \Phi \, \exp\bigg\{i \tilde S - i \frac{\kappa}{2} \int d^4x \, h^{\alpha\beta}(x) \, T_{\alpha\beta}(x) \bigg\}.
\eeqa
The generating functional of connected graphs is then given by
\beq\label{W}\exp{\bigg\{i W[J,J^\dag,J^{\mu},\c,\bar{\c},h_{\a\b}]\bigg\}} =
\frac{Z[J,J^\dag,J^{\mu},\c,\bar{\c},h_{\a\b}]}{Z[0]}\,,
\eeq
normalized respect to the vacuum functional $Z[0]$. From this we obtain the relations
\beqa\label{Legendre1}
\phi_c(x) = \frac{\delta W }{\delta J^\dag(x)}\, , \quad \phi^\dag_c(x) = \frac{\delta W }{\delta J(x)}\, , \quad
\psi_c(x) = \frac{\delta W }{\delta \bar{\c}(x)}\, ,\quad
\bar{\psi}_c(x) = \frac{\delta W}{\delta\c(x)}\, ,\quad
A^{\mu}_c(x) = \frac{\delta W }{\delta J_{\mu}(x)}
\eeqa
for the classical fields of the theory, identified by a subscript "c". The effective action is then defined via the usual Legendre transform of the fields except for the gravitational source $h_{\alpha\beta}$
\beqa\label{EffectiveAction}
\Gamma[\phi_c,\phi^\dag_c,A^{\mu}_c,\psi_c,\bar{\psi}_c,h_{\alpha\beta}]
&=& W[J,J^\dag,J^{\mu},\c,\bar{\c},h_{\a\b}] - \int d^4x\,\bigg[J^\dag(x)\phi_c(x) + \phi_c^\dag(x) J(x)\nn\\
&+& \bar{\psi}_c(x)\c(x) + \bar{\psi}_c(x)\c(x)  + J^{\mu}(x)A_{c\,\mu}(x) \bigg]
\eeqa
which satisfies the relations
\beqa\label{Legendre2}
\frac{\d\Gamma}{\d \phi_c(x)} =  - J^\dag(x) ,\quad
\frac{\d\Gamma}{\d \phi^\dag_c(x)} = - J(x) , \quad
\frac{\d\Gamma}{\d\psi_c(x)} = - \bar{\c}(x) , \quad
\frac{\d\Gamma}{\d \bar{\psi}_c(x)} = - \c(x) ,\quad
\frac{\d\Gamma}{\d A_{c\,\mu}(x)} = - J^{\mu}(x) .
\eeqa
Notice that the functional derivatives of both $W$ and $\Gamma$ respect to the classical background field $h_{\alpha\beta}$ coincide
\beq\label{deltah} \frac{\d W}{\d h_{\a\b}(x)} = \frac{\d\Gamma}{\d h_{\a\b}(x)}\, .\eeq
Therefore, the Ward identity (\ref{Ward}) can be rewritten in terms of the connected functional integral as
\beqa\label{WardW}
\pd_\a\frac{\d W}{\d h_{\a\b}}
&=& -\frac{\kappa}{2}\bigg\{J^\dag\pd^{\b}\frac{\d W}{\d J^\dag} + \pd^{\b}\frac{\d W}{\d J}J  + \pd^{\b}\frac{\d W}{\d J_{\mu}}J_{\mu}
    - \pd_{\a}\bigg( \frac{\d W}{\d J_{\b} }J^{\a}\bigg) \nn \\
&+& \bar{\c}\pd^{\b}\frac{\d W}{\d \bar{\c}} +
    \pd^{\b} \frac{\d W}{\d\c}\c - \frac{1}{2}\pd_{\a}\bigg(\bar \c \s^{\a\b} \frac{\d W}{\d
    \bar{\c}}- \frac{\d W}{\d\c}\s^{\a\b}\c\bigg)
 \bigg\}\, ,
\eeqa
or equivalently in terms of the 1PI generating functional
\beqa\label{WardGamma}
\pd_\a\frac{\d\Gamma}{\d h_{\a\b}}
&=& - \frac{\kappa}{2}\bigg\{- \frac{\d\Gamma}{\d \phi_c}\pd^{\b}\phi_c - \pd^{\b}\phi_c^\dag\frac{\d\Gamma}{\d \phi_c^\dag}
 - \frac{\d\Gamma}{\d A_{c\,\a}}\pd^{\b}A_{c\,\a}
- \pd_{\a}\bigg(\frac{\d\Gamma}{\d A_{c\,\a}}
    A^{\b}_c\bigg)\nn\\
&-&  \pd^{\b}\bar{\psi}_c \frac{\delta\Gamma}{\d\bar{\psi}_c} +
        \frac{\d\Gamma}{\d\psi}_c\psi_c\pd^{\b}\psi_c
        - \frac{1}{2}\pd_{\a}\bigg(\frac{\d\Gamma}{\d\psi}_c\s^{\a\b}\psi_c
    - \bar{\psi}_c\s^{\a\b}\frac{\d \Gamma}{\d\bar{\psi}_c}\bigg)
\bigg\}\, ,
\eeqa
having used  (\ref{Legendre1}), (\ref{Legendre2}), (\ref{deltah}),(\ref{WardW}).\\

\subsection{The Ward identity for $TVV'$}
In the case of the $TVV'$ correlator in the Standard Model the derivation of the Ward identity requires two functional differentiations of (\ref{WardGamma}) (extended to the entire spectrum of SM) respect to the classical fields $V^\a_c(x_1)$ and $V^{'\,\b}_c(x_2)$ where $V$ and $V'$ stand for the two neutral gauge bosons $A$ and $Z$, obtaining
\beqa
\label{WardTVV}
 -i\frac{\kappa}{2}\pd^{\mu}\langle T_{\mu\nu}(x) V_\a (x_1) V'_{\b} (x_2)\rangle_{amp}
&=& - \frac{\kappa}{2}\bigg\{- \pd_\nu\d^{(4)}(x_1-x) P^{-1\,VV'}_{\a\b}(x_2,x) \nn \\
&& \hspace{-6cm} - \pd_\nu\d^{(4)}(x_2-x) P^{-1\,V V'}_{\a\b}(x_1,x)
+ \pd^\mu[ \h_{\a\nu} \d^{(4)}(x_1-x) P^{-1\,V V'}_{\b\mu}(x_2,x) + \h_{\b\nu}\delta^{(4)}(x_2-x) P^{-1\,V V'}_{\a\mu}(x_1,x)]\bigg\}
\eeqa
where we have introduced the (amputated) mixed 2-point function
\beq P^{-1\,VV'}_{\a\b}(x_1,x_2) = \langle 0| T V_{\a}(x_1) V'_{\b}(x_2) | 0 \rangle_{amp}
= \frac{\delta^2 \Gamma}{\delta V^{\alpha}_c(x_1)\d V'^{\b}_c(x_2) }\, .\eeq
After a Fourier transform
\beqa
(2\pi)^4\d^{(4)}(k-p-q)\Gamma^{VV'}_{\mu\nu\a\b}(p,q)&=&
 -i\frac{\kappa}{2}\int d^4zd^4xd^4y\, \langle T_{\mu\nu}(z) V_{\a}(x) V'_{\b}(y)\rangle_{amp}\,
e^{-ikz + ipx + iqy}\, ,\eeqa
Eq. (\ref{WardTVV}) becomes
\beqa\label{WardmomTVV}
k^{\mu}\Gamma^{VV'}_{\mu\nu\a\b}(p,q)
&=& - \frac{\kappa}{2}\bigg\{k^\mu P^{-1\,VV'}_{\a\mu}(p) \h_{\b\nu}
+ k^\mu P^{-1\,VV'}_{\b\mu}(q) \h_{\a\nu}  - q_\nu P^{-1\,VV"}_{\a\b}(p) - p_\nu P^{-1\,VV'}_{\a\b}(q)\bigg\} \, .
\eeqa
The perturbative test of this relation, computationally very involved, as well as of all the other relations that we will derive in the next sections, is of paramount importance for determining the structure of the interaction vertex.

\section{BRST symmetry and Slavnov-Taylor identities}
\label{BRSTsection}
Before coming to the derivation of the STI's which will be crucial for a consistent definition of the $TVV$ correlator
for the Lagrangian of the SM, we give the BRST variation of the EMT in QCD and in the electroweak theory which will be used in the following. \\
The QCD sector gives
\beqa
\delta T_{\mu\nu}^{QCD} = \frac{1}{\xi}\left[ A_{\mu}^i \partial_{\nu} \partial^{\rho}
D^{ij}_{\rho}c^j + A_{\nu}^i \partial_{\mu} \partial^{\rho} D^{ij}_{\rho}c^j -
\eta_{\mu\nu}\partial^{\sigma}(A_{\sigma}^i \partial^{\rho}D^{ij}_{\rho}c^j ) \right] \,,
\eeqa
with $i,j$ being color indices in the adjoint representation of $SU(3)$, while

in the electroweak sector and in the interaction basis we have
\beqa\label{deltaTewbrstI}
\d T_{\mu\nu}^{e.w.}  =  \frac {1}{\xi}\bigg[W^r_\mu\pd_\nu\d \mathcal F^r
                           + W^r_\nu\pd_\mu\d \mathcal F^r
                           + B_\mu\pd_\nu\d \mathcal F^0
                           + B_\nu\pd_\mu \d \mathcal F^0\bigg]
                      - \eta_{\mu\nu}\frac{1}{\xi}\pd^\r\bigg[W^r_\r\d \mathcal F^r
                           + B_\r \d \mathcal F^0\bigg].
\eeqa
Here the indices $r$ and $0$ refer respectively to the $SU(2)$ and $U(1)$ gauge groups and can be expanded directly in the basis of the mass eigenstates (i.e. a=(+,-, A, Z)). We obtain
\beqa\label{deltaTewbrst}
\d T_{\mu\nu}^{e.w.}  &=&  \frac {1}{\xi}\bigg[W^+_\mu\pd_\nu\d \mathcal F^-
                           + W^-_\mu\pd_\nu \d \mathcal F^+
                           + A_\mu\pd_\nu\d \mathcal F^A + Z_\mu\pd_\nu \d \mathcal F^Z
                           + (\mu \leftrightarrow \nu) \bigg] \nn \\
                      &-& \frac {1}{\xi}\h_{\mu\nu}\pd^\r\bigg[W^+_\r\d \mathcal F^-
                           + W^-_\r \d \mathcal F^+
                           + A_\r\d \mathcal F^A + Z_\r\d \mathcal F^Z \bigg].
\eeqa

To proceed with the derivation of the STI's for the SM, we start introducing the generating functional of the theory in the presence of a background gravitational field $h_{\mu\nu}$ (also denoted as "$h$")
\beqa
Z(h,J) =  \int \mD \Phi \, \exp\bigg\{i \tilde S - i \frac{\kappa}{2} \int d^4x \, h^{\mu\nu}(x) \, T_{\mu\nu}(x) \bigg\}
\label{Zhj}
\eeqa
where $\tilde S$ denotes the action of the Standard Model $(S)$ with the inclusion of the external sources $(J, \omega, \xi)$ coupled to the SM fields
\beqa
\tilde S = S + \int d^4 x \, \left( J^{{a}}_\mu A^{\mu\,{a}} + \bar\w^{{a} }\eta^{{a}} + \bar\eta^{{a}} \w^{{a}} + \bar\xi^i \psi^i  + \bar \psi^i \xi^i \right),
\eeqa
with $a = A, Z, +, -$ and i which runs over the fermion fields.
We also define the functional describing the insertion of the EMT on the vacuum amplitude
\beqa
Z^T_{\mu\nu}(J; z) \equiv  \langle T_{\mu\nu}(z) \rangle_J =  \int \mD \Phi \, T_{\mu\nu}(z) \, \exp{i \tilde S}
\eeqa
where $Z^T_{\mu\nu}(J; z)$ is related to $Z(h,J)$ by
\beqa
-i \frac{\kappa}{2} Z^T_{\mu\nu}(J; z) = \frac{\delta}{\delta \, h^{\mu\nu}(z)} Z(h,J) \bigg |_{h = 0} \,.
\eeqa
 The STI's of the theory are obtained by using the invariance of the functional average under a change of integration variables
\beqa \label{ZT}
Z^T_{\mu\nu}(J; z) = \int \mD \Phi \, T_{\mu\nu}(z) \, \exp{i \tilde S} = Z^{T}_{\mu\nu}(J; z)^\prime = \int \mD \Phi' \, T'_{\mu\nu}(z) \, \exp{i \tilde S'}
\eeqa
which leaves invariant the quantum action $S$. These transformations, obviously, are the ordinary BRST variations of the fundamental fields of the theory. The integration measure is clearly invariant under these transformations and one obtains
\beqa \label{ST1}
\int \mD \Phi \, \exp{i \tilde S} \, \bigg\{ \d T_{\mu\nu}(z) + i \, T_{\mu\nu}(z) \int d^4 x \bigg[ J^{a}_\mu \d A^{\mu \, a} + \bar\w^{{a} } \d \eta^{{a}} + \d \bar\eta^{{a}} \w^{{a}} + \bar\xi^i \d \psi^i  + \d \bar \psi^i \xi^i \bigg] \bigg\} = 0 \,,
\eeqa
where the operator $\d$ is the BRST variation of the various fields, which is given in appendix (\ref{appendixBRST}).

The STI' s are then derived by a functional differentiation of the previous identity with respect to the sources. We just remark that since the BRST variations
increase the ghost number of the integrand by 1 unit, we are then forced to differentiate respect to the source of the antighost field in order go back to a zero ghost number in the integrand. This allows to extract correlation functions which are not trivially zero.  This procedure, although correct, may however generate STI' s among different correlators which are rather involved. For this reason we will modify the generating functional $Z^T_{\mu\nu}(J; z)$ by adding to the argument of the exponential extra contributions proportional to the product of the gauge fixing functions $\mathcal F^a(x)$ and of the
corresponding sources  $\chi^a(x)$. Therefore,  we redefine the action $\tilde{S}$ as $\tilde{S}_\chi$
\beqa
\tilde S_\chi \equiv \tilde S + \int d^4 x \,  \chi^{{a} }\mathcal F^{{a}} .
\eeqa
The condition of invariance of the generating functional that will be used below for the extraction of the STI's then becomes
\beq \label{ST2}
\int \mD \Phi \, \exp{i \tilde S} \, \bigg\{ \d T_{\mu\nu}(z) + i \, T_{\mu\nu}(z) \int d^4 x \bigg[ J^{a}_\mu \d A^{\mu \, a} + \bar\w^{{a} } \d \eta^{{a}} + \d \bar\eta^{{a}} \w^{{a}} + \bar\xi^i \d \psi^i  + \d \bar \psi^i \xi^i  + \chi^a \d \mathcal F^a \bigg] \bigg\} = 0 \,.
\eeq
The implications of BRST invariance on the correlator $TVV'$ are obtained
by functional differentiation of (\ref{ST2}) respect to the source $\chi^{a}(x)$ of the gauge-fixing function $\mathcal F^{a}$ and to the source  $\w^{a}(y)$ coupled to the antighost fields $\bar \eta^{a}$. For this reason in the following we set to zero the other external fields.
\subsection{STI for the $TAA$ correlator}
Eq. (\ref{ST2}) can be used in the derivation of the
STI's for the $TAA$ correlator by setting appropriately to zero all the components of the external sources except some of them. For instance, if only the sources in the photon sector $(\omega^A, \chi^A)$ are non-vanishing,
this equation becomes
\beq \label{STTA2}
 \int \mD \Phi \, \exp \left[i \, S + i \int d^4 x \left( \bar \eta^{A}\w^{A}  + \chi^{A} \mathcal F^{A} \right) \right]
 \bigg\{ \d T_{\mu\nu}(z) + i \, T_{\mu\nu}(z) \int d^4 x \, \bigg( - \w^{A} \frac{1}{\xi}\mathcal F^{A} + \chi^{A} \mathcal E^{A} \bigg)\bigg\}=0, \\
\eeq
where the function $\mathcal{E}^A$ denotes the finite part of the BRST variation (with the infinitesimal Grassmann parameter $\lambda$ removed) of the gauge-fixing function of the photon $\mathcal F^A$
\beqa
\label{photoneq}
\mathcal E^A(x) = \d \mathcal F^A(x) = \Box \, \eta^A + i\, e\, \partial^{\mu}\left( W^-_\mu \, \eta^+ - W^+_\mu \, \eta^- \right) \, .
\eeqa
Functional differentiating this relation with respect to $\chi^A(x)$ and $\w^A(y)$ and then setting to zero the external sources,
we obtain the STI for the  $\langle T A A \rangle$ correlator
\beqa \label{STTAA1}
\frac{1}{\xi}\langle T_{\mu\nu}(z)\pd^\a A_\a(x)\pd^\b A_\b(y) \rangle = \langle T_{\mu\nu}(z) \mathcal E^A(x) \bar{\h}^A(y) \rangle + \langle \d T_{\mu\nu}(z)\pd^\a A_\a(x)\bar{\h}^A(y) \rangle \,.
\eeqa
Its right-hand side can be simplified using the fields equation of motion. The BRST variation of $\mathcal F^A$, given by $\mathcal E^A$,  is indeed the equation of motion for the ghost of the photon.
This can be easily derived by computing the change of the action under a small variation of the antighost field of the photon
$\bar \eta^A$
\beqa \label{trasfbareta}
\bar \eta^A(x) \rightarrow \bar \eta^A(x) + \epsilon(x),
\eeqa
which gives, integrating by parts,
\beqa
\mathcal L \rightarrow  \mathcal L + ( \partial^{\mu} \epsilon) \left( \partial_{\mu} \eta^A + i \, e ( W^-_\mu \, \eta^+ - W^+_\mu \, \eta^- ) \right) = \mathcal L - \epsilon \, \d \mathcal F^A \,,
\eeqa
and the equation of motion $\d \mathcal F^A(x) = \mathcal E^A(x) = 0$.\\
The first correlator on the right hand side of Eq. (\ref{STTAA1}) can be expressed in terms of simpler correlation functions using the invariance of the generating functional $Z^T_{\mu\nu}(z)$ given in (\ref{ZT}) under the transformation (\ref{trasfbareta}). One obtains
\beqa \label{varbaretaZT1}
Z^T_{\mu\nu}(z)^\prime  =  \int\mD \Phi \, e^{i \, \tilde S} \exp \bigg\{i \int d^4x \, \epsilon(x) \bigg[ -\mathcal E^A(x) + \w^A(x) \bigg] \bigg\}\bigg(T_{\mu\nu}(z) + \d_{\bar \eta^A}T_{\mu\nu}(z)\bigg) =  Z^T_{\mu\nu}(z)
\eeqa
where $\d_{\bar \eta^A}T_{\mu\nu}(z)$ denotes the variation of the EMT under the transformation (\ref{trasfbareta})
\beqa
\d_{\bar \eta^A} T_{\mu\nu}(z) &=& \pd_{\mu} \epsilon(z) [\pd_\nu\h^A + i e (W_\nu^-\h^+ - W_\nu^+\h^-)](z) +  (\mu \leftrightarrow \nu) \nn \\
&-& \eta_{\mu\nu}\pd^\r \epsilon(z) [\pd_\r\h^A + i e(W^-_\r\h^+ - W^+_\r\h^-)](z)\,.
\eeqa
This equation can be formally rewritten as an integral expression in the form
\beqa \label{varTbareta}
\d_{\bar \eta^A} T_{\mu\nu}(z) &=& \int d^4 x \,  \epsilon(x) \, \bar \delta_{\bar \eta^A} T_{\mu\nu}(z,x) \,,
\eeqa
where $\bar \delta T_{\mu\nu}(z,x)$ has been defined as
\beq
\bar{\d}T_{\mu\nu}(z,x) = \h_{\mu\nu}\pd_x^\r(\d^{(4)}(z - x)\,D_\r^{A}\h^A(x)) - \pd^x_\mu(\d^{(4)}(z - x)\, D^{A}_\nu \h^A(x))
- \pd^x_\nu(\d^{(4)}(z - x)\,D^{A}_\mu \h^A(x)) \,.
\label{deltaT}
\eeq
We have used the notation $D_\r^{A}\h^A$ to denote the covariant derivative of the ghost of the photon
\beqa
D_\r^{A}\h^A(x) = \pd_\r\h^A(x) + i e (W_\r^-\h^+ - W_\r^+\h^-)(x)
\eeqa
and its four-divergence equals the equation of  motion of the ghost $\eta^A$
\beqa
\partial^\r D_\r^{A}\h^A(x) = \mathcal E^A(x).
\eeqa
Using Eq. (\ref{varTbareta}) and expanding to first order in $\epsilon$, the identity in (\ref{varbaretaZT1}) takes the form \beqa \label{EM1}
\int\mD \Phi \, e^{i\, \tilde S}\bigg\{ T_{\mu\nu}(z) \bigg[-\mathcal E^A(x) + \w^A(x) \bigg] - i \, \bar{\d}T_{\mu\nu}(z,x)\bigg\}  =  0.
\eeqa
This relation represents the functional average of the equations of motion of the ghost $\eta^A$. As such, it can be used to derive the implications of the ghost equations on the correlation functions which are extracted from it.

For instance, to derive a relation for the first correlation function appearing on the rhs of Eq. (\ref{STTAA1}), it is sufficient to take a functional derivative of  (\ref{EM1}) respect to $\omega^A(y)$
\beqa
\langle T_{\mu\nu}(z) \mathcal E^A(x) \bar \eta^A(y) \rangle = - i \langle \bar \delta_{\bar \eta^A}
 T_{\mu\nu}(z,x) \bar \eta^A(y) \rangle - i \delta^{(4)}(x-y) \langle T_{\mu\nu}(z) \rangle.
\label{GreenF1}
\eeqa
Notice that the term proportional to $\delta^{(4)}(x-y)$ corresponds to a disconnected diagram and as such can be dropped
in the analysis of connected correlators. We can substitute in (\ref{GreenF1}) the explicit form of $\bar \delta T_{\mu\nu}(z,x)$, rewriting it in terms of the 2-point function of the covariant derivative of the ghost $\eta^A$ ($D^{A}_\r\eta^A$) and of the antighost $\bar\eta^A$
\beqa \label{Firstterm1}
\langle T_{\mu\nu}(z) \mathcal E^A(x) \bar \eta^A(y) \rangle & = &   -  i \bigg\{ 
\eta_{\mu\nu}\, \pd^\r_x \left[\d^{(4)}(z - x) \langle (D^{A}_\r\eta^A(x)\,\bar \eta^A(y) \rangle \right] \nn \\
&& - \left( \pd_\mu^x \left[ \d^{(4)}(z - x) \langle D^{A}_\nu\eta^A(x)\, \bar\eta^A(y)\rangle \right] + (\mu \leftrightarrow \nu) \right) \bigg\}.
\eeqa
The correlation functions involving the covariant derivative of the ghost and of the antighost, appearing on the right-hand side of
(\ref{Firstterm1}), are related - by some STI's - to derivatives of the photon 2-point function. We leave the proof of this point to appendix (\ref{appendixBRST}) and just quote the result. Then Eq. (\ref{Firstterm1}) becomes
\beqa \label{Firstterm2}
\langle T_{\mu\nu}(z) \mathcal E^A(x) \bar \eta^A(y) \rangle & = &   
- \frac{i}{\xi}\bigg\{ \eta_{\mu\nu} \, \pd^\r_x \left[\d^{(4)}(z - x) \partial^\alpha_y \langle A_\r(x) \, A_\alpha(y) \rangle \right] \nn \\
&& - \left( \pd_\mu^x \left[ \d^{(4)}(z - x) \partial^\alpha_y \langle A_\nu(x) \, A_\alpha(y)\rangle \right] + (\mu \leftrightarrow \nu) \right) \bigg\}.
\eeqa

\begin{figure}[t]
\centering
\includegraphics[scale=0.8]{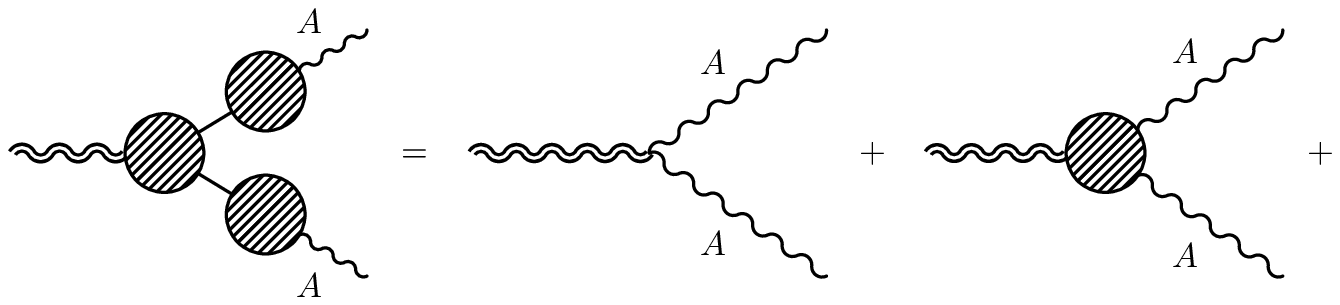}
\includegraphics[scale=0.8]{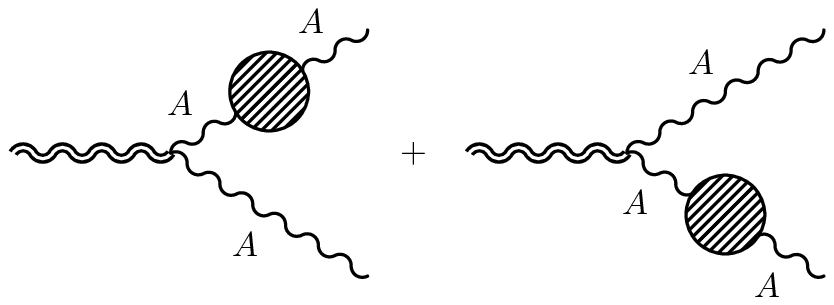}
\caption{One-loop loop decomposition of  $G_{\mu\nu\alpha\beta}^{AA}(p,q)$ in terms of the amputated function $\Gamma_{\mu\nu\alpha\beta}^{AA}(p,q)$ and of 2-point functions on the external legs. \label{Fig.greenAA}}
\end{figure}

Having simplified the first of the two functions on the right hand side of (\ref{STTAA1}), we proceed with the analysis of the second one, containing the BRST variation of the EMT, which can be expressed as a combination of BRST variations of the gauge-fixing functions $\mathcal F^{a}$
\beqa\label{deltaTewbrstM}
\d T_{\mu\nu} & = & \frac {1}{\xi}\bigg[W^+_\mu\pd_\nu\d F^- + W^-_\mu\pd_\nu \d F^+
                           + A_\mu\pd_\nu\d F^A + Z_\mu\pd_\nu \d F^Z + (\mu \leftrightarrow \nu) \bigg]\nn \\
                     && - \frac {1}{\xi}\h_{\mu\nu}\pd^\r\bigg[W^+_\r\d F^- + W^-_\r \d F^+ + A_\r\d F^A + Z_\r\d F^Z \bigg].
\eeqa
Similarly to the photon case, where $\d \mathcal F^A$ is proportional to the equation of motion of the corresponding ghost, also in this more general case we have
\beqa
\d \mathcal F^{{r} }= \mathcal E^{{r}}  \quad {r} = +,-,A,Z \,
\eeqa
and $\delta T_{\mu\nu}$ can be rewritten in the form
\beqa\label{deltaTewbrstM2}
\d T_{\mu\nu} & = & \frac {1}{\xi}\bigg[W^+_\mu\pd_\nu \mathcal E^- + W^-_\mu\pd_\nu \mathcal E^+
                           + A_\mu\pd_\nu \mathcal E^A + Z_\mu\pd_\nu \mathcal E^Z + (\mu \leftrightarrow \nu) \bigg]\nn \\
                     && - \frac {1}{\xi}\h_{\mu\nu}\pd^\r\bigg[W^+_\r \mathcal E^- + W^-_\r \mathcal E^+ + A_\r\mathcal E^A + Z_\r\mathcal E^Z \bigg].
\eeqa
The appearance of the operators $\mathcal E^{r}$ in the expression above suggests that Eq. (\ref{STTAA1}) can
be simplified if we derive STI's involving the equations of motion of the ghost fields. Therefore, we proceed with a functional average of the equation of motions of the ghosts
\beq\label{ghosteom}
\int \mD \Phi \, e^{i \, \tilde S_\chi}\bigg[- \mathcal E^{{r}}(z) + \w^{{r}}(z)\bigg] = 0 \quad r = +,-,A,Z \, .
\eeq
The terms appearing in Eq. (\ref{deltaTewbrstM2}) are obtained by acting on this generating functional with appropriate differentiations. For instance, to reproduce the term $\partial W^+ \mathcal E^-$ we take a functional derivative of (\ref{ghosteom}) with respect to the source $J^{a}_\r(z)$ followed by a differentiation respect to $z^\r$ obtaining
\beqa
\pd^\r_z\,\frac{\d}{\d J^{a}_\r(z)}  \int \mD \Phi \, e^{i \, \tilde S_\chi}\bigg[- \mathcal E^{{r}}(z) + \w^{{r}}(z)\bigg] = i \int \mD \Phi \, e^{i\, \tilde S_\chi}\bigg[- \pd_z^\r \left(A^{a}_\r(z) \mathcal E^{{r}}(z) \right) +
\pd^\r_z \left( A^{a}_\r(z)\w^{{r}}(z) \right)\bigg] = 0\,. \nn \\
\label{eom1}
\eeqa
At this stage we need to take a derivative respect to the source $\chi^A(x)$ and to the source $\omega^A(y)$ of the antighost field $\bar\eta^A$
\beq
\label{eom2}
\int \mD \Phi \, e^{i\, \tilde S_\chi}\bigg[ \pd_z^\r \left( A^a_\r(z) \mathcal E^r(z) \right)  \partial^\alpha A_\alpha(x) \bar \eta^A(y) + i \, \delta^{{{r}}A}\pd^\r_z \left(A_\r(z) \d^{(4)}(z - y) \right)\pd^\a A_\a(x)\bigg] = 0\, .
\eeq
In the expression above the Kronecher $\delta^{{r}A}$  is 1 for ${r}=A$ and 0 for ${r}= +,-, Z$.
This shows that in $\delta T_{\mu\nu}$ in (\ref{deltaTewbrstM2}) only the photon contributes to the $\langle \d T_{\mu\nu}(z)\pd^\a A_\a(x)\bar{\h}^A(y) \rangle$ correlator and gives
\beqa \label{Secondterm}
\langle \d T_{\mu\nu}(z)\pd^\a A_\a(x)\bar{\h}^A(y) \rangle & = &  -\frac{i}{\xi}   \bigg\{\pd_\nu^z \d^{(4)}(z-y)  \pd_x^\a \langle A_\mu(z) A_\a(x) \rangle
 +  \pd_\mu^z \d^{(4)}(z-y)  \pd_x^\a \langle A_\nu(z) A_\a(x) \rangle  \nn\\
&& - \eta_{\mu\nu}\pd^\r_z  \left( \d^{(4)}(z-y) \pd_x^\a  \langle A_\r(z) A_\a(x) \rangle \right) \bigg\} \, .
\eeqa
Using the results of (\ref{Firstterm2}) and (\ref{Secondterm}) in (\ref{STTAA1}) we obtain a simple expression for the STI, just in terms of derivatives of the photon 2-point function
\beqa
 \frac{1}{\xi}\langle T_{\mu\nu}(z)\pd^\a A_\a(x)\pd^\b A_\b(y) \rangle &=& 
- \frac{i}{\xi}\bigg\{ \eta_{\mu\nu} \, \pd^\r_x \left[\d^{(4)}(z - x) \partial^\alpha_y \langle A_\r(x) \, A_\alpha(y) \rangle \right] \nn \\
&& \hspace{-4cm} - \eta_{\mu\nu}\pd^\r_z  \left[ \d^{(4)}(z-y) \pd_x^\a  \langle A_\r(z) A_\a(x) \rangle \right]
 - \bigg( \pd_\mu^x \left[ \d^{(4)}(z - x) \partial^\alpha_y \langle A_\nu(x) \, A_\alpha(y)\rangle \right]  \nn \\
&& \hspace{-4cm} -  \pd_\mu^z \d^{(4)}(z-y)  \pd_x^\a \langle A_\nu(z) A_\a(x) \rangle  + (\mu \leftrightarrow \nu)  \bigg)   \bigg\} \, ,
\eeqa
which in momentum space becomes
\beqa
p^\a \, q^\b \, G_{\mu\nu\alpha\beta}^{AA}(p,q) &=&  \frac{\kappa}{2} \, q^\a \, \bigg\{  p_\mu \, P^{AA}_{\nu\a}(q) + p_\nu \, P^{AA}_{\mu\a}(q)  - \eta_{\mu\nu} p^\r \, P^{AA}_{\r\a}(q) \bigg\} \nn \\
&+&  \frac{\kappa}{2} \, p^\a \, \bigg\{  q_\mu \, P^{AA}_{\nu\a}(p) + q_\nu \, P^{AA}_{\mu\a}(p)  - \eta_{\mu\nu} (p+q)^\r \, P^{AA}_{\r\a}(p) \bigg\}
\label{STAA0}
\eeqa
having defined
\beqa \label{STAA1}
(2 \pi)^4 \d^{(4)}(k-p-q) \, G_{\mu\nu\alpha\beta}^{AA}(p,q) &=& -i \frac{\kappa}{2} \int d^4 z \,d^4 x \,d^4 y \, \langle T_{\mu\nu}(z) A_{\alpha}(x) A_{\beta}(y) \rangle \, e^{-i k \cdot z + i p \cdot x + i q \cdot y} \,, \nn  \\
(2 \pi)^4 \d^{(4)}(p-q) \, P^{AA}_{\a\b}(p) & = &  \int d^4 x \,d^4 y \, \langle  A_{\alpha}(x) A_{\beta}(y) \rangle \, e^{i p \cdot x - i q \cdot y} \,.
\eeqa
The STI given in  (\ref{STAA0}) involves the Green function $G_{\mu\nu\alpha\beta}^{AA}(p,q)$ which differs from the vertex function $\Gamma_{\mu\nu\alpha\beta}^{AA}(p,q)$ for the presence of propagators on the external vector lines. In the one-loop approximation the decomposition of $G_{\mu\nu\alpha\beta}^{AA}(p,q)$ in terms of vertex and external lines corrections simplifies, as illustrated in Fig.(\ref{Fig.greenAA}). In momentum space this takes the form
\beqa \label{TAAGreen}
G_{\mu\nu\alpha\beta}^{AA}(p,q) &=&  V^{hAA}_{\mu\nu\sigma\rho}(p,q) \,
{P^{AA}_0}^{\sigma}_{\alpha}(p) \, {P^{AA}_0}^{\rho}_{\beta}(q)+ \Gamma_{\mu\nu\sigma\rho}^{AA, \,
1}(p,q) \, {P^{AA}_0}^{\sigma}_{\alpha}(p) \, {P^{AA}_0}^{\rho}_{\beta}(q) \nn\\
&+&  V^{hAA}_{\mu\nu\si\rho}(p,q) \, {P^{AA}_1}^{\sigma}_{\alpha}(p) \, {P^{AA}_0}^{\rho}_{\beta}(q)
+ V^{hAA}_{\mu\nu\si\rho}(p,q) \, {P^{AA}_0}^{\sigma}_{\alpha}(p) \, {P^{AA}_1}^{\rho}_{\beta}(q)
\, ,  \eeqa
where $V^{hAA}_{\mu\nu\sigma\rho}(p,q)$ is the tree level graviton-photon-photon interaction vertex defined in appendix \ref{FeynRules}.
The right-hand-side of Eq. (\ref{STAA0}) can be rewritten in the form
\beqa\label{STAA3}
p^\a \, q^\b \, G_{\mu\nu\alpha\beta}^{AA}(p,q)
&=& - i\frac{\kappa}{2}\frac{\xi}{q^2}\bigg\{ p_\mu q_\nu + p_\nu q_\mu - \eta_{\mu\nu} p \cdot q \bigg\}  - i \frac{\kappa}{2} \frac{\xi}{p^2}\bigg\{ q_\mu p_\nu + q_\nu p_\mu
      - \eta_{\mu\nu}( p \cdot q + p^2) \bigg\} \nn\\
&=& \frac{(-i \, \xi)^2}{p^2\, q^2} \, p^\a \, q^\b \, V^{hAA}_{\mu\nu\a\b}(p,q) \eeqa
which implies, together with (\ref{TAAGreen}), that
\beqa \label{STAA2}
p^\a \, q^\b \, \Gamma_{\mu\nu\a\b}^{AA, \, 1}(p,q)  = 0.
\eeqa
This is the Slavnov-Taylor identity satisfied by the one-loop vertex function.
\subsection{STI for the $T A Z $ correlator}\label{BRST TAZ}
The derivation of the STI for $TAZ $ follows a pattern similar to the $TAA$ case.
The starting point is the condition of BRST invariance of the generating functional given in Eq.(\ref{ST1}).
Also in this case we introduce some auxiliary sources $\chi^a(x)$ for the gauge-fixing terms, but we differentiate
(\ref{ST2}) with respect to
$\chi^A(x)$ and to the source $\w^Z(y)$ of the antighost $\bar\h^Z(y)$, and then set all the sources to zero.
We obtain a relation similar to Eq. (\ref{STTA2}), that is
\beqa \label{ST3AZ}
\int \mD \Phi \, \exp \left[i \, S + i \int d^4 x \left( \bar\eta^Z \w^Z
+ \chi^A \mathcal F^A \right) \right]
 \bigg\{ \d T_{\mu\nu}(z) + i \, T_{\mu\nu}(z) \int d^4 x \,
\bigg( - \w^Z \frac{1}{\xi}\mathcal F^Z + \chi^A \mathcal E^A \bigg)\bigg\} = 0 \, , \nn \\
\eeqa
where $\mathcal E^A(x)$, the operator describing the equation of motion of the photon, has been defined in
(\ref{photoneq}). Therefore, by taking a derivative with respect to $\chi^A(x)$ and to $\w^Z(y)$ we obtain

\beq\label{STAZ} \frac{1}{\xi}\langle T_{\mu\nu}(z\mathcal )F^A(x)\mathcal F^Z(y)\rangle =
\langle T_{\mu\nu}(z)\mathcal E^A(x) \bar\h^Z(y)\rangle
+ \langle \d T_{\mu\nu}(z)\mathcal F^A(x)\bar\h^Z(y)\rangle \, .\eeq
The right-hand-side of this equation can be simplified using the equation of motion for the ghost of the photon on
$Z^T_{\mu\nu}(J;z)$.

We start from the first of the two correlators $\langle T_{\mu\nu}(z)\mathcal E^A(x) \bar\h^Z(y)\rangle$.
Using the invariance of $Z^T_{\mu\nu}(J;z)$  respect to the variation (\ref{trasfbareta})
of the antighost of the photon $\bar\h^A$ and expressing $\d_{\bar\h^A}T_{\mu\nu}(z)$ as in Eqs. (\ref{varTbareta})
and (\ref{deltaT}), we obtain Eq. (\ref{EM1}). At this point we differentiate this relation respect to the source $\w^Z(y)$ obtaining
\beq\label{STAZfirstterm} \langle T_{\mu\nu}(z)\mathcal E^A(x) \bar\h^Z(y)\rangle
= -i \langle \bar\d_{\bar\h^A} T_{\mu\nu}(z,x)\bar\h^Z(y)\rangle
- i\d^{(4)}(x-y)\langle T_{\mu\nu}(z)\rangle\, .\eeq
As in the previous case, we omit the term which is proportional to the vev of the EMT, since this generates only disconnected diagrams. The explicit form of $\bar\d_{\bar\h^A} T_{\mu\nu}(z,x)$
allows to express Eq. (\ref{STAZfirstterm}) in the form
\beqa \label{STAZfirstterm1}
\langle T_{\mu\nu}(z) \mathcal E^A(x) \bar \eta^Z(y) \rangle
&=& - i \bigg\{\eta_{\mu\nu}\, \pd^\r_x \left[\d^{(4)}(z - x) \langle D^{A}_\r\eta^A(x)\,\bar
\eta^Z(y) \rangle \right] \nn \\
&& \hspace{-4cm} - \bigg(\pd_\mu^x \left[ \d^{(4)}(z - x) \langle D^{A}_\nu\eta^A(x)\,
\bar\eta^Z(y)\rangle \right]
 +  \pd_\nu^x \left[\d^{(4)}(z - x) \langle D^{A}_\mu\eta^A(x)\,
\bar\eta^Z(y)\rangle\right] \bigg) \bigg\}\, . \eeqa
To express
$\langle T_{\mu\nu}(z)\mathcal E^A(x) \bar\h^Z(y)\rangle$ in terms of 2-point functions and of their derivatives, we use the
identity
\beq \label{FZA}
\langle\bar\h^Z(y) D^A_\alpha\h^A(x)\rangle = \frac{1}{\xi}\langle \mathcal F^Z(y)A_\alpha(x)\rangle = 0\, ,
\eeq
which is proved in appendix (\ref{appendixBRST}).
This equation relates the correlators in Eq. (\ref{STAZfirstterm1}) to two-point functions involving the photon and the
gauge-fixing function of the $Z$ gauge boson $\mathcal F^Z$.
Using (\ref{FZA}), we then conclude that
\beq \langle T_{\mu\nu}(z) \mathcal E^A(x) \bar \eta^Z(y) \rangle = 0\, .\eeq
To complete the simplification of (\ref{STAZ}) we need to re-express
$\langle \d T_{\mu\nu}(z)\mathcal F^A(x)\bar\h^Z(y)\rangle$ in terms of 2-point functions. This correlation function involves
the BRST variation of the EMT, defined in (\ref{deltaTewbrstM2}), which contains a linear combination of operators proportional to the equations of motion of the ghosts. For this reason it is more convenient to start from the same equations
functionally averaged as in (\ref{ghosteom}), and then proceed with further differentiations, as shown in Eq. (\ref{eom1}). Finally, we perform a functional differentiation of (\ref{eom1}) respect to the sources
$\chi^A(x)$ and $\w^Z(y)$, analogously to Eq. (\ref{eom1}), thereby obtaining the relation
\beqa
\int \mD \Phi \, e^{i\, \tilde S_\chi}\,\bigg[ \pd_z^\r \left( A^a_\r(z) \mathcal E^r(z) \right)
\partial^\alpha A_\alpha(x) \bar \eta^Z(y) + i \, \delta^{rZ}\pd^\r_z \left(A_\r(z) \d^{(4)}(z - y)
\right)\pd^\a A_\a(x)\bigg] = 0 \, .
\eeqa
Following this procedure for all the terms of $\d T_{\mu\nu}(z)$ we obtain
\beqa\label{SecondTermAZ} \langle\d T_{\mu\nu}(z)\pd^\a A_\a(x)\bh^Z(y)\rangle
&=& -\frac{i}{\xi} \bigg\{-\h_{\mu\nu}\pd^\s_z\bigg[\d^{(4)}(z-y)\langle
    Z_\s(z)\pd^\a A_\a(x)\rangle\bigg] \nn\\
&-& \pd_\nu^z\d^{(4)}(z-y) \langle Z_\mu(z)\pd^\a A_\a(x)\rangle  - \pd_\mu^z\d^{(4)}(z-y) \langle Z_\nu(z)\pd^\a A_\a(x)\rangle \bigg\}\, . \eeqa
Given that this is the only non-vanishing correlator on the right-hand-side of Eq. (\ref{STAZ}),
we conclude that the BRST relation that we have been searching for can be expressed in the form
\beqa\label{BRSTfinalcoordAZ}
\frac{1}{\xi}\langle T_{\mu\nu}(z) \mathcal F^A(x)\mathcal F^Z(y)\rangle
&=& - \frac{i}{\xi}\bigg\{-\h_{\mu\nu}\pd^\s_z\bigg[\d^{(4)}(z-y)\langle
       Z_\s(z)\pd^\a A_\a(x)\rangle\bigg] \nn\\
&+& \pd_\nu^z\d^{(4)}(z-y) \langle Z_\mu(z)\pd^\a A_\a(x)\rangle + \pd_\mu^z\d^{(4)}(z-y) \langle Z_\nu(z)\pd^\a A_\a(x)\rangle\bigg\}\, . \eeqa
Notice that on the left-hand-side of this identity, differently from the case of $T A A$, appear the gauge fixing functions of the photon and of the $Z$ gauge bosons
\beqa
\mathcal F^A  =  \pd^\s A_\s\, ,\qquad \qquad
\mathcal F^Z  =  \pd^\s Z_\s - \xi M_Z \f\, ,
\eeqa
which give
\beq
\langle T_{\mu\nu}(z) \mathcal F^A(x)\mathcal F^Z(y)\rangle =
\langle T_{\mu\nu}(z)\pd^\a A_\a(x)\pd^\b Z_\b(y)\rangle
-\xi M_Z\langle T_{\mu\nu}(z) \pd^\a A_\a(x) \phi(y)\rangle\, ,\eeq
where  $\phi$ is the Goldstone of the $Z$.
Going to momentum space, with the inclusion of an overall $-i{\kappa}/{2}$ factor we define
\beqa
\label{FourierAZ1}
(2 \pi)^4 \d^{(4)}(k-p-q) \, G^{AZ}_{\mu\nu\alpha\beta}(p,q) &=&
 -i \frac{\kappa}{2} \int d^4 z \,d^4 x \,d^4 y \,\langle T_{\mu\nu}(z)
A_{\alpha}(x) Z_{\beta}(y) \rangle \, e^{-i k \cdot z + i p \cdot x + i q \cdot y}\nn \\
(2 \pi)^4 \d^{(4)}(p-q) \, P^{AZ}_{\a\b}(p) &=& \int d^4 x \,d^4 y \,
\langle A_{\alpha}(x) Z_{\beta}(y)\rangle \, e^{- i p \cdot x + i q \cdot y}\label{FourierAZ2}\,, \nn\\
(2 \pi)^4 \d^{(4)}(k-p-q)\, G^{A\phi}_{\mu\nu\a}(p,q) &=& -i \frac{\kappa}{2}\int d^4 x \,d^4 y \,
\langle  T_{\mu\nu}(z)A_{\alpha}(x)\phi(y) \rangle \, e^{-i kz + i p \cdot x + i q \cdot y} \,,
\label{FourierAZ3}
\eeqa
\begin{figure}[t]
\centering
\includegraphics[scale=0.8]{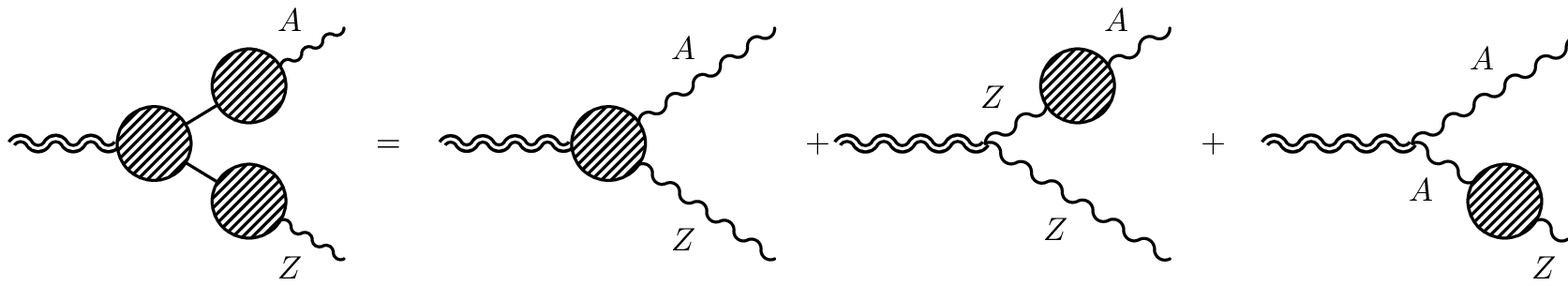}
\caption{One loop decomposition of  $G_{\mu\nu\alpha\beta}^{AZ}(p,q)$ in
terms of the amputated funtion $\Gamma_{\mu\nu\alpha\beta}^{AZ}(p,q)$ and of the corrections on the external lines. \label{Fig.greenAZ}}
\end{figure}
\begin{figure}[t]
\centering
\includegraphics[scale=0.8]{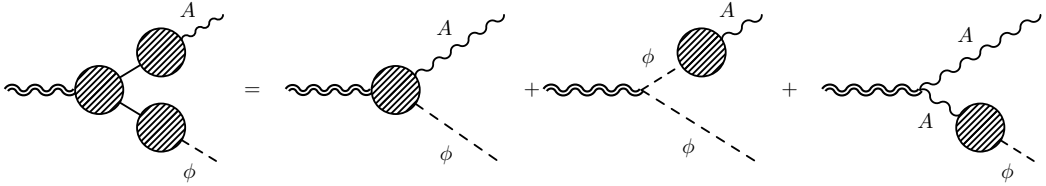}
\caption{Decomposition of $G_{\mu\nu\alpha}^{A\phi}(p,q)$ at 1-loop
in terms of the amputated correlator $\Gamma_{\mu\nu\alpha}^{A\phi,\,1}(p,q)$ and of the corrections on the external legs.\label{Fig.greenAphi}}
\end{figure}
and the final STI (\ref{BRSTfinalcoordAZ}) in momentum space reads as
\beqa\label{STAZ1}
p^\a q^\b G^{AZ}_{\mu\nu\a\b}(p,q) - i\xi M_Z p^\a G^{A\phi}_{\mu\nu\a}(p,q)
&=& \frac{\kappa}{2}\, p^\a\bigg\{q_\nu P^{ZA}_{\mu\a}(p) + q_\mu P^{ZA}_{\nu\a}(p) - \eta_{\mu\nu}(p+q)^\rho P^{ZA}_{\rho\a}(p)\bigg\}. \,\,
\eeqa
At this point, we are interested in the identification of a STI for amputated Green functions. For this purpose we perform a decomposition on the left-hand-side of this equation similarly to Eq. (\ref{TAAGreen}) for $G^{AA}_{\mu\nu\a\b}(p,q)$, working in the
1-loop approximation.
In this case, the decomposition of the $G^{AZ}_{\mu\nu\a\b}(p,q)$ correlator, shown in Fig.
(\ref{Fig.greenAZ}), is given by
\beqa \label{TAZGreen}
G_{\mu\nu\alpha\beta}^{AZ}(p,q) &=&  \Gamma_{\mu\nu\sigma\rho}^{AZ, \, 1}(p,q) \,
{P^{AA}_0}^{\sigma}_{\alpha}(p) \, {P^{ZZ}_0}^{\rho}_{\beta}(q)  + V^{hZZ}_{\mu\nu\si\rho}(p,q) \, {P^{ZA}_1}^{\sigma}_{\alpha}(p) \, {P^{ZZ}_0}^{\rho}_{\beta}(q) \nn \\
&+& V^{hAA}_{\mu\nu\si\rho}(p,q) \, {P^{AA}_0}^{\sigma}_{\alpha}(p) \, {P^{AZ}_1}^{\rho}_{\beta}(q)
\, .
\eeqa
This decomposition, differently from the one in Eq. (\ref{TAAGreen}), does not contain a tree-level contribution $V^{hAZ}_{\mu\nu\a\b}(p,q)$ since this vertex is zero at the lowest order.\\
A similar procedure has to be followed for the correlator $G^{A\phi}_{\mu\nu\a}(p,q)$. Also in this case the vertices
 $V^{hA\phi}_{\mu\nu\a}(p,q)$,
$V^{hAZ}_{\mu\nu\a\b}(p,q)$ and $V^{h\phi Z}_{\mu\nu\b}(p,q)$ are zero at tree-level.
The 3-point function
$G^{A\phi}_{\mu\nu\a}(p,q)$, shown in Fig.(\ref{Fig.greenAphi}), is then decomposed into the form
\beqa
\label{TAGoldGreen}
G^{A\phi}_{\mu\nu\a}(p,q)
&=& \Gamma^{A\phi,\, 1}_{\mu\nu\rho}(p,q)\,{P^{AA}_0}^\rho_\a(p)\, P^{\phi\phi}_0(q)  + V^{hAA}_{\mu\nu\rho\si}\,{P^{AA}_0}^\rho_\a(p)\,{P^{A\phi}_1}^\si(q)
 +  V^{h\phi\phi}_{\mu\nu}\,{P^{\phi A}_1}_\a (p)\,P^{\phi\phi}_0(q) \, .
 \eeqa
 The tree-level vertices used in Eq. (\ref{TAZGreen}) and (\ref{TAGoldGreen}) are defined in appendix \ref{FeynRules}.
The STI for this correlator is then obtained from (\ref{STAZ1}) using the decompositions in (\ref{TAZGreen}) and
(\ref{TAGoldGreen}). \\
One can show that the terms generated on the left-hand-side of (\ref{STAZ1}) by contracting tree-level vertices with the 1-loop insertions on the external legs, coincide with those generated from the right-hand side at the same order. For this one can use the expressions given in appendix (\ref{propagators}).
The result is summarized by the equation
\beq
p^\a q^\b \Gamma^{AZ,\,1}_{\mu\nu\a\b}(p,q)+ i\xi M_Zp^\a \Gamma^{A\phi,\, 1}_{\mu\nu\a}(p,q) = 0
\, , \eeq
which gives the STI at 1-loop for the amputated functions.

\subsection{STI for the $ T Z Z $ correlator}
The derivation of the STI for the $T Z Z$ follows a similar pattern.
We perform a functional derivative of (\ref{ST2}) respect to the source $\chi^Z(x)$ of
the gauge-fixing function $\mathcal F^Z(x)$ and to the source for the antighost $\bar\h^Z(y)$, which is $\w^Z(y)$.
We obtain a result quite similar to Eq.  (\ref{ST3AZ})
\beqa \label{ST3ZZ}
 \int \mD \Phi \, \exp \left[i \, S + i \int d^4 x \left( \bar\eta^Z \w^Z
+ \chi^Z \mathcal F^Z \right) \right] \bigg\{ \d T_{\mu\nu}(z) + i \, T_{\mu\nu}(z) \int d^4 x \,
\bigg( - \w^Z \frac{1}{\xi}\mathcal F^Z + \chi^Z \mathcal E^Z \bigg)\bigg\} = 0\, . \nn \\
\eeqa
Here, clearly, $\mathcal E^Z(x)$ is the operator of the equations
of motion of the ghost
$\h^Z(x)$, derived from the BRST variation of the gauge-fixing function of the $Z$ gauge boson,
\bea \d\mathcal F^Z(x) = \mathcal E^Z(x)
&=& \Box\h^Z(x) + i e\frac{\cos\th_W}{\sin\th_W}\pd^\r(W^-_\r\h^+ - W^+_\r\h^-) \nn \\
&+& \frac{\xi e M_Z}{\sin2\th_W}[(v+H)\h^Z + i\cos\th_W(\f^+\h^- - \f^-\h^+)]  \nn \\
&=& \pd^\r D_\r^Z\h^Z(x)  + \frac{\xi e M_Z}{\sin2\th_W}[(v+H)\h^Z + i\cos\th_W(\f^+\h^-
- \f^-\h^+)],\,  \eea
where we have introduced, for convenience, the covariant derivative of the ghost $\h^Z(x)$, $D_\r^Z\h^Z(x)$,
which is given by
\beq
D^Z_\r\h^Z(x) = \pd_\r\h^Z(x) + ie\frac{\cos\th_W}{\sin\th_W}(W^-_\r\h^+ - W^+_\r\h^-)(x). \eeq
Performing a functional derivative of (\ref{ST3ZZ}) respect to $\chi^Z(x)$ and $\w^Z(y)$ we obtain the equivalent of Eq. (\ref{STAZ}), which is
\beq\label{STZZ}
\frac{1}{\xi}\langle T_{\mu\nu}(z\mathcal )F^Z(x)\mathcal F^Z(y)\rangle = \langle T_{\mu\nu}(z)\mathcal E^Z(x) \bar\h^Z(y)\rangle
+ \langle \d T_{\mu\nu}(z)\mathcal F^Z(x)\bar\h^Z(y)\rangle \, .
\eeq
At this point, the correlation functions on the right-hand-side of (\ref{STZZ}) must be re-expressed in terms of
2-point functions and of their derivatives. Also in this case we use a functional average of the equations of motion of the
ghost of the $Z$ gauge boson,  $\h^Z$, on the generating functional $Z^T_{\mu\nu}(J;z)$.
For this reason we start from the correlator $\langle T_{\mu\nu}(z)\mathcal E^Z(x) \bar\h^Z(y)\rangle$
and exploit the invariance of $Z^T_{\mu\nu}(J;z)$ under the BRST variation of the antighost field $\bar\h^Z(x)$,
\beq \label{trasfbaretaZ}\bar\h^Z(x) \rightarrow \bar\h^Z(x) + \e(x)\, ,\eeq
and express the variation of the EMT $\d_{\bar\h^Z}T_{\mu\nu}(z)$ as an integral,  having factorized the parameter $\e(x)$,
\beq \d_{\bar\h^Z} T_{\mu\nu}(z) = \int d^4 x\, \e(x)\bar\d_{\bar\h^Z} T_{\mu\nu}(z;x).\, \eeq
In this case
\beqa\label{bardeltaTEI ZZ}
\bar\d_{\bar\h^Z} T_{\mu\nu}(z,x)
&=& -\pd_\mu^x[\d^{(4)}(x-z)D^Z_\nu\h^Z(x)] - \pd_\nu^x[\d^{(4)}(x-z)D^Z_\mu\h^Z(x)]\nn\\
&& + \h_{\mu\nu}\mathcal \d^{(4)}(x-z)\mathcal E^Z(x) + \h_{\mu\nu}\pd^\r_x[\d^{(4)}(x-z)]D_\r^Z\h^Z(x)\,
.\eeqa
The equation obtained by the requirement of BRST invariance of  $Z^T_{\mu\nu}(J;z)$ is
\beq  \label{eq.motghZ}
\int\mD\Phi\,e^{i\tilde S}\bigg\{T_{\mu\nu}(z)\bigg[-\mathcal E^Z(x) + \w^Z(x)\bigg] - i\bar\d T_{\mu\nu}(z,x)\bigg\} = 0 \, .
\eeq
At this point we take a functional derivative of (\ref{eq.motghZ}) respect to $\w^Z(y)$ and then set all the sources to zero, obtaining
\beq
\label{STZZfirstterm}
\langle T_{\mu\nu}(z)\mathcal E^Z(x)\bar\h^Z(y)\rangle
= -i \langle \bar\d_{\bar\h^Z} T_{\mu\nu}(z,x)\bar\h^Z(y)\rangle
-i\d^{(4)}(x-y)\langle T_{\mu\nu}(z)\rangle  \, .
\eeq
Notice that if we are looking for a STI of connected graphs, then the term $-i\langle T_{\mu\nu}(z)\rangle$ does not contribute, being a disconnected part. Expressing $\bar\d_{\bar\h^Z}T_{\mu\nu}(z;x)$ according to (\ref{bardeltaTEI ZZ}), we conclude that Eq. (\ref{STZZfirstterm}) takes the form
\beqa
\langle T_{\mu\nu}(z)\mathcal E^Z(x)\bar\h^Z(y)\rangle
&=& - i\bigg\{\h_{\mu\nu}\langle\mathcal E^Z(x)\bar\h^Z(y)\rangle\d^{(4)}(x-z) + \h_{\mu\nu}\pd^\r_x[\d^{(4)}(x-z)]\langle D_\r^Z\h^Z(x)\bar\h^Z(y)\rangle\nn\\
&-&  \pd_\mu^x\bigg[\d^{(4)}(x-z)\langle D^Z_\nu\h^Z(x)\bar\h^Z(y)\rangle\bigg] - \pd_\nu^x\bigg[\d^{(4)}(x-z)\langle D^Z_\mu\h^Z(x)\bar\h^Z(y)\rangle\bigg]\bigg\} .\,\,
\eeqa
This equation can be simplified using the identities
\beqa
\langle\bar\h^Z(y)D^Z_\r\h^Z(x)\rangle = \frac{1}{\xi}\langle\mathcal F^Z(y)Z_\r(x)\rangle, \qquad \qquad
\langle\mathcal F^Z(x)\mathcal F^Z(y)\rangle = -i\xi\delta^{(4)}(x-y) \, ,\eeqa
which are proven in appendix (\ref{appendixBRST}) and we finally obtain the relation
\beqa\label{firsttermSTZZ}
\langle T_{\mu\nu}(z)\mathcal E^Z(x)\bar\h^Z(y)\rangle
&=& - \frac{i}{\xi}\bigg\{-i\xi^2\h_{\mu\nu}\d^{(4)}(x-y)\d^{(4)}(x-z)  + \h_{\mu\nu}\pd^\r_x[\d^{(4)}(x-z)]\langle Z_\r(x)\mathcal F^Z(y)\rangle\nn\\
&-&  \pd_\mu^x\bigg[\d^{(4)}(x-z)\langle Z_\nu(x)\mathcal F^Z(y)\rangle\bigg] - \pd_\nu^x\bigg[\d^{(4)}(x-z)\langle Z_\mu(x)\mathcal F^Z(y)\rangle\bigg]\bigg\}.\,
\eeqa
To complete the simplification of Eq. (\ref{STZZ}) an appropriate reduction of the correlator $\langle \d T_{\mu\nu}(z)\mathcal F^Z(x)\bar\h^Z(y)\rangle$ is needed.
This can be achieved working as in the previous cases. We start from the equations of motion of the ghosts averaged with the functional integral $Z^T_{\mu\nu}$, and then take appropriate functional derivatives respect to the sources in order to reproduce all the terms of Eq. (\ref{deltaTewbrstM2}) containing $\mathcal F^Z(x)$ ed $\bar\h^Z(y)$.  
We obtain the intermediate relation
\beq \int \mD \Phi \, e^{i\, \tilde S}\bigg[ \pd_z^\r \left( A^a_\r(z) \mathcal E^r(z) \right)
\mathcal F^Z(x) \bar \eta^Z(y)
+ i \, \delta^{rZ}\pd^\r_z \left(A_\r(z) \d^{(4)}(z - y)\right)\mathcal F^Z(x)\bigg] = 0\, ,\eeq
$ a = A, Z, +, -$,
while the final identity is given by
\beqa\label{secondtermSTZZ}
\langle\d T_{\mu\nu}(z)\mathcal F^Z(x)\bar\h^Z(y)\rangle
&=& -\frac{i}{\xi}\bigg\{\pd_\mu^z[\d^{(4)}(z-y)]\langle Z_\nu(z)\mathcal F^Z(x)\rangle + \pd_\nu^z[\d^{(4)}(z-y)]\langle Z_\nu(z)\mathcal F^Z(x)\rangle\nn\\
&-&  \h_{\mu\nu}\pd^\r_z\bigg[\d^{(4)}(z-y)\langle Z_\r(z)\mathcal F^Z(x)\rangle\bigg]\bigg\}\, .
\eeqa
Finally, inserting into (\ref{STZZ}) the results of (\ref{firsttermSTZZ}) and (\ref{secondtermSTZZ}), we obtain
\beqa\label{STZZfinalcoord}
\frac{1}{\xi}\langle T_{\mu\nu}(z)\mathcal F^Z(x)\mathcal F^Z(y)\rangle
&=& -\frac{i}{\xi}\bigg\{-i\xi^2\h_{\mu\nu}\d^{(4)}(x-y)\d^{(4)}(x-z) + \h_{\mu\nu}\pd^\r_x\bigg[\d^{(4)}(x-z)\bigg]\langle Z_\r(x)\mathcal F^Z(y)\rangle\nn\\
&-&  \pd_\mu^x\bigg[\d^{(4)}(x-z)\langle Z_\nu(x)\mathcal F^Z(y)\rangle\bigg] - \pd_\nu^x\bigg[\d^{(4)}(x-z)\langle Z_\mu(x)\mathcal F^Z(y)\rangle\bigg]\nn\\
&+&  \pd_\mu^z\bigg[\d^{(4)}(z-y)\bigg]\langle Z_\nu(z)\mathcal F^Z(x)\rangle + \pd_\nu^z\bigg[\d^{(4)}(z-y)\bigg]\langle Z_\mu(z)\mathcal F^Z(x)\rangle\nn\\
&-&  \h_{\mu\nu}\pd^\r_z\bigg(\d^{(4)}(z-y)\langle Z_\r(z)\mathcal F^Z(x)\rangle\bigg)\bigg\} \, .
\eeqa
We then move to momentum space introducing 2 and 3-point functions, generically defined as
\beqa
(2\pi)^4\d^{(4)}(p-q)P^{\phi_l \phi_m}(p) &=& \int d^4x d^4y\,\langle \phi_l(x)\phi_m(y)\rangle e^{-i px + i qy}\,, \\
(2\pi)^4\d^{(4)}(k-p-q)G^{\phi_l \phi_m}_{\mu\nu\a\b}(p,q) &=& -i\frac{\kappa}{2}\int d^4z d^4x d^4y \, \langle
T_{\mu\nu}(z)\phi_l(x)\phi_m(y)\rangle
e^{-i kz +i px +i qy}\, ,\
\eeqa
for generic fields $\phi_l=(Z, \phi)$,  and rewrite (\ref{STZZfinalcoord}) in the form
\beqa
&& p^\a q^\b G^{ZZ}_{\mu\nu\a\b}(p,q) - i\xi M_Z p^\a G^{Z\f}_{\mu\nu\a}(p,q)
 - i\xi M_Z q^\b G^{\f Z}_{\mu\nu\b}(p,q) - \xi^2 M_Z^2 G^{\f\f}_{\mu\nu}(p,q) = \nn \\
&& \frac{\kappa}{2}\bigg\{
  i p_\mu[-i q^\b P^{ZZ}_{\nu\b}(q) - \xi M_Z P_\nu^{Z\f}(q)]
 +  i p_\nu[-i q^\b P^{ZZ}_{\mu\b}(q) - \xi M_Z P_\mu^{Z\f}(q)]\nn\\
&& + \, i q_\mu[- i p^\a P^{ZZ}_{\a\nu}(p) - \xi M_Z P^{Z\f}_\nu(p)]
 +  i q_\nu[- i p^\a P^{ZZ}_{\a\mu}(p) - \xi M_Z P^{Z\f}_\mu(p)]\nn\\
&& \, - i \h_{\mu\nu}k_\r[- iq_\b P^{\r\a}_{ZZ}(q) - \xi M_Z P^\r_{Z\f}(q)]
 -  i \h_{\mu\nu}k_\r[-ip_\a P^{\r\a}_{ZZ}(p) - \xi M_Z P^\r_{Z\f}(p)] - i\xi^2\h_{\mu\nu}\bigg\}\label{STZZmom}\, .
\eeqa
\begin{figure}[t]
\centering
\includegraphics[scale=0.8]{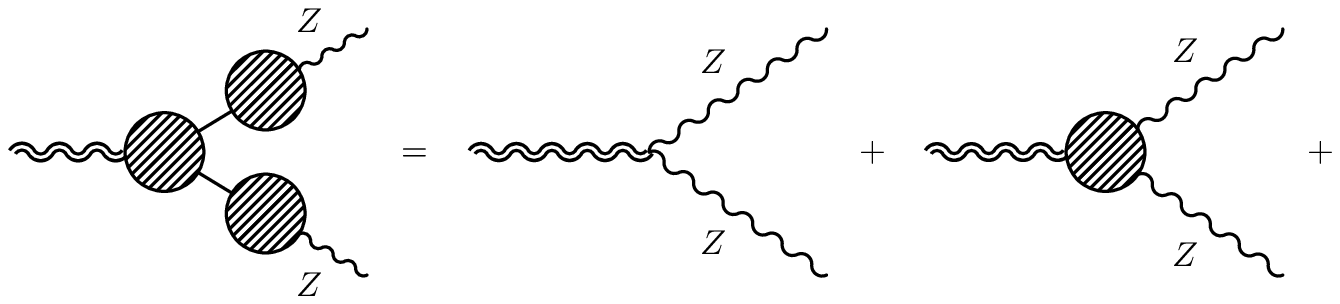}
\includegraphics[scale=0.8]{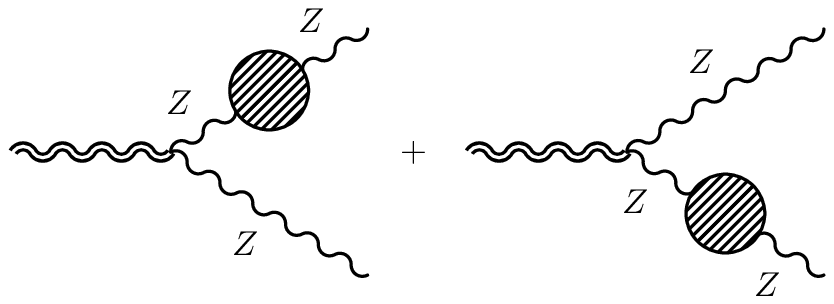}
\caption{Decomposition of $G_{\mu\nu\alpha\beta}^{ZZ}(p,q)$ in terms of the amputated $\Gamma_{\mu\nu\alpha\beta}^{ZZ}(p,q)$ and of the corrections on the external legs. \label{Fig.greenZZ}}
\end{figure}
\begin{figure}[t]
\centering
\includegraphics[scale=0.8]{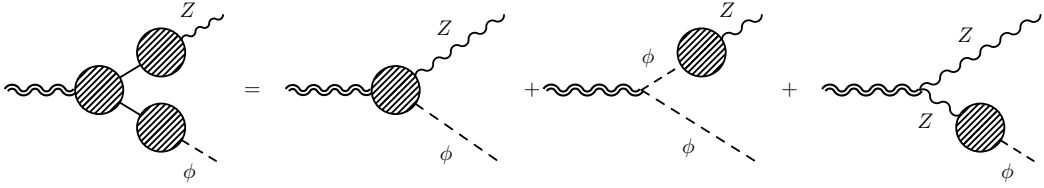}
\caption{Decomposizion of Green $G_{\mu\nu\alpha}^{Z\phi}(p,q)$ in
terms of the amputated function $\Gamma_{\mu\nu\alpha}^{Z\phi}(p,q)$ and of the corrections on the external lines \label{Fig.greenZphi}}
\end{figure}
\begin{figure}[t]
\centering
\includegraphics[scale=0.8]{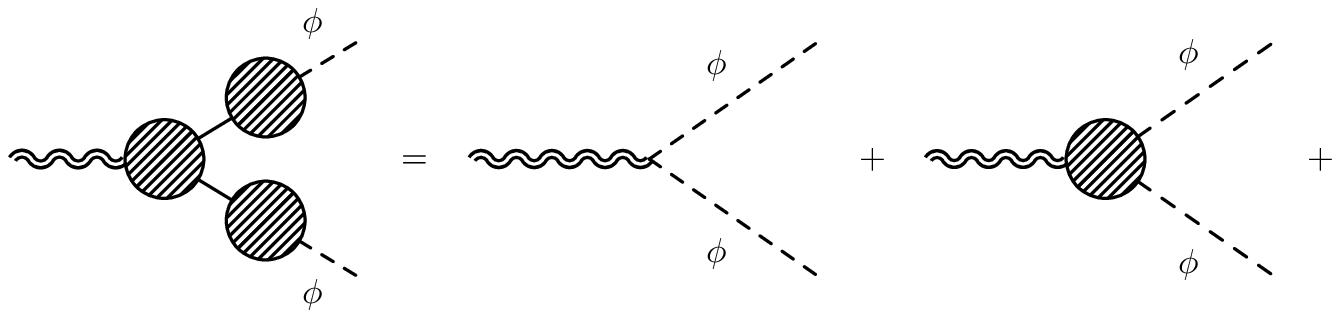}
\includegraphics[scale=0.8]{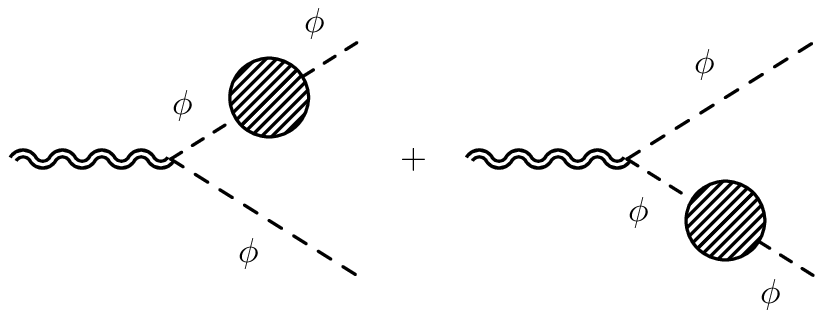}
\caption{One-loop decomposition of $G_{\mu\nu\alpha\beta}^{\phi\phi}(p,q)$
in terms of the amputated function $\Gamma_{\mu\nu}^{\phi\phi}(p,q)$ and of the corrections on the external lines.  \label{Fig.greenphiphi}}
\end{figure}
As in the cases of $ TAA$ and $ T A Z$, we are interested in deriving the form of the STI for amputated correlators.
From the left-hand-side of (\ref{STZZmom}) it is clear that there are 3 correlators which need to be decomposed, i.e. $G^{ZZ}_{\mu\nu\a\b}(p,q)$,
$G^{Z\phi}_{\mu\nu\a}(p,q)$ and $G^{\phi\phi}_{\mu\nu}(p,q)$. We have illustrated pictorially their decompositions at one loop order in
Figs. (\ref{Fig.greenZZ}), (\ref{Fig.greenZphi}) and (\ref{Fig.greenphiphi}),
while their explicit expressions are given by
\beqa
G_{\mu\nu\alpha\beta}^{ZZ}(p,q)
&=&  V^{hZZ}_{\mu\nu\sigma\rho}(p,q) \, {P^{ZZ}_0}^{\sigma}_{\alpha}(p) \,
     {P^{ZZ}_0}^{\rho}_{\beta}(q)+ \Gamma_{\mu\nu\sigma\rho}^{ZZ, \,1}(p,q) \,
     {P^{ZZ}_0}^{\sigma}_{\alpha}(p) \, {P^{ZZ}_0}^{\rho}_{\beta}(q) \nn\\
&+&  V^{hZZ}_{\mu\nu\si\rho}(p,q) \, {P^{ZZ}_1}^{\sigma}_{\alpha}(p) \, {P^{ZZ}_0}^{\rho}_{\beta}(q)
   + V^{hZZ}_{\mu\nu\si\rho}(p,q) \, {P^{ZZ}_0}^{\sigma}_{\alpha}(p) \, {P^{ZZ}_1}^{\rho}_{\beta}(q)  \, , \label{TZZGreen} \\
\nn \\
G_{\mu\nu\alpha}^{Z\phi}(p,q)
&=& \Gamma^{hZ\phi,\,1}_{\mu\nu\rho}(p,q)\,{P^{ZZ}_0}^\rho_\alpha(p)\,P^{\phi\phi}_0(q)
+  V^{hZZ}_{\mu\nu\rho}(p,q)\,{P^{ZZ}_0}^\rho_\alpha(p)\,P^{Z\phi}_1(q) \nn \\
&+&  V^{h\phi\phi}_{\mu\nu\rho}(p,q)\,{P^{\phi Z}_1}^\rho_\alpha(p)\,P^{\phi\phi}_0(q)\, , \label{TZphiGreen} \\
\nn \\
G_{\mu\nu}^{\phi\phi}(p,q)
&=& V^{h\phi\phi}_{\mu\nu}(p,q)\,P^{\phi\phi}_0(p)\,P^{\phi\phi}_0(q)
 +  \Gamma^{h\phi\phi,\,1}_{\mu\nu}(p,q)\,P^{\phi\phi}_0(p)\,P^{\phi\phi}_0(q)\nn\\
&& + V^{h\phi\phi}_{\mu\nu\rho}(p,q)\,P^{\phi\phi}_0(p)\,P^{\phi\phi}_1(q)
 +  V^{h\phi\phi}_{\mu\nu\rho}(p,q)\,P^{\phi\phi}_1(p)\,P^{\phi\phi}_0(q)\,. \label{TphiphiGreen}
 \eeqa
Eq. (\ref{STZZmom}), after the insertion of (\ref{TZZGreen}), (\ref{TZphiGreen}) and
(\ref{TphiphiGreen}), gives the STI for amputated functions that we have been looking for.
One can explicitly verify that the contributions on the left-hand-side of Eq. (\ref{STZZmom}) - generated both by the tree-level vertices and by the contraction of these with 1-loop 2-point functions on the external legs - are equal to the right-hand-side of the same equation. These checks are far from being obvious since they require a complete and explicit computation of all the correlators, as will be discussed next. Here we just conclude by quoting the STI for amputated functions, which takes the simpler form
\beq
p^\a q^\b \Gamma^{ZZ, \, 1}_{\mu\nu\alpha\b}(p,q) + i\xi M_Z p^\alpha \Gamma^{Z\phi, \, 1}_{\mu\nu\alpha}(p,q)
+ i\xi M_Z q^\beta \Gamma^{\phi Z, \, 1}_{\mu\nu\beta}(p,q) - \xi^2 M_Z^2 \Gamma^{\phi\phi, \, 1}_{\mu\nu}(p,q)
= 0\, .\eeq
This and the previous similar STI's  are fundamental relations which define consistently the coupling of one graviton to the neutral sector of the SM.

\section{ Perturbative results for all the correlators}
\label{resultsection}
\begin{figure}[t]
\centering
\subfigure[]{\includegraphics[scale=0.8]{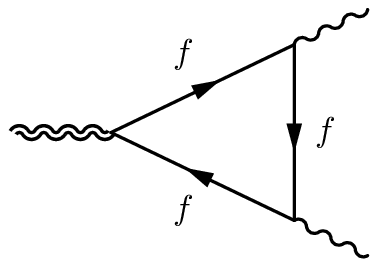}} \hspace{.5cm}
\subfigure[]{\includegraphics[scale=0.8]{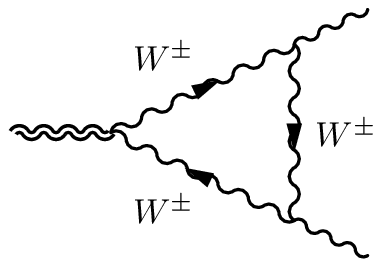}} \hspace{.5cm}
\subfigure[]{\includegraphics[scale=0.8]{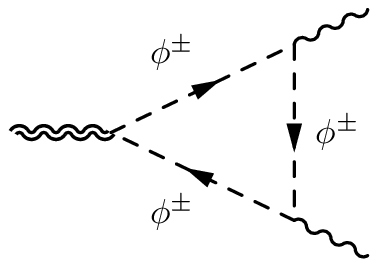}} \hspace{.5cm}
\subfigure[]{\includegraphics[scale=0.8]{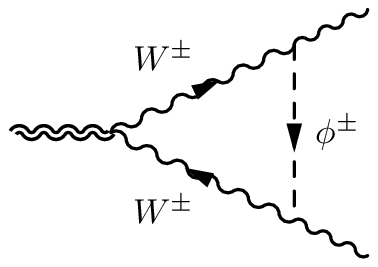}}
\\
\vspace{.5cm}
\subfigure[]{\includegraphics[scale=0.8]{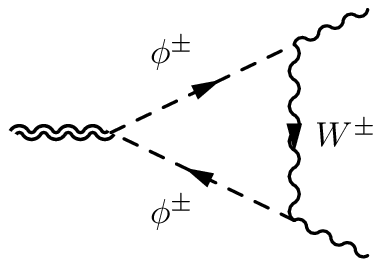}} \hspace{.5cm}
\subfigure[]{\includegraphics[scale=0.8]{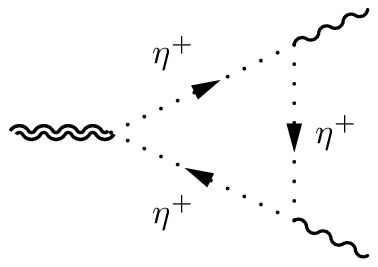}} \hspace{.5cm}
\subfigure[]{\includegraphics[scale=0.8]{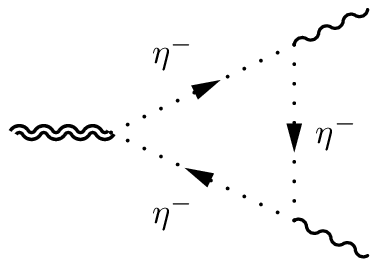}} \hspace{.5cm}
\caption{Amplitudes with the triangle topology for the three correlators $TAA$, $TAZ$ and $TZZ$. \label{triangles}}
\end{figure}

\begin{figure}[t]
\centering
\subfigure[]{\includegraphics[scale=0.75]{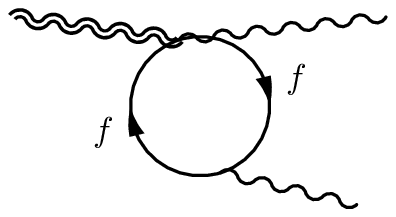}}\hspace{.5cm}
\subfigure[]{\includegraphics[scale=0.75]{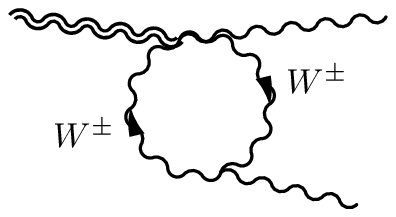}}\hspace{.5cm}
\subfigure[]{\includegraphics[scale=0.75]{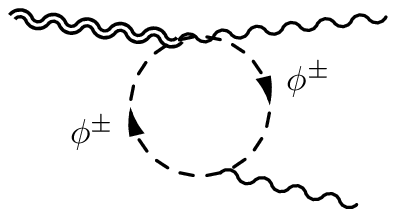}}\hspace{.5cm}
\subfigure[]{\includegraphics[scale=0.75]{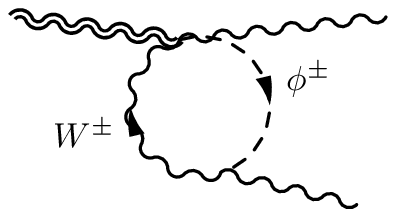}}
\\
\vspace{.5cm}
\subfigure[]{\includegraphics[scale=0.75]{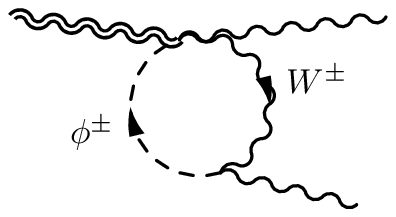}}\hspace{.5cm}
\subfigure[]{\includegraphics[scale=0.75]{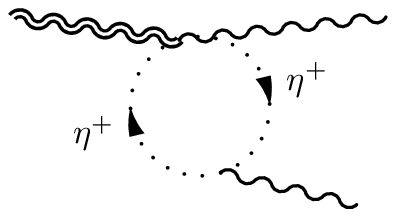}}\hspace{.5cm}
\subfigure[]{\includegraphics[scale=0.75]{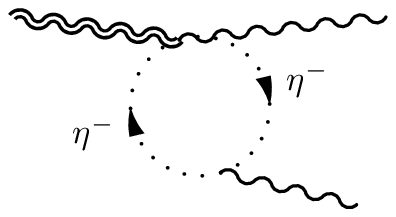}}\hspace{.5cm}
\caption{Amplitudes with t-bubble topology for the three correlators $TAA$, $TAZ$ and $TZZ$. \label{t-bubble}}
\end{figure}

\begin{figure}[t]
\centering
\subfigure[]{\includegraphics[scale=0.75]{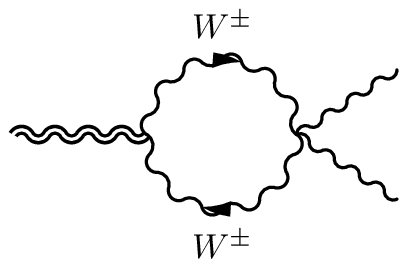}}\hspace{.5cm}
\subfigure[]{\includegraphics[scale=0.75]{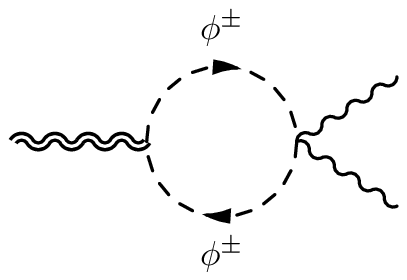}}
\caption{Amplitudes with s-bubble topology for the three correlators $TAA$, $TAZ$ and $TZZ$. \label{s-bubble}}
\end{figure}

\begin{figure}[t]
\centering
\subfigure[]{\includegraphics[scale=0.8]{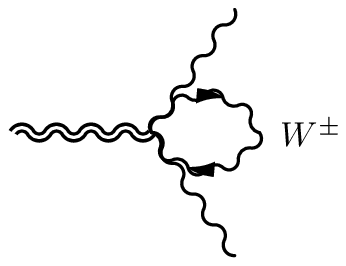}}\hspace{.5cm}
\subfigure[]{\includegraphics[scale=0.8]{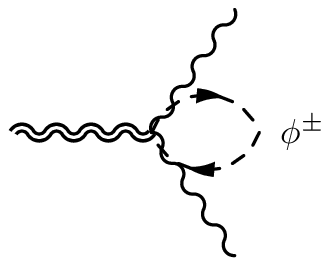}}
\caption{Amplitudes with the tadpole topology for the three correlators $TAA$, $TAZ$ and $TZZ$.\label{tadpoles}}
\end{figure}

\begin{figure}[t]
\centering
\subfigure[]{\includegraphics[scale=0.8]{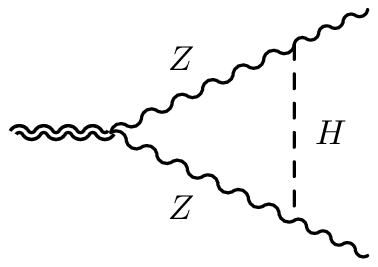}}\hspace{.5cm}
\subfigure[]{\includegraphics[scale=0.8]{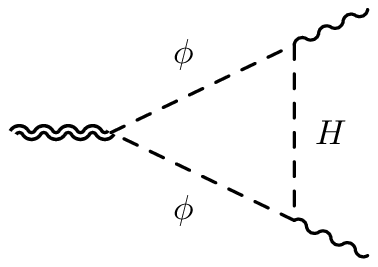}}\hspace{.5cm}
\subfigure[]{\includegraphics[scale=0.8]{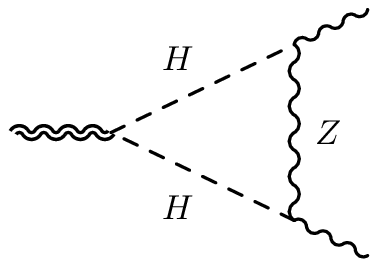}}\hspace{.5cm}
\subfigure[]{\includegraphics[scale=0.8]{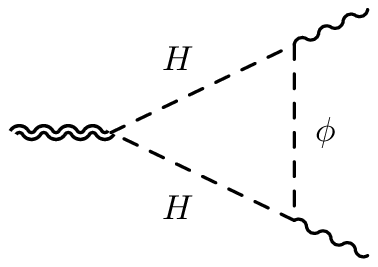}}
\caption{Amplitudes with the triangle topology for the correlator $TZZ$. \label{triangles1}}
\end{figure}

\begin{figure}[t]
\centering
\subfigure[]{\includegraphics[scale=0.75]{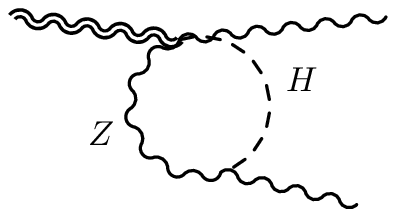}}\hspace{.5cm}
\subfigure[]{\includegraphics[scale=0.75]{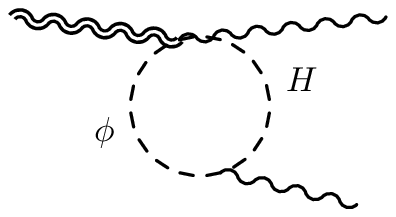}}
\caption{Amplitudes with the t-bubble topology for the correlator $TZZ$.\label{t-bubble1}}
\end{figure}

\begin{figure}[t]
\centering
\subfigure[]{\includegraphics[scale=0.75]{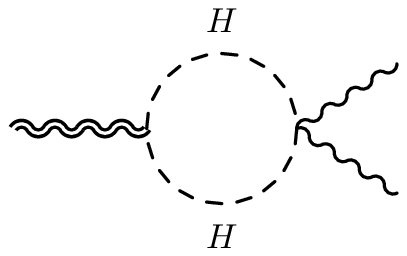}}\hspace{.5cm}
\subfigure[]{\includegraphics[scale=0.75]{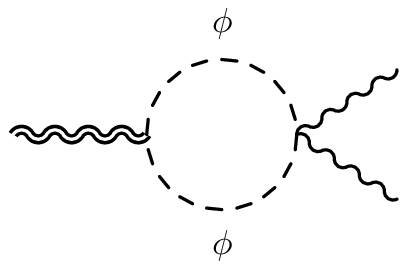}}
\caption{Amplitudes with the s-bubble topology for the correlator $TZZ$.\label{s-bubble1}}
\end{figure}

\begin{figure}[t]
\centering
\subfigure[]{\includegraphics[scale=0.8]{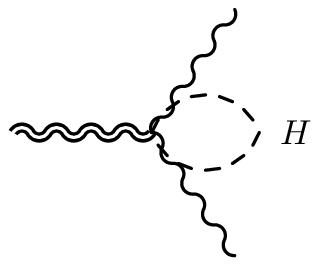}}\hspace{.5cm}
\subfigure[]{\includegraphics[scale=0.8]{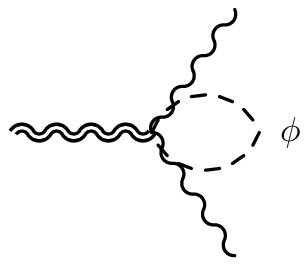}}
\caption{Amplitudes with the tadpole topology for the correlator $TZZ$.\label{tadpoles1}}
\end{figure}

\begin{figure}[t]
\centering
\subfigure[]{\includegraphics[scale=0.8]{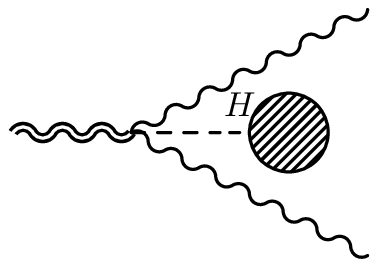}}\hspace{.5cm}
\subfigure[]{\includegraphics[scale=0.8]{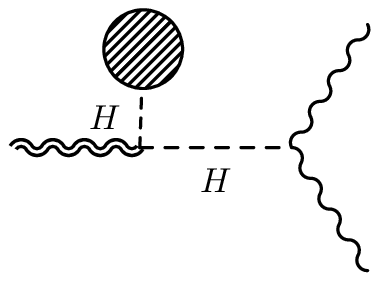}}
\caption{Amplitudes with the Higgs tadpole for the correlator $TZZ$ which vanish after renormalization.\label{tadpolesHiggs}}
\end{figure}

In this section we illustrate the various diagrammatic contributions appearing in the perturbative expansion of the $TV V'$ vertex.
We show in Figs. (\ref{triangles}-\ref{tadpolesHiggs}) all the basic diagrams involved, for which we are going to present explicit results. Figs. (\ref{triangles}) and (\ref{triangles1}) are characterized by a typical triangle topology, while (\ref{t-bubble}) and (\ref{t-bubble1}) denote typical terms where the point of insertion of the EMT coincides with that of a gauge current. We will refer to these last contributions with the term "t-bubbles", while those characterized by two gauge bosons emerging from a single vertex, such as in Figs. (\ref{s-bubble}) and (\ref{s-bubble1}), are called "s-bubble" diagrams. Other contributions are those with a topology of tadpoles, shown in Figs.
(\ref{tadpoles}), (\ref{tadpoles1}) and (\ref{tadpolesHiggs}).

The two sectors $TAA$ and $TAZ$ involve 32 diagrams each, while the $TZZ$ correlator includes 70 diagrams.
The computation of these diagrams is rather involved and has been performed in DR using the on-shell renormalization scheme  \cite{Ross:1973fp} and the t'Hooft-Veltman prescription for $\gamma_5$ matrix. We have used a
reduction of tensor integrals to the scalar form and checked explicitly all the Ward and STI's derived in the previous sections.  The reduction involves non-standard rank-4 integrals (due to the momenta coming from the insertion of the EMT on the triangle
topology) with 3 propagators.

One of the non trivial points of the computation concerns the treatment of diagrams containing fermion loops and insertions of the EMT on correlators with both vector ($J_V$) and axial-vector ($J_A$) currents. This problem has been analyzed and solved in
a related work \cite{Armillis:2010pa} to which we refer for more details. In particular, it has been shown that there are no mixed chiral and trace anomalies in diagrams of this type even in the presence of explicit mass corrections, due to the
vanishing of the $TJ_V J_A$ vertex mediated by fermion loops. This result has been obtained in a  simple $U(1)_V\times U(1)_A$ gauge model, with an explicit breaking of the gauge symmetry due to a fermion mass term. The result remains true both for global and local currents, being the gauge fields (vector and axial-vector) in the treatment of \cite{Armillis:2010pa} purely external fields. This preliminary analysis has been instrumental in all the generalizations discussed in this work.

At this point few more comments concerning the number of form factors introduced in our analysis are in order.
We recall, from a previous study \cite{Giannotti:2008cv}, that the number of original tensor structures which can be built out of the metric and of the two momenta $p$ and $q$ of the two gauge lines is 43  before imposing the Ward and the STI's of the theory. These have been classified in \cite{Giannotti:2008cv} and \cite{Armillis:2009pq}.
In particular, the form factors appearing in the fermion sector can be expressed (in the off-shell case) in terms of 13 tensor structures for the case of vector currents and of 22 structures for the axial-vector current, as shown in \cite{Armillis:2010pa}.

In the on-shell case, the fermion loops with external photons are parameterized just by 3 independent form factors. This analysis has been generalized more recently to QCD, with the computation of the graviton-gluon-gluon ($hgg$)
vertex in full generality \cite{Armillis:2009pq}. The entire vertex in the on-shell QCD case - which includes fermion and gluon loops -
is also parameterized just by 3 form factors. A similar result holds for the $TAA$ in the electroweak case. On the other hand the $TZZ$ and the $TAZ$ correlators have been expressed in terms of 9 form factors. A special comment deserves the handling of the symbolic computations. These have been performed using some software entirely written by us and implemented in the symbolic manipulation program {\em  MATHEMATICA}. This allows the reduction to scalar form of tensor integrals for correlators of rank-4 with the triangle topology. The software alllows to perform direct tests of all the Ward and Slavnov-Taylor identities on the correlator, 
which are crucial in order to secure the correctness of the result.

\subsection{ $\Gamma^{\mu\nu\alpha\beta}(p,q)$ and the terms of improvement}
Before giving the results for the anomalous correlators, we pause for some comments.

In our computations the gravitational field is non-dynamical and the analysis of the Ward and STI's shows that these can be
consistently solved only if we include the graviton-Higgs mixing on the graviton line. In other words, the graviton line is uncut. We will denote with $\Delta^{\mu\nu\alpha\beta}(p,q)$ these extra contributions and with $\Sigma^{\mu\nu\alpha\beta}(p,q)$ the completely cut vertex. These two contributions appear on the right-hand-side of the
expression of the correlation function $\Gamma^{\mu\nu\alpha\beta}(p,q)$
\bea
\Gamma^{\mu\nu\alpha\beta}(p,q) = \Sigma^{\mu\nu\alpha\beta}(p,q) + \Delta^{\mu\nu\alpha\beta}(p,q).
\eea
Finally, we just mention that we have excluded from the final expressions of the vertices all the contributions at tree-level. For this reason our results
are purely those responsible for the generation of the anomaly.

\begin{figure}[t]
\centering
\includegraphics[scale=0.8]{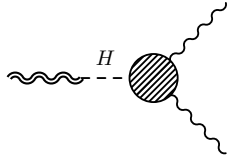}
\caption{Amplitude with the graviton - Higgs mixing vertex generated by the term of improvement. The blob represents the SM Higgs -VV' vertex at one-loop. \label{HVVImpr}}
\end{figure}
\begin{figure}[t]
\centering
\subfigure[]{\includegraphics[scale=0.8]{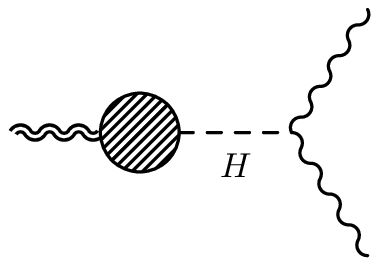}}\hspace{.5cm}
\subfigure[]{\includegraphics[scale=0.8]{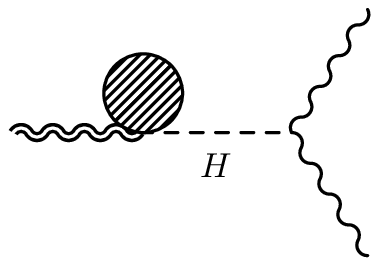}}\hspace{.5cm}
\subfigure[]{\includegraphics[scale=0.8]{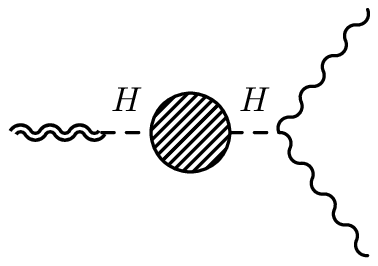}}\hspace{.5cm}
\caption{Leg corrections to the external graviton for the $TZZ$ correlator.\label{legcorr12}}
\end{figure}

\subsection{Results for the TAA correlator}
In this section we present the one-loop result of the computation of these correlators for on-shell vector bosons lines and discuss some of their interesting features, such as the appearance of massless anomaly poles in all the gauge invariants subsectors of the perturbative expansion.\\
We start from the case of the $TAA$ vertex and then move to the remaining ones.
In this case the full irreducible contribution $\Sigma^{\mu\nu\alpha\beta}(p,q)$ is written in the form
\bea
\Sigma ^{\mu\nu\a\b}(p,q) = \Sigma_{F}^{\mu\nu\a\b}(p,q) + \Sigma_{B}^{\mu\nu\a\b}(p,q) + \Sigma_{I}^{\mu\nu\a\b}(p,q),
\eea
where each term can be expanded in a tensor basis
\bea
\Sigma ^{\mu\nu\alpha\beta}_{F}(p,q) &=&  \, \sum_{i=1}^{3} \Phi_{i\,F} (s,0, 0,m_f^2) \, \phi_i^{\mu\nu\alpha\beta}(p,q)\,, \\
\Sigma ^{\mu\nu\alpha\beta}_{B}(p,q) &=&  \, \sum_{i=1}^{3} \Phi_{i\,B} (s,0, 0,M_W^2) \, \phi_i^{\mu\nu\alpha\beta}(p,q)\,, \\
\Sigma ^{\mu\nu\alpha\beta}_{I}(p,q) &=&  \Phi_{1\,I} (s,0, 0,M_W^2) \, \phi_1^{\mu\nu\alpha\beta}(p,q) + \Phi_{4\,I} (s,0, 0,M_W^2) \, \phi_4^{\mu\nu\alpha\beta}(p,q) \,.
\eea
The tensor basis on which we expand the on-shell vertex is given by
\bea
\label{phis}
  \phi_1^{\, \mu \nu \a \b} (p,q) &=&
 (s \, \eta^{\mu\nu} - k^{\mu}k^{\nu}) \, u^{\a \b} (p,q),
 \label{widetilde1} \nn \\
\phi_2^{\, \mu \nu \a \b} (p,q) &=& - 2 \, u^{\a \b} (p,q) \left[ s \, \eta^{\mu \nu} + 2 (p^\mu \, p^\nu + q^\mu \, q^\nu )
- 4 \, (p^\mu \, q^\nu + q^\mu \, p^\nu) \right],
\label{widetilde2} \nn \\
\phi^{\, \mu \nu \alpha \beta}_{3} (p,q) &=&
\big(p^{\mu} q^{\nu} + p^{\nu} q^{\mu}\big)\eta^{\alpha\beta}
+ \frac{s}{2} \left(\eta^{\alpha\nu} \eta^{\beta\mu} + \eta^{\alpha\mu} \eta^{\beta\nu}\right) - \eta^{\mu\nu}  \, u^{\a \b} (p,q) \nn \\
&&
-\left(\eta^{\beta\nu} p^{\mu}
+ \eta^{\beta\mu} p^{\nu}\right)q^{\alpha}
 - \big (\eta^{\alpha\nu} q^{\mu}
+ \eta^{\alpha\mu} q^{\nu }\big)p^{\beta}, \nonumber \\
\phi_4^{\mu\nu\alpha\beta}(p,q) &=& (s \, \eta^{\mu\nu} - k^{\mu}k^{\nu}) \, \eta^{\alpha\beta}.
\label{widetilde3}
\eea
where $u^{\a \b} (p,q)$ has been defined as
\beq
u^{\alpha\beta}(p,q) \equiv (p\cdot q) \,  \eta^{\alpha\beta} - q^{\alpha} \, p^{\beta},\,\\
\label{utensor}
\eeq
among which only  $\phi_1^{\, \mu \nu \a \b}$ shows manifestly a trace, the remaining ones being traceless.

The one loop vertex $\Sigma ^{\mu\nu\a\b}(p,q)$ with two on-shell photons is expressed as a sum of a fermion sector (F) (Fig. \ref{triangles}(a), Fig. \ref{t-bubble}(a)) ,
a gauge boson sector (B) (Fig. \ref{triangles}(b)-(g), Fig. \ref{t-bubble}(b)-(g), Fig. \ref{s-bubble}, Fig. \ref{tadpoles}) and a term of improvement denoted as $\Sigma^{\mu\nu\a\b}_{I}$. The contribution from the term of improvement is given by the diagrams depicted in Fig. \ref{triangles}(c), (d) and Fig. \ref{s-bubble}(b), with the graviton - scalar - scalar vertices determined by the $T_I^{\mu\nu}$. \\
The first three arguments of the form factors stand for the three independent kinematical invariants $k^2 = (p+q)^2 = s$, $p^2 = q^2 = 0$ while the remaining ones denote the particle masses circulating in the loop.

As already shown for QED and QCD, in the massless limit (i.e. before electroweak symmetry breaking), the entire contribution to the trace anomaly comes from the first tensor structure $\phi_1$ both for the fermion and for the gauge boson cases.\\
In the fermion sector the form factors are given by
\bea
\Phi_{1\, F} (s,\,0,\,0,\,m_f^2) &=& - i \frac{\kappa}{2}\, \frac{\alpha}{3 \pi \, s} \, Q_f^2 \bigg\{
- \frac{2}{3} + \frac{4\,m_f^2}{s} - 2\,m_f^2 \, \mathcal C_0 (s, 0, 0, m_f^2, m_f^2, m_f^2)
\bigg[1 - \frac{4 m_f^2}{s}\bigg] \bigg\}\, ,  \nn \\\\
\Phi_{2\, F} (s,\,0,\,0,\,m_f^2)  &=& - i \frac{\kappa}{2}\, \frac{\alpha}{3 \pi \, s} \, Q_f^2 \bigg\{
-\frac{1}{12} - \frac{m_f^2}{s} - \frac{3\,m_f^2}{s} \mathcal D_0 (s, 0, 0, m_f^2, m_f^2)  \nn \\
&-&  m_f^2 \mathcal C_0(s, 0, 0, m_f^2, m_f^2, m_f^2 )\, \left[ 1 + \frac{2\,m_f^2}{s}\right] \bigg\}\, , \\
\Phi_{3\,F} (s,\,0,\,0,\,m_f^2) &=&  - i \frac{\kappa}{2}\, \frac{\alpha}{3 \pi \, s} \, Q_f^2 \bigg\{
\frac{11\,s}{12}+ 3 m_f^2 +  \mathcal D_0 (s, 0, 0, m_f^2, m_f^2)\left[5 m_f^2 + s \right]\nn\\
&+&  s \, \mathcal B_0 (0, m_f^2, m_f^2) + 3 \, m_f^2 \, \mathcal C_0(s, 0, 0, m_f^2 , m_f^2, m_f^2) \left[s + 2m_f^2 \right] \bigg\}\, .
\eea
The form factor $\Phi_{1\, F}$ is characterized by the presence of an anomaly pole
\beq
\Phi^F_{1\, pole}\equiv i \kappa \frac{\alpha}{9 \pi \, s} \, Q_f^2
\eeq
which is responsible for the generation of the anomaly in the massless limit. This $1/s$ behaviour of the amplitude is also clearly identifiable in a $m_f^2/s$ (asymptotic) expansion ($s \gg m_f^2$), where $m_f$ denotes generically any fermion of the
SM.
In this second case, the scaleless contribution associated with the exchange of a massless state (i.e. the $1/s$ term) is corrected by other terms which are suppressed as powers of $m_f^2/s$. This pattern, as we are going to show, is general.

The other gauge-invariant sector of the $TAA$ vertex is the one mediated by the exchange of bosons and ghosts in the loop. In this
sector the form factors are given by
\bea
\Phi_{1\, B} (s,\,0,\,0,\,M_W^2) &=& - i \frac{\kappa}{2}\, \frac{\alpha}{\pi \, s} \bigg\{
\frac{5}{6} - \frac{2\,M_W^2}{s} + 2\,M_W^2 \, \mathcal C_0 (s, 0, 0, M_W^2, M_W^2, M_W^2)
\bigg[1 - \frac{2 M_W^2}{s}\bigg] \bigg\},  \nn \\\\
\Phi_{2\, B} (s,\,0,\,0,\,M_W^2)  &=& - i \frac{\kappa}{2}\, \frac{\alpha}{\pi \, s} \bigg\{
\frac{1}{24} + \frac{M_W^2}{2\,s} + \frac{3\,M_W^2}{2\,s} \mathcal D_0 (s, 0, 0, M_W^2, M_W^2)  \nn \\
&+&  \frac{M_W^2}{2} \mathcal C_0(s, 0, 0, M_W^2, M_W^2, M_W^2 )\, \left[ 1 + \frac{2\,M_W^2}{s}\right] \bigg\}\, , \\
\Phi_{3\,B} (s,\,0,\,0,\,M_W^2) &=& - i \frac{\kappa}{2}\, \frac{\alpha}{\pi \, s} \bigg\{
-\frac{15\,s}{8}-\frac{3\,M_W^2}{2} - \frac{1}{2}\, \mathcal D_0 (s, 0, 0, M_W^2, M_W^2)\left[5 M_W^2+7\,s \right]\nn\\
&-& \frac{3}{4}s\, \mathcal B_0 (0, M_W^2, M_W^2) - \mathcal C_0(s, 0, 0, M_W^2 , M_W^2, M_W^2) \left[s^2 + 4 M_W^2\,s + 3\,M_W^4\right]
\bigg\}.
\eea
As in the previous case, we focus our attention on $\Phi_{1\, B}$, which multiplies the tensor structure $\phi_1$, responsible for the generation of the anomalous trace. In this case the contribution of the anomaly pole is isolated in the form
\beq
\Phi_{1\, B, pole}\equiv - i \frac{\kappa}{2}\, \frac{\alpha}{\pi \, s} \frac{5}{6}\, .
\eeq
It is clear, also in this case, that in the massless limit ($M_W=0$), i.e. in the symmetric phase of the theory, this pole is completely responsible for the generation of the anomaly. At the same time, at high energy  (i.e. for $s\gg M_W^2$) the massless exchange can be easily exposed as a dominant contribution to the trace part of the correlator. Notice that, in general, the correlator has other $1/s$ singularities in the remaining form factors and even constant terms which are unsuppressed for a large $s$, but these are not part of the trace.

The contributions coming from the term of improvement are characterized just by two form factors
\bea
\Phi_{1\, I} (s,\,0,\,0,\,M_W^2) &=& - i \frac{\kappa}{2}\frac{\alpha}{3 \pi \, s} \bigg\{ 1 + 2 M_W^2 \,C_0 (s, 0, 0, M_W^2, M_W^2, M_W^2)\bigg\} \,,\\
\Phi_{4\, I} (s,\,0,\,0,\,M_W^2) &=&  i \frac{\kappa}{2}\frac{\alpha}{6 \pi }  M_W^2 \,C_0 (s, 0, 0, M_W^2, M_W^2, M_W^2) \,.
\eea
Now we consider the external graviton leg corrections $\Delta^{\mu\nu\alpha\beta}(p,q)$. In this case only the term of improvement contributes with the diagram depicted in Fig. \ref{HVVImpr}
\bea
\Delta^{\mu\nu\alpha\beta}(p,q) &=& \Delta^{\mu\nu\alpha\beta}_I (p,q) \nn \\
&=&  \Psi_{1\, I} (s,0, 0,m_f^2,M_W^2,M_H^2) \, \phi_1^{\mu\nu\alpha\beta}(p,q) + \Psi_{4 \, I} (s,0, 0,M_W^2)  \, \phi_4^{\mu\nu\alpha\beta}(p,q)\, .
\label{DAA}
\eea
This is built by combining the tree level vertex for graviton/Higgs mixing  - coming from the improved EMT -  and the Standard Model Higgs/photon/photon correlator at one-loop
\bea
\Psi_{1\, I} (s,\,0,\,0,\,m_f^2,M_W^2,M_H^2) &=& - i \frac{\kappa}{2} \frac{\alpha}{3 \pi \, s (s - M_H^2)} \bigg\{ 2 m_f^2 \, Q_f^2 \bigg[ 2 + (4 m_f^2 -s) C_0 (s, 0, 0, m_f^2, m_f^2, m_f^2) \bigg] \nn \\
&& \hspace{-2cm} + M_H^2 + 6 M_W^2 + 2 M_W^2 (M_H^2 + 6 M_W^2 - 4 s) C_0(s,0,0, M_W^2,M_W^2,M_W^2) \bigg\} \,, \\
\Psi_{4\, I} (s,\,0,\,0,M_W^2) &=& - \Phi_{4\, I} (s,\,0,\,0,\,M_W^2)  \, .
\eea
\subsection{Results for the TAZ correlator}
We proceed with the analysis of the $TAZ$ correlator, in particular we start with the irreducible vertex $\Sigma ^{\mu\nu\alpha\beta}(p,q)$ that can be defined, as in the previous case, as a sum of the three gauge invariant contributions: the fermion sector (F), (Fig. \ref{triangles}(a), Fig. \ref{t-bubble}(a)), the gauge boson sector (B), (Fig. \ref{triangles}(b)-(g), Fig. \ref{t-bubble}(b)-(g), Fig. \ref{s-bubble}, Fig. \ref{tadpoles}) and the improvement term (I) given by the diagrams depicted in Fig. \ref{triangles}(c), (d) and Fig. \ref{s-bubble}(b), with the graviton - scalar - scalar vertices determined by the $T_I^{\mu\nu}$
\bea
\Sigma ^{\mu\nu\a\b}(p,q) = \Sigma_{F}^{\mu\nu\a\b}(p,q) + \Sigma_{B}^{\mu\nu\a\b}(p,q) + \Sigma_{I}^{\mu\nu\a\b}(p,q).
\eea
Each of these terms can be expanded in the on-shell case ($p^2 = 0$, $q^2 = M_Z^2$) on a tensor basis $f_i^{\mu\nu\alpha\beta}(p,q)$
\bea
\Sigma ^{\mu\nu\alpha\beta}_{F}(p,q) &=&  \, \sum_{i=1}^{7} \Phi_{i\,F} (s,0, M_Z^2,m_f^2) \, f_i^{\mu\nu\alpha\beta}(p,q)\,, \\
\Sigma ^{\mu\nu\alpha\beta}_{B}(p,q) &=&  \, \sum_{i=1}^{9} \Phi_{i\,B} (s,0, M_Z^2,M_W^2) \, f_i^{\mu\nu\alpha\beta}(p,q)\,, \\
\Sigma ^{\mu\nu\alpha\beta}_{I}(p,q) &=&  \Phi_{1\,I} (s,0, M_Z^2,M_W^2) \, f_1^{\mu\nu\alpha\beta}(p,q) + \Phi_{8\,I} (s,0, M_Z^2,M_W^2) \, f_8^{\mu\nu\alpha\beta}(p,q) \,.
\eea
For the on-shell $TAZ$ correlator the tensor structures are explicitly defined as
\bea
f_1^{\mu\nu\alpha\beta}(p,q) &=&  (s \, \eta^{\mu\nu} - k^{\mu}k^{\nu}) [ \frac{1}{2}\left( s-M_Z^2 \right)\eta^{\alpha\beta} - q^\alpha p^\beta ]  \,,\nn \\
f_2^{\mu\nu\alpha\beta}(p,q) &=& p^{\mu} p^{\nu} [ \frac{1}{2}\left( s-M_Z^2 \right)\eta^{\alpha\beta} - q^\alpha p^\beta ]  \,,\nn \\
f_3^{\mu\nu\alpha\beta}(p,q) &=& (M_Z^2 \, \eta^{\mu\nu} - 4 q^{\mu} q^{\nu})[ \frac{1}{2}\left( s-M_Z^2 \right)\eta^{\alpha\beta} - q^\alpha p^\beta]  \,,\nn \\
f_4^{\mu\nu\alpha\beta}(p,q) &=& [ \frac{1}{2}\left( s-M_Z^2 \right) \eta^{\mu\nu} - 2 ( q^{\mu} p^{\nu} + p^{\mu} q^{\nu})][ \frac{1}{2}\left( s-M_Z^2 \right)\eta^{\alpha\beta} - q^\alpha p^\beta]  \,,\nn \\
f_5^{\mu\nu\alpha\beta}(p,q) &=& p^{\beta}[  \frac{1}{2}\left( s-M_Z^2 \right) (\eta^{\alpha\nu} q^{\mu} + \eta^{\alpha\mu} q^{\nu}) - q^{\alpha}(q^{\mu} p^{\nu} + p^{\mu} q^{\nu})] \,,\nn  \\
f_6^{\mu\nu\alpha\beta}(p,q) &=& p^{\beta}[ \frac{1}{2}\left( s-M_Z^2 \right) (\eta^{\alpha\nu} p^{\mu} + \eta^{\alpha\mu} p^{\nu}) -2 q^{\alpha} p^{\mu} p^{\nu}] \,,\nn \\
f_7^{\mu\nu\alpha\beta}(p,q) &=& (p^{\mu} q^{\nu} + p^{\nu} q^{\mu}) \eta^{\alpha \beta} +  \frac{1}{2}\left( s-M_Z^2 \right) (\eta^{\alpha\nu} \eta^{\beta\mu} + \eta^{\alpha\mu} \eta^{\beta\nu}) - \eta^{\mu\nu}  [ \frac{1}{2}\left( s-M_Z^2 \right)\eta^{\alpha\beta} - q^\alpha p^\beta ] \nn \\
&-&  (\eta^{\beta\nu} p^{\mu} + \eta^{\beta\mu} p^{\nu}) q^{\alpha} - (\eta^{\alpha\nu} q^{\mu} + \eta^{\alpha\mu} q^{\nu}) p^{\beta} \,,\nn \\
f_8^{\mu\nu\alpha\beta}(p,q) &=&  (s \, \eta^{\mu\nu} - k^{\mu}k^{\nu}) \eta^{\alpha\beta}  \,,\nn \\
f_9^{\mu\nu\alpha\beta}(p,q) &=& q^{\alpha}[ 3 s (\eta^{\beta\mu}p^{\nu} + \eta^{\beta\nu}p^{\mu}) - p^{\beta}(s \, \eta^{\mu\nu} + 2 k^{\mu}k^{\nu} ) ].
\eea
We collect here just the form factors in the fermion and boson sectors which contribute to the trace anomaly, while the remaining ones are given in appendix \ref{formfactors}
\bea
\Phi_1^{(F)}(s,0,M_Z^2,m_f^2) &=& - i \frac{\kappa}{2} \frac{\alpha}{3 \pi s\, s_w \, c_w } C_v^f \, Q_f \bigg\{
-\frac{1}{3} +\frac{2 m_f^2}{s-M_Z^2}
+\frac{2 m_f^2 \, M_Z^2 }{(s-M_Z^2)^2} \mathcal D_0\left(s,M_Z^2,m_f^2,m_f^2\right) \nn \\
&& - m_f^2  \bigg[1- \frac{4 m_f^2}{s-M_Z^2}\bigg] \mathcal C_0\left(s,0,M_Z^2,m_f^2,m_f^2,m_f^2\right)
  \bigg\}  \,, \\
\Phi_1^{(B)}(s,0,M_Z^2,M_W^2) &=& - i \frac{\kappa}{2} \frac{\alpha}{3 \pi s \, s_w \, c_w }  \bigg\{
\frac{1}{12}(37 - 30 s_w^2) - \frac{M_Z^2}{2(s-M_Z^2)}(12 s_w^4 - 24 s_w^2 + 11) \nn \\
&& \hspace{-3cm} -\frac{M_Z^2}{2 \left(s-M_Z^2\right)^2} \left(2 M_Z^2 \left(6 s_w^4-11 s_w^2+5\right)-2 s_w^2 s +s \right) \mathcal D_0\left(s,M_Z^2,M_W^2,M_W^2\right) \nn \\
&& \hspace{-3cm} -\frac{M_Z^2 c_w^2}{s-M_Z^2} \left(2 M_Z^2 \left(6 s_w^4-15 s_w^2+8\right)+s \left(6 s_w^2-5\right)\right) \mathcal C_0\left(s,0,M_Z^2,M_W^2,M_W^2,M_W^2\right) \bigg\} \, .
\eea
Moreover, the improvement term is defined by the following two form factors
\bea
\Phi_{1\,I} (s,0, M_Z^2,M_W^2) &=&- i \frac{\kappa}{2} \frac{\alpha \, (c_w^2 - s_w^2)}{6 \pi \, s_w \, c_w \, (s-M_Z^2)}\bigg\{ 1 + 2 M_W^2 \, \mathcal C_0\left(s,0,M_Z^2,M_W^2,M_W^2,M_W^2\right) \,, \nn \\
&+&  \frac{M_Z^2}{s-M_Z^2}  \mathcal D_0\left(s,M_Z^2,M_W^2,M_W^2\right)  \bigg\}\, , \\
\Phi_{2\,I} (s,0, M_Z^2,M_W^2) &=& - i \frac{\kappa}{2} \frac{\alpha}{6 \pi}  s_w^2 \, M_Z^2 \, \mathcal C_0\left(s,0,M_Z^2,M_W^2,M_W^2,M_W^2\right).
\eea
Now we consider the external graviton leg corrections $\Delta^{\mu\nu\alpha\beta}(p,q)$. In this case only the improvement term contributes with the diagram shown in Fig. \ref{HVVImpr}
\bea
\Delta^{\mu\nu\alpha\beta}(p,q) &=& \Delta^{\mu\nu\alpha\beta}_I (p,q) \nn \\
&=&  \Psi_{1\, I} (s,0, M_Z^2,m_f^2,M_W^2,M_H^2) \, \phi_1^{\mu\nu\alpha\beta}(p,q) + \Psi_{4 \, I} (s,0, M_Z^2,M_W^2)  \, \phi_4^{\mu\nu\alpha\beta}(p,q).
\label{DAZ}
\eea
This is built by joining the graviton/Higgs mixing tree level vertex - coming from the improved energy-momentum tensor - and the Standard Model Higgs/photon/Z boson one-loop correlator.
\bea
\Psi_{1\, I} (s,0, M_Z^2,m_f^2,M_W^2,M_H^2) &=& - i \frac{\kappa}{2} \frac{\alpha}{6 \pi \, s_w \, c_w \, (s-M_H^2)(s-M_Z^2)} \bigg\{ 2 m_f^2 \, C_v^f \, Q_f \bigg[ 2 + 2 \frac{M_Z^2}{s-M_Z^2} \, \mathcal D_0\left(s,M_Z^2,m_f^2,m_f^2\right)   \nn \\
&& \hspace{-3cm}  + (4 m_f^2 + M_Z^2 -s) \mathcal C_0\left(s,0,M_Z^2,m_f^2,m_f^2,m_f^2\right) \bigg] + M_H^2 (1 - 2 s_w^2) + 2 M_Z^2 (6 s_w^4 - 11 s_w^2 + 5) \nn \\
&&  \hspace{-3cm}  + \frac{M_Z^2}{s-M_Z^2} \left(M_H^2 (1 - 2 s_w^2) + 2 M_Z^2 (6 s_w^4 - 11 s_w^2 + 5)\right) \mathcal D_0\left(s,M_Z^2,M_W^2,M_W^2\right) \nn \\
&&  \hspace{-3cm}  + 2 M_W^2  \, \mathcal C_0\left(s,0,M_Z^2,M_W^2,M_W^2,M_W^2\right) \left( M_H^2 (1-2s_w^2) + 2 M_Z^2 (6 s_w^4 - 15 s_w^2 + 8) + 2 s (4 s_w^2-3) \right)
\bigg\}, \nn \\ \\
\Psi_{4\, I} (s,0, M_Z^2,M_W^2) &=& - i \frac{\kappa}{2} \frac{\alpha \, c_w}{6 \pi \, s_w} M_Z^2 \bigg\{ \frac{2}{s - M_H^2} \mathcal B_0\left( 0, M_W^2,M_W^2\right) - s_w^2 \, \mathcal C_0\left(s,0,M_Z^2,M_W^2,M_W^2,M_W^2\right) \bigg\}. \nn \\
\eea
\subsection{Results for the TZZ correlator}

Our analysis starts with the irreducible amplitude and then we move to the insertions on the external graviton leg.\\
The irreducible vertex $\Sigma ^{\mu\nu\alpha\beta}(p,q)$ of the TZZ correlator for on-shell Z bosons can be separated into 
three contributions defined by the mass of the particles circulating in the loop, namely the fermion mass $m_f$ (fermion sector
(F)  with diagrams depicted in Fig. \ref{triangles}(a), Fig. \ref{t-bubble}(a)), the $W$ gauge boson mass $M_W$ (the $W$ gauge
boson sector (W) with diagrams Fig. \ref{triangles}(b)-(g), Fig. \ref{t-bubble}(b)-(g), Fig. \ref{s-bubble}, Fig.
\ref{tadpoles}),the $Z$ and the Higgs bosons masses, $M_Z$ and $M_H$ ($(Z,H)$ sector with the contributions represented in Figs.
\ref{triangles1} - \ref{tadpoles1}), which cannnot be separated because of scalar integrals with both masses in their internal
lines. There is also a diagram proportional to a Higgs tadpole (Fig. \ref{tadpolesHiggs}(a)) which vanishes after 
renormalization and so it is not included in the results given below. Finally there is the improvement term (I) given by 
the diagrams depicted in Fig. \ref{triangles}(c), (d), Fig. \ref{s-bubble}(b), Fig. \ref{triangles1}(b), (c), (d) and Fig.
\ref{s-bubble1} with the graviton - scalar - scalar vertices given by the $T_I^{\mu\nu}$. We obtain
\bea
\Sigma ^{\mu\nu\alpha\beta}(p,q) = \Sigma ^{\mu\nu\alpha\beta}_F (p,q)  + \Sigma ^{\mu\nu\alpha\beta}_W (p,q) + \Sigma
^{\mu\nu\alpha\beta}_{Z,H} (p,q) +  \Sigma ^{\mu\nu\alpha\beta}_I (p,q).
\eea
These four on-shell contributions can be expanded on a tensor basis given by 9 tensors
\bea
t_1^{\mu\nu\alpha\beta}(p,q) &=&  (s g^{\mu\nu} - k^{\mu}k^{\nu}) \left[ \left( \frac{s}{2}-M_Z^2 \right)g^{\alpha\beta} -
q^\alpha p^\beta \right] \,, \nn \\
t_2^{\mu\nu\alpha\beta}(p,q) &=&  (s g^{\mu\nu} - k^{\mu}k^{\nu}) g^{\alpha\beta} \,, \nn \\
t_3^{\mu\nu\alpha\beta}(p,q) &=&  g^{\mu\nu} g^{\alpha\beta} - 2 \left( g^{\mu\alpha} g^{\nu\beta} + g^{\mu\beta} g^{\nu\alpha}  \right) \,, \nn \\
t_4^{\mu\nu\alpha\beta}(p,q) &=& \left(p^{\mu}p^{\nu} + q^{\mu}q^{\nu} \right) g^{\alpha\beta} - M_Z^2 \left( g^{\mu\alpha} g^{\nu\beta} + g^{\mu\beta} g^{\nu\alpha}  \right) \,,\nn
\eea
\bea
t_5^{\mu\nu\alpha\beta}(p,q) &=&  \left(p^{\mu}q^{\nu} + q^{\mu}p^{\nu} \right) g^{\alpha\beta} - \left( \frac{s}{2} - M_Z^2 \right) \left( g^{\mu\alpha} g^{\nu\beta} + g^{\mu\beta} g^{\nu\alpha}  \right) \,, \nn \\
t_6^{\mu\nu\alpha\beta}(p,q) &=& \left(g^{\mu\alpha}q^{\nu} + g^{\nu\alpha}q^{\mu} \right) p^{\beta} +  \left(g^{\mu\beta}p^{\nu} + g^{\nu\beta}p^{\mu} \right) q^{\alpha} - g^{\mu\nu}p^{\beta}q^{\alpha} \,, \nn \\
t_7^{\mu\nu\alpha\beta}(p,q) &=& \left(g^{\mu\alpha}p^{\nu} + g^{\nu\alpha}p^{\mu} \right) p^{\beta} +  \left(g^{\mu\beta}q^{\nu} + g^{\nu\beta}q^{\mu} \right) q^{\alpha} \,, \nn \\
t_8^{\mu\nu\alpha\beta}(p,q) &=& \left[ 2 \left( p^{\mu} p^{\nu} + q^{\mu} q^{\nu} \right) - M_Z^2 g^{\mu\nu} \right] p^{\beta}q^{\alpha} \,, \nn \\
t_9^{\mu\nu\alpha\beta}(p,q) &=& \left[ 2 \left( p^{\mu} q^{\nu} + q^{\mu} p^{\nu} \right) - \left(\frac{s}{2}- M_Z^2 \right) g^{\mu\nu} \right] p^{\beta}q^{\alpha} \,,
\eea
and can be written in terms of form factors $\Phi_i$
\bea
\Sigma^{\mu\nu\alpha\beta}_F (p,q) &=& \sum_{i = 1}^9 {\Phi^{(F)}_i (s,M_Z^2,M_Z^2,m_f^2) \, t_i^{\mu\nu\alpha\beta}(p,q)} \,, \\
\Sigma^{\mu\nu\alpha\beta}_W (p,q) &=& \sum_{i = 1}^9 {\Phi^{(W)}_i (s,M_Z^2,M_Z^2,M_W^2) \, t_i^{\mu\nu\alpha\beta}(p,q)} \,, \\
\Sigma^{\mu\nu\alpha\beta}_{Z,H} (p,q) &=& \sum_{i = 1}^9 {\Phi^{(Z,H)}_i (s,M_Z^2,M_Z^2,M_Z^2,M_H^2) \, t_i^{\mu\nu\alpha\beta}(p,q)} \,, \\
\Sigma^{\mu\nu\alpha\beta}_{I} (p,q) &=& \Phi^{(I)}_1 (s,M_Z^2,M_Z^2,M_W^2,M_Z^2,M_H^2) \, t_1^{\mu\nu\alpha\beta}(p,q) +  \Phi^{(I)}_2 (s,M_Z^2,M_Z^2,M_W^2,M_Z^2,M_H^2) \, t_2^{\mu\nu\alpha\beta}(p,q) \,, \nn \\
\eea
where the first three arguments of the $\Phi_i$ represent the mass-shell and virtualities of the external lines $k^2 = s, \, p^2 = q^2 = M_Z^2$, while the remaining ones give the masses in the internal lines.\\ Moreover, we expand each form factor into a basis of independent scalar integrals.
\subsubsection{The fermion sector}
We start from the fermion contribution to $TZZ$ and then move to those coming from a $W$ running inside the loop (W loops) or a
$Z$ and a Higgs ($Z, H$ loops). We expand each form factor in terms of coefficients ${C_{(F)}}_j^i$
\bea
\Phi^{(F)}_i (s,M_Z^2,M_Z^2,m_f^2) = \sum_{j = 0}^4 {C_{(F)}}_j^i(s,M_Z^2,M_Z^2,m_f^2) \, \mathcal I^{(F)}_j
\label{fermionhZZ}
\eea
where $\mathcal I^{(F)}_j$ are a set of scalar integrals given by
\bea
\mathcal I^{(F)}_0 &=& 1 \, , \nn \\
\mathcal I^{(F)}_1 &=& \mathcal A_0(m_f^2) \, , \nn \\
\mathcal I^{(F)}_2 &=& \mathcal B_0(s, m_f^2,m_f^2) \, , \nn \\
\mathcal I^{(F)}_3 &=& \mathcal D_0(s, M_Z^2, m_f^2,m_f^2) \, , \nn \\
\mathcal I^{(F)}_4 &=& \mathcal C_0(s, M_Z^2,M_Z^2 ,m_f^2,m_f^2,m_f^2) \, . \label{IntF}
\eea
As in the previous case, only $\Phi^{(F)}_1$ contributes to the anomaly, and we will focus our attention only on this form factor. The expressions of all the coefficients ${C_{(F)}}_j^i$ for $(i\neq 1)$ can be found in appendix \ref{formfactors}. We obtain
\bea
{C_{(F)}}_0^1 &=&   -\frac{i  \kappa\, \alpha  \, m_f^2}{6 \pi  s^2 c_w^2 s_w^2 \left(s-4 M_Z^2\right)}
\left(s-2 M_Z^2\right) \left(C_a^{f \, 2}+C_v^{f\, 2}\right)
+ \frac{i \alpha \, \kappa}{36 \pi c_w^2 s_w^2 \, s}\left(C_a^{f \, 2}+C_v^{f \, 2}\right)\, ,  \nn \\
{C_{(F)}}_1^1 &=&   {C_{(F)}}_2^1 = 0 \, , \nn \\
{C_{(F)}}_3^1 &=&     -\frac{i \kappa\, \alpha  \, m_f^2}{3 \pi  s^2 c_w^2 \left(s-4 M_Z^2\right){}^2 s_w^2}
\left(\left(2 M_Z^4-3 s M_Z^2+s^2\right) C_a^{f \, 2}+C_v^{f \, 2} M_Z^2 \left(2 M_Z^2+s \right)\right) , , \nn \\
{C_{(F)}}_4^1 &=&   -\frac{i \kappa\, \alpha \, m_f^2}{12 \pi  s^2 c_w^2 \left(s-4 M_Z^2\right){}^2 s_w^2}
\left(s-2 M_Z^2\right) \bigg(\left(4 M_Z^4-2 \left(8 m_f^2+s \right) M_Z^2+s \left(4 m_f^2+s \right)\right) C_a^{f \, 2}\nn\\
&& +C_v^{f \, 2} \left(4 M_Z^4+2 \left(3 s-8 m_f^2\right) M_Z^2-s \left(s-4 m_f^2\right)\right)\bigg)\, .
\eea
The anomaly pole of $\Phi^{(F)}_1$ is entirely contained in ${C_{(F)}}_0^1$ and it is given by
\beq
\Phi^{(F)}_{1\, pole}\equiv\frac{i \alpha \, \kappa  \left(C_a^{f \, 2}+C_v^{f \, 2}\right)}{36 \pi c_w^2 s_w^2 \, s}\, .
\eeq
\subsubsection{The $W$ boson sector}
As we move to the contributions coming from loops of $W$'s, the 9 form factors are expanded as
\bea
\Phi^{(W)}_i (s,M_Z^2,M_Z^2,M_W^2) = \sum_{j = 0}^4 {C_{(W)}}_j^i(s,M_Z^2,M_Z^2,M_W^2) \, \mathcal I^{(W)}_j
\label{boson1hZZ}
\eea
where $\mathcal I^{(W)}_j$ are now given by
\bea
\mathcal I^{(W)}_0 &=& 1 \, , \nn \\
\mathcal I^{(W)}_1 &=& \mathcal A_0(M_W^2) \, , \nn \\
\mathcal I^{(W)}_2 &=& \mathcal B_0(s, M_W^2,M_W^2) \, , \nn \\
\mathcal I^{(W)}_3 &=& \mathcal D_0(s, M_Z^2, M_W^2,M_W^2) \, , \nn \\
\mathcal I^{(W)}_4 &=& \mathcal C_0(s, M_Z^2,M_Z^2 ,M_W^2,M_W^2,M_W^2) \, . \label{IntW}
\eea
The anomaly pole is extracted from the expansion of $\Phi^{(W)}_1 $, whose coefficients are
\bea
{C_{(W)}}_0^1 &=&  \frac{-i \kappa\, \alpha}{2 s_w^2 \, c_w^2 \, \pi \, s}
\bigg\{ \frac{M_Z^2}{6 s \left(s-4 M_Z^2\right)} \bigg[ 2 M_Z^2 \left(-12 s_w^6+32 s_w^4-29 s_w^2+9\right)\nn\\
&& +s \left(12 s_w^6-36 s_w^4+33 s_w^2-10\right)\bigg]+  \frac{(60s_w^4-148s_w^2+81)}{72}  \bigg\}\, , \nn \\
{C_{(W)}}_1^1 &=& {C_{(W)}}_2^1 = 0\, , \nn \\
{C_{(W)}}_3^1 &=&  \frac{- i \kappa \, \alpha \, M_Z^2}{ 12 s_w^2 \, c_w^2 \, \pi \, s^2 \, (s-4 M_Z)^2}   \bigg( 4 M_Z^4 (12 s_w^6 -32 s_w^4 +29 s_w^2 - 9) \nn \\
 && + 2 M_Z^2 s (s_w^2 - 2)(12 s_w^4 - 12 s_w^2 +1) + s^2 (-4 s_w^4+8s_w^2-5)   \bigg)\, , \nn \\
{C_{(W)}}_4^1 &=&   \frac{- i \kappa \, \alpha \, M_Z^2}{ 12 s_w^2 \, c_w^2 \, \pi \, s^2 \, (s-4 M_Z)^2}  \bigg(-4M_Z^6(s_w^2-1)(4 s_w^2-3)(12 s_w^4-20s_w^2+9) \nn \\
&& + 2 M_Z^4 s (18 s_w^4-34s_w^2 + 15)(4(s_w^2-3)s_w^2+7) - 2M_Z^2 s^2 (12 s_w^8-96s_w^6 \nn \\
&& +201s_w^4-157s_w^2+41)+ s^3(-12 s_w^6+32s_w^4-27s_w^2+7)  \bigg) \, .
\eea
As one can immediately see, the pole is entirely contained in ${C_{(W)}}_0^1$, and we obtain
\beq
\Phi^{(W)}_{1\, pole} \equiv  - i \frac{\kappa}{2} \frac{\alpha}{s_w^2 \, c_w^2 \, \pi \, s} \frac{(60s_w^4-148s_w^2+81)}{72}\, .
\eeq
\subsubsection{The $(Z,H)$ sector}
Finally, the last contribution to investigate in the $TZZ$ vertex is the one coming from a Higgs
($H$) or a Z boson ($Z$) running in the loops. Also in this case we obtain
\bea
\Phi^{(Z,H)}_i (s,M_Z^2,M_Z^2,M_Z^2,M_H^2) = \sum_{j = 0}^8 {C_{(Z,H)}}_j^i(s,M_Z^2,M_Z^2,M_Z^2,M_H^2) \, \mathcal I^{(Z,H)}_j
\label{boson2hZZ}
\eea
with the corresponding $\mathcal I^{(Z,H)}_j$ given by
\bea
\mathcal I^{(Z,H)}_0 &=& 1 \, , \nn \\
\mathcal I^{(Z,H)}_1 &=& \mathcal A_0(M_Z^2) \, , \nn \\
\mathcal I^{(Z,H)}_2 &=& \mathcal A_0(M_H^2) \, , \nn \\
\mathcal I^{(Z,H)}_3 &=& \mathcal B_0(s, M_Z^2,M_Z^2) \, , \nn \\
\mathcal I^{(Z,H)}_4 &=& \mathcal B_0(s, M_H^2,M_H^2) \, , \nn \\
\mathcal I^{(Z,H)}_5 &=& \mathcal B_0(M_Z^2, M_Z^2,M_H^2) \, , \nn \\
\mathcal I^{(Z,H)}_6 &=& \mathcal C_0(s, M_Z^2,M_Z^2 ,M_Z^2,M_H^2,M_H^2) \, , \nn \\
\mathcal I^{(Z,H)}_7 &=& \mathcal C_0(s, M_Z^2,M_Z^2 ,M_H^2,M_Z^2,M_Z^2) \, . \label{IntZH}
\eea
Again, as before, the contributions to $\Phi^{(Z,H)}_1$ are those responsible for a non vanishing trace in the massless limit. These are given by
\bea
{C_{(Z,H)}}_0^1 &=& \frac{i \kappa \, \alpha}{24 \pi  s^2 c_w^2 s_w^2 \left(s-4 M_Z^2\right)}
\left(M_H^2 \left(s-2 M_Z^2\right)+3 s M_Z^2- 2 M_Z^4\right) + \frac{7 i \alpha  \kappa }{144 \pi  s c_w^2 s_w^2}\, ,  \nn \\
{C_{(Z,H)}}_1^1 &=& \frac{i \kappa \, \alpha}{12 \pi  s^2 c_w^2 s_w^2 \left(s-4 M_Z^2\right)}
\left(M_Z^2-M_H^2\right) \, ,\nn \\
{C_{(Z,H)}}_2^1 &=& - {C_{(Z,H)}}_1^1  \, ,\nn \\
{C_{(Z,H)}}_3^1 &=&  \frac{i \kappa \, \alpha}{24 \pi  s^2 c_w^2 s_w^2 \left(s-4 M_Z^2\right){}^2}
   \left(2 M_H^2 \left(s M_Z^2-2 M_Z^4+s^2\right)+3 s^2 M_Z^2-14 s M_Z^4+8 M_Z^6\right)  \, ,\nn \\
{C_{(Z,H)}}_4^1 &=&  -\frac{i \kappa \, \alpha}{24 \pi  s^2 c_w^2 s_w^2 \left(s-4 M_Z^2\right){}^2}
   \left(2 M_H^2+s\right) \left(2 M_H^2 \left(s-M_Z^2\right)-3 s M_Z^2\right) \, ,  \nn \\
{C_{(Z,H)}}_5^1 &=&  \frac{i \kappa \, \alpha}{12 \pi  s^2 c_w^2 \left(s-4 M_Z^2\right){}^2 s_w^2}
\left(s M_H^4+6 \left(s-M_H^2\right) M_Z^4+\left(2 M_H^4-3 s M_H^2-3 s^2\right) M_Z^2\right)\, ,\nn \\
{C_{(Z,H)}}_6^1 &=&  \frac{i \kappa \, \alpha}{24 \pi  s^2 c_w^2 s_w^2 \left(s-4 M_Z^2\right){}^2}
\left(2 M_H^2+s\right) \bigg(M_Z^2 \left(-8 s M_H^2-2 M_H^4+s^2\right)\nn\\
&& +2 M_Z^4 \left(4 M_H^2+s\right)+2 s M_H^4\bigg) \, ,\nn  \\
{C_{(Z,H)}}_7^1 &=& \frac{i \kappa \, \alpha  M_H^2}{24 \pi  s^2 c_w^2 s_w^2
   \left(s-4 M_Z^2\right){}^2} \left(2 M_H^2 \left(s M_Z^2-2
   M_Z^4+s^2\right)-20 s M_Z^4+16 M_Z^6+s^3\right)\, ,
\eea
with the anomaly pole, extracted from ${C_{(Z,H)}}_0^1$, given by
\beq
\Phi^{(Z,H)}_{1\, pole} \equiv \frac{7 i \alpha  \kappa }{144 \pi  s c_w^2 s_w^2}\, .
\eeq
\subsubsection{ Terms of improvement and external leg corrections}
The expression of form factors $\Phi^{(I)}_1$ and $\Phi^{(I)}_2$ coming from the terms of improvement for the $\Sigma^{\mu\nu\alpha\beta}_I(p,q)$ vertex are given in appendix \ref{imprformfactors}.

The next task is to analyze the external leg corrections to the $TZZ$ correlator. This case is much more involved than the previous one because there are contributions coming from the minimal EMT (i.e. without the improvement terms) Fig. \ref{tadpolesHiggs}(b), Fig. \ref{legcorr12}(a)-(b) and from the improved $T^{\mu\nu}_I$. This last contribution can be organized into three sectors: the first is characterized by a contribution from the one-loop graviton/Higgs two-point function Fig. \ref{tadpolesHiggs}(b), Fig. \ref{legcorr12}(a). The second is constructed with the Higgs self-energy Fig. \ref{legcorr12}(c) and the last is built with the Standard Model Higgs/Z/Z one-loop vertex Fig. \ref{HVVImpr}. Furthermore, it is important to note that the diagram depicted in Fig. \ref{tadpolesHiggs}(b) is proportional to the Higgs tadpole and vanishes in our renormalization scheme. \\
The $\Delta^{\mu\nu\alpha\beta}(p,q)$ correlator is decomposed as
\bea
\Delta^{\mu\nu\alpha\beta}(p,q) &=& \bigg[ \Sigma^{\mu\nu}_{Min, \, hH}(k) + \Sigma^{\mu\nu}_{I, \, hH}(k) \bigg] \frac{i}{k^2 - M_H^2} V_{HZZ}^{\alpha\beta} + V_{I, \, hH}^{\mu\nu}(k)  \frac{i}{k^2 - M_H^2}  \Sigma_{HH}(k^2) \frac{i}{k^2 - M_H^2} V_{HZZ}^{\alpha\beta}  \nn \\
&+& \Delta^{\mu\nu\alpha\beta}_{I, \, HZZ}(p,q)
\eea
where $\Sigma_{HH}(k^2)$ is the Higgs self-energy given in appendix (\ref{propagators}) for completeness, $V_{HZZ}^{\alpha\beta}$ and $ V_{I, \, hH}^{\mu\nu}$ are tree level vertices defined in appendix (\ref{FeynRules}) and $\Delta^{\mu\nu\alpha\beta}_{I, \, HZZ}(p,q)$ is expanded into the two form factors of improvement as
\bea
\Delta^{\mu\nu\alpha\beta}_{I, \, HZZ}(p,q) = \Psi^{(I)}_1 (s,M_Z^2,M_Z^2,m_f^2,M_W^2,M_Z^2,M_H^2) \, t_1^{\mu\nu\alpha\beta}(p,q) +  \Psi^{(I)}_2 (s,M_Z^2,M_Z^2,m_f^2,M_W^2,M_Z^2,M_H^2) \, t_2^{\mu\nu\alpha\beta}(p,q) \, ,\nn
\eea
\bea
\Psi^{(I)}_i (s,M_Z^2,M_Z^2,m_f^2,M_W^2,M_Z^2,M_H^2)  &=& \sum^4_{j=0} \, {C^{(I)}_{(F)}}^i_j \left( s,M_Z^2,M_Z^2,m_f^2 \right) \, \mathcal I^{(F)}_j + \sum^4_{j=0} \, {C^{(I)}_{(W)}}^i_j \left( s,M_Z^2,M_Z^2,M_W^2 \right) \, \mathcal I^{(W)}_j \nn \\
&+& \sum^7_{j=0} \, {C^{(I)}_{(Z,H)}}^i_j \left( s,M_Z^2,M_Z^2,M_Z^2,M_H^2 \right) \, \mathcal J^{(Z,H)}_j
\eea
where the basis of scalar integrals $\mathcal I^{(F)}_j$ and $\mathcal I^{(W)}_j$ have been defined respectively in Eq. \ref{IntF} and \ref{IntW}. The $(Z,H)$ sector is expanded into a different set (instead of Eq. \ref{IntZH}) which is given by
\bea
\mathcal J^{(Z,H)}_0 &=& 1 \,, \nn \\
\mathcal J^{(Z,H)}_1 &=& \mathcal A_0 \left( M_Z^2 \right) \,, \nn \\
\mathcal J^{(Z,H)}_2 &=& \mathcal A_0 \left( M_H^2 \right) \,, \nn \\
\mathcal J^{(Z,H)}_3 &=& \mathcal B_0 \left( s, M_Z^2, M_Z^2 \right) \,, \nn \\
\mathcal J^{(Z,H)}_4 &=& \mathcal B_0 \left( s, M_H^2, M_H^2 \right) \,, \nn \\
\mathcal J^{(Z,H)}_5 &=& \mathcal B_0 \left( M_Z^2, M_Z^2, M_H^2 \right) \,, \nn \\
\mathcal J^{(Z,H)}_6 &=& \mathcal C_0 \left( s, M_Z^2, M_Z^2, M_Z^2,M_H^2,M_H^2 \right) \,, \nn \\
\mathcal J^{(Z,H)}_7 &=& \mathcal C_0 \left( s, M_Z^2, M_Z^2, M_H^2,M_Z^2,M_Z^2 \right) \,.
\eea
The expressions of these coefficients together with the graviton-Higgs mixing $\Sigma^{\mu\nu}_{Min, \, hH}(k)$,  $\Sigma^{\mu\nu}_{I, \, hH}(k)$ can be found in appendix \ref{externalleg}.

\section{Renormalization}
In this section we discuss the renormalization of the correlators. This is based on the identification of the
1-loop counterterms to the Standard Model Lagrangian which, in turn, allow to extract a counterterm vertex for the improved EMT.
We have checked that the renormalization of all the parameters of the Lagrangian is indeed sufficient to cancel all the singularities of all the vertices, as expected. We have used the on-shell scheme which is widely used in the electroweak theory. In this scheme the
renormalization conditions are fixed in terms of the physical parameters of the theory to all orders in perturbation theory. These are the masses of physical particles  $M_W, M_Z, M_H, m_f$, the electric charge $e$ and the quark mixing matrix $V_{ij}$. The renormalization conditions on the fields - which allow to extract the renormalization constants of the wave functions - are obtained by requiring a unit residue of the full 2-point functions on the physical particle poles.

We start by defining the relations
\bea
e_0 &=& (1+ \delta Z_e) e \,, \nn \\
M^2_{W,0} &=& M^2_W + \delta M^2_W \, , \nn \\
M^2_{Z,0} &=& M^2_Z + \delta M^2_Z \, , \nn \\
M^2_{H,0} &=& M^2_H + \delta M^2_H \, , \nn \\
\left( \begin{array}{c} Z_0 \\ A_0 \end{array} \right) &=& \left( \begin{array}{cc} 1+\frac{1}{2} \delta Z_{ZZ} & \frac{1}{2} \delta Z_{ZA} \\  \frac{1}{2} \delta Z_{AZ} & 1+ \frac{1}{2} \delta Z_{AA} \end{array}\right) \left( \begin{array}{c} Z \\ A \end{array} \right) \,, \nn \\
H_0 &=& \left( 1 + \frac{1}{2}\delta Z_H \right) H.
\eea
At the same time we need the counterterms for the sine of the Weinberg angle $s_w$ and of the vev of the Higgs field $v$
\bea
s_{w\,,0} = s_w + \delta s_w \,, \qquad v_0 = v + \delta v, \,
\eea
which are defined to all orders by the relations
\bea
s_w^2 = 1 - \frac{M_W^2}{M_Z^2} \,, \qquad v^2 = 4 \frac{M_W^2 \, s_w^2}{e^2} \,,
\eea
and are therefore linked to the renormalized masses and gauge couplings. Specifically, one obtains
\bea
\frac{\delta s_w}{s_w} = - \frac{c_w^2}{2 s_w^2} \left( \frac{\delta M_W^2}{M_W^2} - \frac{\delta M_Z^2}{M_Z^2} \right) \,, \qquad \frac{\delta v}{v} = \left( \frac{1}{2} \frac{\delta M_W^2}{M_W^2} + \frac{\delta s_w}{s_w} - \delta Z_e \right),
\eea
while electromagnetic gauge invariance gives
\bea
\delta Z_e = - \frac{1}{2}\delta Z_{AA} + \frac{s_w}{2 c_w}\delta Z_{ZA}.
\eea
We also recall that the wave function renormalization constants are defined in terms of the 2-point functions of the fundamental fields as
\bea
&& \delta Z_{AA} = - \frac{\partial \Sigma_T^{AA}(k^2)}{\partial k^2} \bigg |_{k^2=0} \,, \quad
\delta Z_{AZ} = - 2 Re \frac{\Sigma_T^{AZ}(M_Z^2)}{M_Z^2} \,, \quad
\delta Z_{ZA} = 2 \frac{\Sigma_T^{AZ}(0)}{M_Z^2} \,, \nn\\
&&\delta Z_{ZZ} = - Re \frac{\partial \Sigma_T^{ZZ}(k^2)}{\partial k^2} \bigg |_{k^2 = M_Z^2} \,, \quad
\delta Z_H = - Re \frac{\partial \Sigma_{HH}(k^2)}{\partial k^2} \bigg |_{k^2 = M_H^2} \,,\quad
\delta M_Z^2 = Re \, \Sigma_T^{ZZ}(M_Z^2) \,, \nn\\
&&\delta M_W^2 = \widetilde{Re} \, \Sigma_T^{WW}(M_W^2) \,, \quad \delta M_H^2 = Re \, \Sigma_{HH}(M_H^2). \,
\eea
From the counterterms Lagrangian defined in terms of the $Z_{V V'}$ factors given above, we compute the corresponding counterterm to the EMT $\delta T^{\mu\nu}$ and renormalized EMT
\bea
T^{\mu\nu}_0 = T^{\mu\nu} + \delta T^{\mu\nu}
\eea
which is sufficient to cancel all the divergences of the theory. One can also verify from the explicit computation that the terms of improvement, in the conformally coupled case, are necessary to renormalize the vertices containing an intermediate scalar with an external bilinear mixing (graviton/Higgs).
The vertices extracted from the counterterms are given by
\bea
\delta [TAA]^{\mu\nu\alpha\beta}(k_1,k_2) &=& - i \frac{\kappa}{2} \bigg\{ k_1 \cdot k_2 \, C^{\mu\nu\alpha\beta} + D^{\mu\nu\alpha\beta}(k_1,k_2)\bigg\} \, \delta Z_{AA} \,, \\
\delta [TAZ]^{\mu\nu\alpha\beta}(k_1,k_2) &=& - i \frac{\kappa}{2} \bigg\{ \left( \delta c_1^{AZ} \, k_1 \cdot k_2 + \delta c_2^{AZ} \, M_Z^2 \right) \, C^{\mu\nu\alpha\beta} + \delta c_1^{AZ} \, D^{\mu\nu\alpha\beta}(k_1,k_2) \bigg\} \,, \\
\delta [TZZ]^{\mu\nu\alpha\beta}(k_1,k_2) &=& - i \frac{\kappa}{2} \bigg\{ \left( \delta c_1^{ZZ} \, k_1 \cdot k_2 + \delta c_2^{ZZ} \, M_Z^2 \right) \, C^{\mu\nu\alpha\beta} + \delta c_1^{ZZ} \, D^{\mu\nu\alpha\beta}(k_1,k_2) \bigg\} \,,
\eea
where the coefficients $\delta c$ are defined as
\bea
\delta c_1^{AZ} = \frac{1}{2}\left( \delta Z_{AZ} + \delta Z_{ZA} \right) \,, \quad \delta c_2^{AZ} = \frac{1}{2} \delta Z_{ZA} \,, \quad
\delta c_1^{ZZ} = \delta Z_{ZZ} \,, \quad  \delta c_2^{ZZ} = M_Z^2 \, \delta Z_{ZZ} + \delta M_Z^2 \,.
\eea
These counterterms are sufficient to remove the divergences of the completely cut graphs ($\Sigma^{\mu\nu\alpha\beta}(p,q)$) which do not contain a bilinear mixing, once we set
on-shell  the external gauge lines. This occurs both for those diagrams which do not involve the terms of improvement and for those involving $T_I$. Regarding those contributions which involve the bilinear mixing on the external graviton line, we encounter two situations. For instance, the insertion of the bilinear mixing on the $TAA$ vertex generates a reducible diagram of the form Higgs/photon/photon which does not require any renormalization, being finite. Its contribution has been denoted as $\Delta^{\mu\nu\alpha\beta}_I(p,q)$ in Eq. (\ref{DAA}). In the case of the $TAZ$ vertex the corresponding contribution is given in Eq. (\ref{DAZ}). In this second case the renormalization is guaranteed, within the Standard Model, by the use of the Higgs/photon/Z counterterm
\bea
\delta [HAZ]^{\alpha\beta} = i \frac{e \, M_Z}{2 s_w c_w} \delta Z_{ZA}\, \eta^{\alpha\beta} \,.
\eea
As a last case, we discuss the contribution to $TZZ$ coming from the bilinear mixing. The corrections on the graviton line involve the graviton/Higgs mixing $i \Sigma^{\mu\nu}_{hH}(k)$, the Higgs self-energy $i \Sigma_{HH}(k^2)$ and the term of improvement $\Delta^{\mu\nu\alpha\beta}_{I\,,HZZ}(p,q)$, which introduces the Higgs/Z/Z vertex (or $HZZ$) of the Standard Model. The Higgs self-energy and the $HZZ$ vertex, in the Standard Model, are renormalized with the counterterms
\bea
\delta [HH](k^2) &=& i (\delta Z_H \, k^2 - M_H^2 \delta Z_H - \delta M_H^2) \, ,\\
\delta [HZZ]^{\alpha\beta} &=& i \frac{e \, M_Z}{s_w \, c_w} \bigg[ 1 + \delta Z_e + \frac{2 s_w^2 - c_w^2}{c_w^2} \frac{\delta s_w}{s_w} + \frac{1}{2} \frac{\delta M_W^2}{M_W^2} + \frac{1}{2} \delta Z_H + \delta Z_{ZZ}  \bigg] \, \eta^{\alpha\beta} \,.
\eea
The self-energy $i \Sigma^{\mu\nu}_{hH}(k)$ is defined by the minimal contribution generated by $T_{\mu\nu}^{Min}$ and by a second term derived from $T_{\mu\nu}^I$. This second term is necessary in order to ensure the renormalizability of the graviton/Higgs mixing.
In fact, the use of the minimal EMT in the computation of this self-energy involves a divergence of the form
\bea
\delta [hH]^{\mu\nu}_{Min} = i \frac{\kappa}{2} \delta t \, \eta^{\mu\nu} \,, \label{CThH}
\eea
with $\delta t$ fixed by the condition of cancellation of the Higgs tadpole $T_{ad}$ ($\delta t + T_{ad} = 0$) and hence of any linear term in $H$ within the 1-loop effective Lagrangian of the Standard Model.
A simple analysis of the divergences in $i \Sigma^{\mu\nu}_{Min, \, hH}$ shows that the counterterm given in Eq. \ref{CThH} is not sufficient to remove all the singularities of this correlator unless we also include the renormalization of the term of improvement which is given by
\bea
\delta [hH]^{\mu\nu}_{I}(k) = - i \frac{\kappa}{2} \left( - \frac{1}{3} \right) i  \bigg[ \delta v + \frac{1}{2} \delta Z_H \bigg] v \, (k^2 \, \eta^{\mu\nu} - k^{\mu}k^{\nu}).
\eea
One can show explicitly that this counterterm indeed ensures the finiteness of $i \Sigma^{\mu\nu}_{hH}(k)$.
\section{Comments}
\label{discussionsection}
Before coming to our conclusions, we pause for some comments on the meaning and the implications of the current computation in a more general context. This concerns the superconformal anomaly and its coupling to supergravity, aspects that we will address more completely in the near future.

 The study of the mechanism of anomaly mediation between the Standard Model and gravity has several interesting features which for sure will require further analysis in order to be put on a more rigorous basis. However, here we have preliminarily shown that the perturbative structure of a correlator - obtained by the insertion of a gravitational field on 2-point  functions of gauge fields - can be organized in terms of a rather minimal set of fundamental form factors. Their expressions have been given in this work, generalizing previous results in the QED and QCD cases. The trace anomaly can be attributed, in all the cases, just to one specific tensor structure, as discussed in the previous analysis.

We have also seen that at high energy the breaking of conformal invariance, in a theory with a Higgs mechanism, has two sources, one of them being radiative. This can be attributed to the exchange of anomaly poles in each gauge invariant sector of the graviton/gauge/gauge vertex, while the second one is explicit. As discussed in \cite{Coriano:2011ti} this result has a simple physical interpretation,
since it is an obvious consequence of the fact that at an energy much larger than any scale of the theory, we should recover the role of the anomaly and its pole-like behaviour. 

In turn, this finding sheds some light on the significance of the anomaly cancellation mechanism in 4-dimensional field theory - discussed in the context of supersymmetric theories coupled to gravity - based on the subtraction of an anomaly pole in superspace  \cite{LopesCardoso:1991zt}. Let's briefly see why. 

The theory indeed becomes conformally invariant at high energy and, in presence of supersymmetric interactions, this invariance is promoted to a superconformal invariance.  In a superconformal theory, such as an ${\cal N}=1$ super Yang-Mills theory,  the superconformal anomaly multiplet, generated by the radiative corrections, puts on the same role the trace anomaly, the chiral anomaly of the corresponding $U(1)_R$ current and the gamma trace of the corresponding supersymmetric current. Notice that these three anomalies are "gauged" if they are coupled to a conformal gravity supermultiplet and all equally need to be cancelled. The role of the Green-Schwarz (GS) mechanism, in this framework, if realized as a pole subtraction, is then to perform a subtraction of these pole-like contributions which show up in the UV region, and has to be realized in superspace \cite{LopesCardoso:1991zt,Bagger:1999rd} for obvious reasons. Then, one can naturally ask what is the nature of the pole that is indeed cancelled by the mechanism, if this is acting in the UV. The answer, in a way, is obvious, since the mechanism works as an ultraviolet completion: the "poles" found in the perturbative analysis are a manifestation of the anomaly in the UV.

As we have explained at length in  \cite{Coriano:2011ti} these poles extracted in each gauge invariant sector do not couple in the infrared region, since the theory is massive and conformal invariance is lost in the broken electroweak phase. Looking for a residue of these poles in the IR, in the case of a massive theory, is simply meaningless. Indeed their role is recuperated in the UV, 
where they describe an effective  massless exchange present in the amplitude at high energy.

Therefore, the $1/s$ behaviour found in these correlators at high energy is the unique signature of the anomaly (they saturate the anomaly) in the same domain, and is captured within an asymptotic expansion in $v^2/s$  \cite{Coriano:2011ti}. Thus, the anomalous nature of the theory reappears as we approach a (classically) conformally invariant theory, with $s$ going to infinity.

 Obviously, this picture is only approximate, since the cancellation of the trace anomaly by the subtraction of a pole in superspace 
remains an open issue, given the fact that the trace anomaly takes contribution at all orders both in $G_N$ and in the gauge coupling.  
 The resolution of this point would require computations similar to the one that we have just performed for correlators of higher order. 
  Indeed, this is another aspect of the "anomaly puzzle" in supersymmetric theories when (chiral) gauge anomalies and trace anomalies appear on the same level, due to their coupling with gravity.

\section{Conclusions and Perspectives}
  \label{conclusions}
We have presented a complete study of the interactions between gravity and the fields of the Standard Model which are responsible for the generation of a trace anomaly in the corresponding effective action. The motivations in favour of these type of studies are several and cover both the cosmological domain and collider physics. In this second case these corrections are important especially in the phenomenological analysis of theories with a low gravity scale/large extra dimensions. We have defined rigorously the structure of these correlators, via an appropriate set of Ward and Slavnov-Taylor identities that we have derived from first principles. We have given the explicit expressions of these corrections, extending to the neutral current sector of the SM previous analysis performed in the QED and QCD cases. We hope to return in the near future with a study of the charged current sector and a complete characterization of the effective Lagrangian of the SM. Here we have made a first step
in that direction.

\vspace{1cm}
\centerline{\bf Acknowledgements}
We thank Alan White for discussions.

\appendix
\section{Appendix}
\label{conventions}
We summarize here some of our conventions used in the computation of the various contributions to the total EMT of the SM.\\
The definitions of the field strengths are
\bea
F^a_{\mu\nu} & = & \pd_\mu G^a_\nu - \pd_\nu G^a_\mu + g_s f^{abc}G^b_\mu G^c_\nu \,, \\
F^A_{\mu\nu} & = & \pd_\mu A_\nu - \pd_\nu A_\mu + g\sin\th_W\c_{\mu\nu} \,, \\
Z_{\mu\nu} & = & \pd_\mu Z_\nu - \pd_\nu Z_\mu + g\cos\th_W\c_{\mu\nu} \,,\\
W^+_{\mu\nu} & = & \pd_\mu W^+_\nu - \pd_\nu W^+_\mu - i g\left[cos\th_W Z_\mu W^+_\nu + \sin\th_W A_\mu W^+_\nu - (\mu \leftrightarrow \nu)\right] \,, \\
W^-_{\mu\nu} & = & \pd_\mu W^-_\nu - \pd_\nu W^-_\mu + i g\left[cos\th_W Z_\mu W^-_\nu + \sin\th_W A_\mu W^-_\nu - (\mu \leftrightarrow \nu)\right] \,,
\eea
with $\c_{\mu\nu}$ given by  $\c_{\mu\nu} = i[W^-_\mu W^+_\nu - W^+_\mu W^-_\nu]$. As usual, we have denoted with $f^{abc}$ the structure constants of $SU(3)_C$, while $e = g \sin\th_W$.
The fermionic Lagrangian is
\bea
\mathcal{L}_{ferm.}
&=& i\bar\psi_{\nu_e}\g^\mu\pd_\mu \psi_{\nu_e} + i\bar \psi_e\g^\mu\pd_\mu \psi_e
    + i\bar \psi_u\g^\mu\pd_\mu \psi_u + i\bar \psi_d\g^\mu\pd_\mu \psi_d  + \frac{e}{\sqrt{2}\sin\th_W}\bigg(\bar\psi_{\nu_e} \g^\mu \frac{1-\g^5}{2} \psi_e\, W^+_\mu \nn \\
&+& \bar \psi_e\g^\mu\frac{1-\g^5}{2}\psi_{\nu_e}\, W^-_\mu\bigg)
 + \frac{e}{\sin2\th_W}\bar\psi_{\nu_e} \g^\mu\frac{1-\g^5}{2}\psi_{\nu_e} Z_\mu  - \frac{e}{\sin2\th_W}\bar \psi_e\g^\mu\bigg(\frac{1-\g^5}{2} \nn \\
 &-& 2\sin^2\th_W\bigg)\psi_e\,Z_\mu
 + \frac{e}{\sqrt{2}\sin\th_W}\bigg(\bar \psi_u\g^\mu\frac{1-\g^5}{2}\psi_d\,W^+_\mu + \bar \psi_d\g^\mu
\frac{1-\g^5}{2}\psi_u\,W^-_\mu \bigg)\nn\\
&+&  \frac{e}{\sin2\th_W}\bar \psi_u\g^\mu\bigg(\frac{1-\g^5}{2} -
2\sin^2\th_W\frac{2}{3}\bigg)\psi_u\,Z_\mu
 - \frac{e}{\sin2\th_W}\bar \psi_d\g^\mu\bigg(\frac{1-\g^5}{2} -
2\sin^2\th_W\frac{1}{3}\bigg)\psi_d\,Z_\mu\bigg]\nn\\
&+& eA_\mu\bigg(-\bar \psi_e\g^\mu \psi_e + \frac{2}{3}\,\bar \psi_u\g^\mu \psi_u - \frac{1}{3}\,\bar \psi_d\g^\mu
\psi_d\bigg) + g_s G^a_\mu\bigg(\bar \psi_u\g^\mu t^a \psi_u + \bar \psi_d\g^\mu t^a \psi_d \bigg) \, .
\eea
The gauge-fixing Lagrangian is given by
\beq
\mathcal L_{g. fix.} = -\frac{1}{2\xi}(\mathcal F^A)^2 -\frac{1}{2\xi}(\mathcal F^Z)^2 -\frac{1}{\xi}(\mathcal F^+)(\mathcal F^-) - \frac{1}{2\xi}(\mathcal F^G)^2 \,,
\eeq
where the gauge-fixing functions in the $R_\xi$ gauge are defined by
\bea
\mathcal F^{G,i} & = & \pd^\s G^i_\s\, ,\nn\\
\mathcal F^A & = & \pd^\s A_\s\, ,\nn\\
\mathcal F^Z & = & \pd^\s Z_\s - \xi M_Z \f\, ,\nn\\
\mathcal F^+ & = & \pd^\s W^+_\s - \frac{1}{2}\xi g v \f^+\, ,\nn\\
\mathcal F^- & = & \pd^\s W^-_\s - \frac{1}{2}\xi g v \f^-\,,
\eea
and we have used for simplicity the same gauge-fixing parameter $\xi$ for all the gauge fields. Finally we give the ghost Lagrangian
\bea
\mathcal{L}_{ghost}
& = & \pd^\mu\bar{c}^a \left(\pd_\mu\d^{ac} + g_s f^{abc}G^b_\mu\right)c^c
 + \pd^\mu\bar{\h}^Z\pd_\mu\h^Z + \pd^\mu\bar{\h}^A\pd_\mu\h^A + \pd^\mu\bar{\h}^+\pd_\mu\h^+ +   \pd^\mu\bar{\h}^-\pd_\mu\h^- \nn \\
&+&  i g \bigg\{\pd^\mu\bar{\h}^+\bigg[W_\mu^+ (\cos\th_W \h^Z + \sin\th_W\h^A) - (\cos\th_W Z_\mu + \sin\th_W A_\mu)\h^+ \bigg] \nn \\
&+& \pd^\mu\bar{\h}^-\bigg[\h^- (\cos\th_W Z_\mu + \sin\th_W A_\mu) - (\cos\th_W\h^Z + \sin\th_W \h^A)W_\mu^-\bigg]\nn \\
&+&  \pd^\mu(cos\th_W\bar{\h}^Z + \sin\th_W\bar{\h}^A )\bigg[W^+_\mu\h^- -   W^-_\mu\h^+\bigg]\bigg\}  - \frac{e \, \xi \, M_W}{\sin2\th_W} \bigg\{-i\f^+\bigg[\cos2\th_W\bar{\h}^+\h^Z \nn \\
&+&   \sin2\th_W\bar{\h}^+\h^A\bigg] + i \f^-\bigg[\cos2\th_W\bar{\h}^-\h^Z + \sin2\th_W\bar{\h}^-\h^A\bigg]\bigg\} - \frac{e\xi}{2\sin\th_W} M_W \bigg[(v + h + i \f)\bar{\h}^+\h^+ \nn \\
&+& (v + h - i \f) \bar{\h}^-\h^-\bigg] - i \frac{e\xi}{2\sin\th_W} M_Z ( - \f^-\bar{\h}^Z\h^+ + \f^+\bar{\h}^Z\h^-) - \frac{e\,\xi \, M_Z}{\sin2\th_W}(v + h)\bar{\h}^Z\h^Z\, .
\eea
\section{Appendix. Ward identities}
\label{ward}
For the derivation of the Ward identities, the transformations of the fields are given by (we have absorbed a factor $\sqrt{-g}$ in their definitions)
\beqa
V^{'\,\underline{a}}_\mu(x)           &=& V^{\underline{a}}_\mu(x) -\int d^4y\,[\d^{(4)}(x-y)\pd_\nu V^{\underline{a}}_\mu(x)
                               + [\pd_\mu\d^{(4)}(x-y)]V^{\underline{a}}_\nu]\e^\nu(y)\, ,\nn\\
J'(x)                      &=& J(x) -\int d^4y\,\pd_\nu[\d^{(4)}(x-y)J(x)]\e^\nu(y)\, ,\nn\\
\c'(x)                     &=& \c(x) -\int d^4y\,\pd_\nu[\d^{(4)}(x-y)\c(x)]\e^\nu(y)\, .
\eeqa
The term which appears in the first line in the integrand of Eq. (\ref{preWard}) can be re-expressed in the following form
\bea \label{WIfirstterm}
&& - \int d^4 x \, V\Theta^\mu_{\,\,\,\underline{a}}\bigg[-\d^{(4)}(x-y)\pd_\nu V^{\underline{a}}_\mu(x) - [\pd_\mu\d^{(4)}(x-y)]V^{\underline{a}}_\nu\bigg] = - V\Theta^\mu_{\,\,\,\nu\,;\mu} + V\Theta^\mu_{\,\,\,\underline{a}}V^{\underline{a}}_{\mu\, ;\nu} \nn \\
&& =  - V\bigg[\Theta^\mu_{\,\,\,\nu\,;\mu} +  V_{\underline{a}\r} V^{\underline{a}}_{\mu\,;\nu}\frac{\Theta^{\mu\r} - \Theta^{\r\mu}}{2}\bigg] \,,
\eea
where in the last expression we used the covariant conservation of the metric tensor expressed in terms of the vierbein
\bea
g_{\mu\nu\,;\r} = 0 \Rightarrow V^{\underline{a}}_{\mu\,;\r}V_{\underline{a}\nu} = - V^{\underline{a}}_\mu V_{\underline{a}\nu\,;\r} = - V_{\underline{a}\mu} V^{\underline{a}}_{\nu\,;\r}.
\eea
Other simplifications are obtained using the invariance of the action under local Lorentz transformations \cite{Caracciolo:1989pt},
parameterized as
\beqa
\d V^{\underline{a}}_\mu = {\w^{\underline{a}}}_{\underline{b}} V^{\underline{b}}_\mu\, , \qquad
\d\psi     = \frac{1}{2}\s^{a b}\w_{{\underline{a}\underline{b}}}\psi\, , \qquad
\d\bar{\psi} = -\frac{1}{2}\bar\psi\s^{\underline{a} \underline{b}}\w_{\underline{a b}}\, ,
\eeqa
that gives, using the antisymmetry of $\w^{\underline{ a b}}$
\beq\label{vincolofermioni}
\frac{\d S}{\d\psi}\s^{\underline{a b}}\psi - \bar\psi\s^{\underline{a b}}\frac{\d S}{\d\bar{\psi}}
- \frac{\d S}{\d V^{\underline{b}}_\mu}V^{\underline{a}}_\mu + \frac{\d S}{\d V^{\underline{b}}_\mu} V^{\underline{a}}_\mu = 0\, .
\eeq
The previous equation can be reformulated in terms of the energy-momentum tensor $\Theta^{\mu\nu}$
\beq
V(\Theta^{\mu\r} - \Theta^{\r\mu}) = \bar{\psi}\s^{\mu\r}\frac{\d S}{\d\bar{\psi}}
- \frac{\d S}{\d\psi}\s^{\mu\r}\psi\,,
\eeq
which is useful to re-express Eq. (\ref{WIfirstterm}) in terms of the symmetric energy-momentum tensor $T^{\mu\nu}$ and to obtain finally, in the flat space-time limit, Eq. (\ref{Ward}).
\section{Appendix. BRST transformations and identities}
\label{appendixBRST}
Here we illustrate the derivation of some identities involving 2-point functions using the BRST invariance of
the generating functional
\beq Z[J, \mathcal{F}] = \int \mD\Phi\, e^{i\tilde S}\, ,\eeq
with
\beqa
\tilde S &=& S_{SM} \, + \, \int d^4x\bigg[ J_\mu(x)A^\mu(x) + \bar\eta^A(x)\w^A(x) + \ldots \nn\\
&+&  \chi^A(x)\mathcal F^A(x) + \chi^Z(x)\mathcal F^Z(x) + \chi^+(x)\mathcal F^+(x) + \chi^-(x)\mathcal F^-(x)\bigg]\, .
\eeqa
For convenience we have summarised the BRST transformation of the fundamental fields of the SM Lagrangian used in the derivations of the various STI's in section \ref{BRSTsection}
\beqa\label{brstQCD}
&& \d A^a_\mu  = \l D^{ab}_ \mu c^b, \qquad \d c^a   =  -\frac{1}{2}g \l f^{abc}c^b c^c, \qquad \d\bar{c}^a  =  -\frac{1}{\xi} \mathcal F^a \, \l = -\frac{1}{\xi}(\pd^\mu A^a_\mu)\l \,,\nn \\
&& \d\psi   =  i g \l c^a t^a\psi, \qquad \d\bar\psi   = - i g\bar\psi t^a\l c^a ,
\eeqa
for an unbroken non abelian gauge theory, and
\beqa
&& \d B_\mu        =  \l \, \pd_\mu\h_Y \qquad \d W^a_\mu    =  \l D^{ab}_{\mu}\eta^b_L = \l(\pd_\mu\h^a_L + \e^{abc}W^b_\mu\h^c_L), \nn \\
&& \d\bar{\eta}_Y   =  -\frac{\l}{\xi}\mathcal F^0, \qquad \d\bar{\eta^a_L}  =  -\frac{\l}{\xi}\mathcal F^a, \qquad \d\eta_Y  =  0, \qquad \d\eta^a_L  = \frac{\l}{2}g\e^{abc}\h^b_L\h^c_L,\nn \\
&& \d H   =   i g'Y H\l\h_Y + i g T^a H\l\h^a_L, \qquad \d H^\dag    =  -i g' H^\dag Y\l \h_Y - i g H^\dag T^a\l\h^a_L,
\eeqa
for the electroweak theory.\\
We require that $\delta_{BRST} Z[J,\mathcal{F}]=0$ under a variation of all the fields and gauge-fixing functions. We then differentiate the resulting equation with respect to the sources of the photon and of the antighost to obtain
\beq
 \label{dbrst}
 \frac{\d^2}{\d J^{A\,\mu}(x)\d\w^A(y)}\delta_{BRST} Z[J,\mathcal{F}] =\int\mD\Phi\, e^{i\tilde S}\bigg\{\bar\h^A(y)\d A_\mu(x)
+ \d\bar\eta^A(y)A_\mu(x)\bigg\} = 0\, .\eeq
Introducing the explicit BRST variation of the antighost field $\bar\eta^A(y)$ and of the gauge field $A_\mu(x)$ we obtain
\beq
\langle\bar\h^A(y) D^A_\mu\h^A(x)\rangle = \frac{1}{\xi}\langle\pd^\b A_\b(y) A_\mu(x)\rangle\, .
\eeq
Similarly, in the case of the $Z$ gauge boson, we take two functional derivatives of the condition of BRST invariance of $Z[J, \mathcal{F}]$, as in Eq. (\ref{dbrst}), but now respect to $J^{Z\,\mu}(x)$ and to $\w^Z(y)$, to obtain the relation
\beq
\langle\bar\h^Z(y)D_\r^Z\h^Z(x)\rangle = \frac{1}{\xi}\langle \mathcal F^Z(y)Z_\r(x)\rangle\,.
\eeq
On the other hand, two functional derivatives of the same invariance condition, now with respect to $J^{A\,\mu}(x)$ and to $\w^Z(y)$, give
\beq
\langle D^A_\r\h^A(x)\bar\h^Z(y)\rangle = \frac{1}{\xi}\langle\mathcal F^Z(y) A_\r(x)\rangle \, .
\eeq
\subsection{Identities from the ghost equations of motion}
A second class of identities is based on the equations of motion of the ghosts.
Differentiating $\delta_{BRST} Z[J,\mathcal F]$ respect to the source of the photon antighost
$\w^A(x)$ and to the source of the corresponding gauge-fixing function $\chi^A(y)$ gives
\beq
\frac{1}{\xi}\langle\pd^\a A_\a(x)\pd^\b A_\b(y)\rangle = \langle\bar\h^A(x)\mathcal E^A(y)\rangle\, .
\eeq
At this point we consider the functional average of the equation of motion of the ghost of the photon
\beq
\int\mD F\, e^{i\tilde S}\bigg\{-\mathcal E^A(y) + \w^A(y)\bigg\} = 0 \,
\eeq
and take a functional derivative of this expression respect to the source $\w^A(x)$ of the antighost  $\bar\h^A(x)$, obtaining the equation
\beq
\int \mD F\, e^{i\tilde S}\bigg\{-i \mathcal E^A(y)\bar\h^A(x) + \d^{(4)}(x-y)\bigg\} = 0\, ,
\eeq
or, in terms of Green's functions
\beq
\frac{1}{\xi}\langle\mathcal F^A(x)\mathcal F^A(y)\rangle
=\frac{1}{\xi}\langle\pd^\a A_\a(x)\pd^\b A_\b(y)\rangle
= \langle\bar\h^A(x)\mathcal E^A(y)\rangle = -i\d^{(4)}(x-y) \,,
\eeq
which involves the correlation function of the photon gauge-fixing function. \\
It is not hard to show, using the same method, the following identities
\bea
\langle \mathcal F^Z(x)\pd^\a A_\a(y)\rangle &=& 0  \Rightarrow \langle \mathcal F^Z(x) A_\a(y) \rangle = 0 \, , \nn \\
\langle \mathcal F^Z(x) \mathcal F^Z(y)\rangle  &=&  -i\xi\d^{(4)}(x-y)\,.
\eea
\section{ Appendix. Feynman Rules}
\label{FeynRules}
We collect here all the Feynman rules used in this work. All the momenta are incoming

\begin{itemize}
\item{ graviton - gauge boson - gauge boson vertex}
\\ \\
\begin{minipage}{95pt}
\includegraphics[scale=1.0]{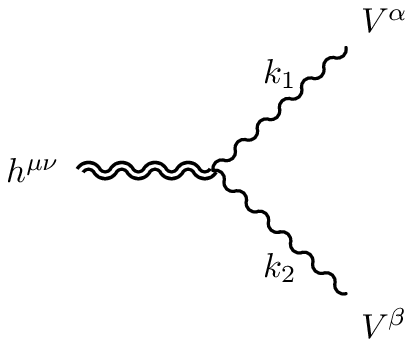}
\end{minipage}
\begin{minipage}{70pt}
\bea
= - i \frac{\kappa}{2} \bigg\{ \left( k_1 \cdot k_2  + M_V^2 \right) C^{\mu\nu\alpha\beta}
+ D^{\mu\nu\alpha\beta}(k_1,k_2) + \frac{1}{\xi}E^{\mu\nu\alpha\beta}(k_1,k_2) \bigg\}
\nn
\eea
\end{minipage}
\bea
\label{FRhVV}
\eea
where $V$ stands for the vector gauge bosons $A, Z$ and $W^{\pm}$.
\item{graviton - fermion - fermion vertex}
\\ \\
\begin{minipage}{95pt}
\includegraphics[scale=1.0]{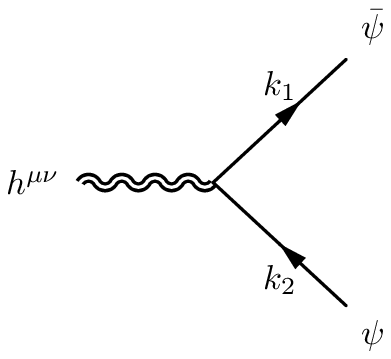}
\end{minipage}
\begin{minipage}{70pt}
\bea
=i \frac{\kappa}{8} \bigg\{ \gamma^\mu \, (k_1 - k_2)^\nu + \gamma^\nu \,(k_1 - k_2)^\mu - 2 \, \eta^{\mu\nu} \left( \ksl_1 - \ksl_2 + 2 m_f \right)\bigg\}
\nn
\eea
\end{minipage}
\bea
\label{FRhFF}
\eea
\item{graviton - ghost - ghost vertex }
\\ \\
\begin{minipage}{95pt}
\includegraphics[scale=1.0]{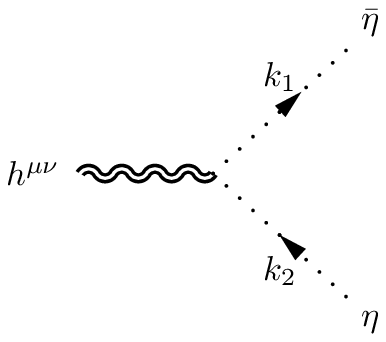}
\end{minipage}
\begin{minipage}{70pt}
\bea
=  i \frac{\kappa}{2} \bigg\{ k_{1\, \rho} \, k_{2 \, \sigma} \, C^{\mu\nu\rho\sigma}
- M_{\eta}^2 \, \eta^{\mu\nu} \bigg\}
\nn
\eea
\end{minipage}
\bea
\label{FRhUU}
\eea
where $\eta$ denotes the ghost fields $\eta^{+}$, $\eta^{-}$ ed $\eta^Z$.
\item{graviton - scalar - scalar vertex}
\\ \\
\begin{minipage}{95pt}
\includegraphics[scale=1.0]{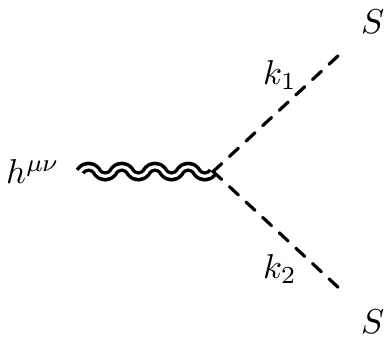}
\end{minipage}
\begin{minipage}{70pt}
\bea
&=&  i \frac{\kappa}{2} \bigg\{ k_{1\, \rho} \, k_{2 \, \sigma} \, C^{\mu\nu\rho\sigma}  - M_S^2 \, \eta^{\mu\nu} \bigg\} \nn \\
&=&  -\frac{i}{3} \frac{\kappa}{2} \bigg\{ (k_1+k_2)^{\mu}(k_1+k_2)^{\nu} - \eta^{\mu\nu} (k_1+k_2)^2 \bigg\} \nn
\eea
\end{minipage}
\bea
\label{FRhSS}
\eea
where $S$ stands for the Higgs $H$ and the Goldstones $\phi$ and  $\phi^{\pm}$. The first expression is the contribution coming from the minimal energy-momentum tensor while the second is due to the term of improvement for a conformally coupled scalar.
\item{graviton - Higgs vertex}
\\ \\
\begin{minipage}{95pt}
\includegraphics[scale=1.0]{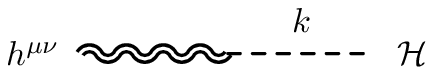}
\end{minipage}
\begin{minipage}{70pt}
\bea
\qquad &=&  i \frac{\kappa}{2} \frac{2 s_w M_W}{3 e} \bigg\{ k^\mu k^\nu - \eta^{\mu\nu}k^2 \bigg\} \nn
\eea
\end{minipage}
\bea
\label{FRhH}
\eea
This vertex is derived from the term of improvement of the energy-momentum tensor and it is a feature of the electroweak symmetry breaking because it is proportional to the Higgs vev.
\item{ graviton - three gauge boson vertex}
\\ \\
\begin{minipage}{95pt}
\includegraphics[scale=1.0]{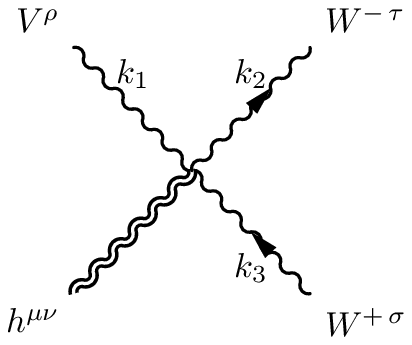}
\end{minipage}
\begin{minipage}{70pt}
\bea
= - i \, e \, \mathcal C_V \frac{\kappa}{2} &\bigg\{&  C^{\mu\nu\rho\sigma} \left(k_3^{\tau}
- k_1^{\tau} \right)  +   C^{\mu\nu\rho\tau} \left(k_1^{\sigma} - k_2^{\sigma} \right) \nn \\
&& + C^{\mu\nu\sigma\tau} \left(k_2^{\rho} - k_3^{\rho} \right)
+\,  F^{\mu\nu\rho\tau\sigma}(k_1,k_2,k_3) \bigg\}
\nn
\eea
\end{minipage}
\bea
\label{FRhVWW}
\eea
where $\mathcal C_A = 1$ and $\mathcal C_Z = \frac{c_w}{s_w}$.
\item{graviton - gauge boson - scalar - scalar vertex }
\\ \\
\begin{minipage}{95pt}
\includegraphics[scale=1.0]{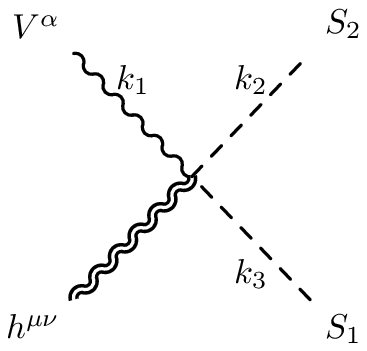}
\end{minipage}
\begin{minipage}{70pt}
\bea
= i \, e \, \mathcal C_{V S_1 S_2} \, \frac{\kappa}{2} \bigg\{ \left( k_{2\, \sigma} - k_{3 \,
\sigma} \right) C^{\mu\nu\alpha\si} \bigg\}
\nn
\eea
\end{minipage}
\bea
\label{FRhVSS}
\eea
with $\mathcal C_{V S_1 S_2}$ given by
\bea
\mathcal C_{A\phi^{+}\phi^{-}} = 1 \qquad
\mathcal C_{Z\phi^{+}\phi^{-}} = \frac{c_w^2 - s_w^2}{2 s_w \, c_w} \qquad
\mathcal C_{Z H \phi} =  \frac{i}{2 s_w \, c_w}. \nn
\eea
\item{graviton - gauge boson - ghost - ghost vertex}
\\ \\
\begin{minipage}{95pt}
\includegraphics[scale=1.0]{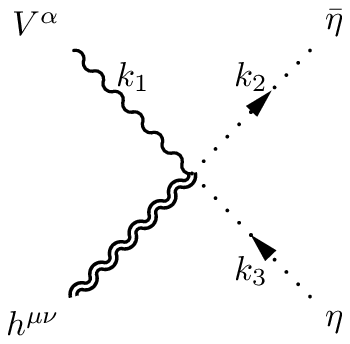}
\end{minipage}
\begin{minipage}{70pt}
\bea
= i \, e \, \mathcal C_{V \eta} \, \frac{\kappa}{2} \, \k_{2 \, \sigma} \,
C^{\mu\nu\alpha\sigma}
\nn
\eea
\end{minipage}
\bea
\label{FRhVUU}
\eea
where $V$ denotes the $A$, $Z$ gauge bosons and $\eta$ the two ghosts $\eta^{+}$
and $\eta^{-}$.
The coefficients $\mathcal C$ are defined as
\bea
\mathcal C_{A \eta^{+}} = 1 \qquad
\mathcal C_{A \eta^{-}} = -1 \qquad
\mathcal C_{Z \eta^{+}} =  \frac{c_w}{s_w} \qquad
\mathcal C_{Z \eta^{-}} =  -\frac{c_w}{s_w}. \nn
\eea
\item{graviton - gauge boson - gauge boson - scalar vertex}
\\ \\
\begin{minipage}{95pt}
\includegraphics[scale=1.0]{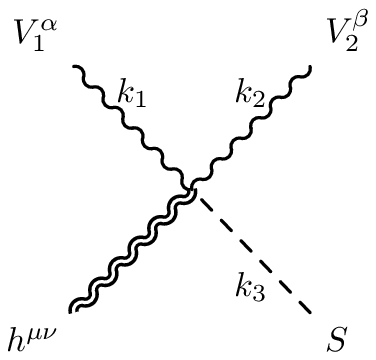}
\end{minipage}
\begin{minipage}{70pt}
\bea
= e \, \mathcal C_{V_1 V_2 S} \, \frac{\kappa}{2} \, M_W \, C^{\mu\nu\alpha\beta}
\nn
\eea
\end{minipage}
\bea
\label{FRhVVS}
\eea
where $V$ stands for $A$, $Z$ o $W^{\pm}$ and $S$ for $\phi^{\pm}$
and $H$. The coefficients are defined as

\bea
\mathcal C_{A W^{+} \phi^{-}} = 1 \qquad
\mathcal C_{A W^{-} \phi^{+}} = -1 \qquad
\mathcal C_{Z W^{+} \phi^{-}} = - \frac{s_w}{c_w} \qquad \nn \\
\mathcal C_{Z W^{-} \phi^{+}} = \frac{s_w}{c_w} \qquad
\mathcal C_{Z Z H} = - \frac{i}{s_w \, c_w^2} \qquad
\mathcal C_{W^{+} W^{-} H} = - \frac{i}{c_w}. \nn
\eea
\item{graviton - scalar - ghost - ghost vertex}
\\ \\
\begin{minipage}{95pt}
\includegraphics[scale=1.0]{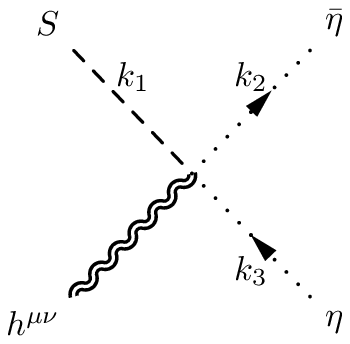}
\end{minipage}
\begin{minipage}{70pt}
\bea
= - i \, e \, \mathcal C_{S \eta} \, \frac{\kappa}{2} \, M_W \, \eta^{\mu\nu}
\nn
\eea
\end{minipage}
\bea
\label{FRhSUU}
\eea
where $S = H$ and $\eta$ denotes $\eta^{+}$, $\eta^{-}$ and
$\eta^{z}$. The vertex is defined with the coefficients
\bea
\mathcal C_{H \eta^{+}} = \mathcal C_{H \eta^{-}} = \frac{1}{2 s_w} \qquad \mathcal C_{H
\eta^{z}} = \frac{1}{2 s_w \, c_w}. \nn
\eea
\item{graviton - three scalar vertex}
\\ \\
\begin{minipage}{95pt}
\includegraphics[scale=1.0]{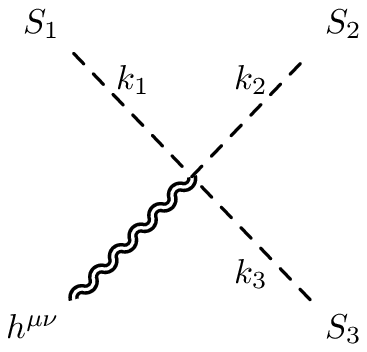}
\end{minipage}
\begin{minipage}{70pt}
\bea
= - i \, e \, \mathcal C_{S_1 S_2 S_3} \, \frac{\kappa}{2} \, \eta^{\mu\nu}
\nn
\eea
\end{minipage}
\bea
\label{FRhSSS}
\eea
with $S$ denoting $H$, $\phi$ and
$\phi^{\pm}$. We have defined the coefficients
\bea
\mathcal C_{H \phi \phi} = \mathcal C_{H \phi^{+} \phi^{-}} = \frac{1}{2 s_w \, c_w}
\frac{M_H^2}{M_Z} \qquad \mathcal C_{H H H} = \frac{3}{2 s_w \, c_w} \frac{M_H^2}{M_Z}. \nn
\eea
\item{graviton - scalar - fermion - fermion vertex}
\\ \\
\begin{minipage}{95pt}
\includegraphics[scale=1.0]{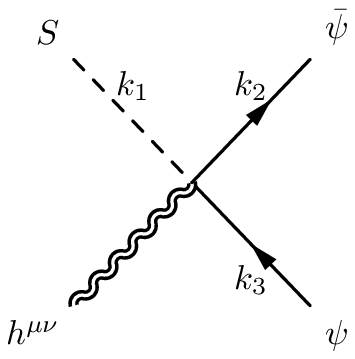}
\end{minipage}
\begin{minipage}{70pt}
\bea
= - i  \frac{\kappa}{2} \frac{e}{2 s_w \, c_w} \frac{m_f}{M_Z} \, \eta^{\mu\nu}
\nn
\eea
\end{minipage}
\bea
\label{FRhSFF}
\eea
where  $S$ is only the Higgs scalar $H$.
\item{graviton - photon - fermion - fermion vertex}
\\ \\
\begin{minipage}{95pt}
\includegraphics[scale=1.0]{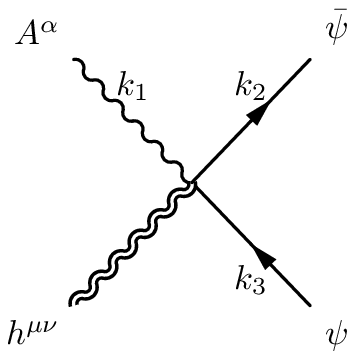}
\end{minipage}
\begin{minipage}{70pt}
\bea
= - i Q_f \, e \frac{\kappa}{4} \bigg\{ \gamma^\mu \, \eta^{\nu\alpha}  + \gamma^\nu \,
\eta^{\mu\alpha} - 2 \, \eta^{\mu\nu} \, \gamma^\alpha  \bigg\}
\nn
\eea
\end{minipage}
\bea
\label{FRhAFF}
\eea
where $Q_f$ is the fermion charge expressed in units of $e$.
\item{graviton - Z - fermion - fermion vertex}
\\ \\
\begin{minipage}{95pt}
\includegraphics[scale=1.0]{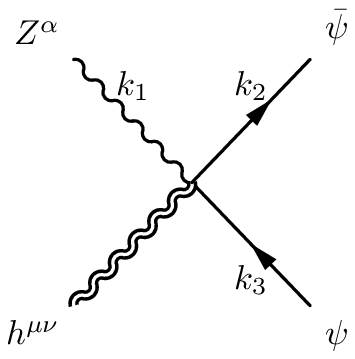}
\end{minipage}
\begin{minipage}{70pt}
\bea
= - i \, e \frac{\kappa}{8 s_w \, c_w} (C_v^f - C_a^f \gamma^5) \bigg\{ \gamma^\mu \,
\eta^{\nu\alpha}  + \gamma^\nu \, \eta^{\mu\alpha} - 2 \, \eta^{\mu\nu} \, \gamma^\alpha
\bigg\}
\nn
\eea
\end{minipage}
\bea
\label{FRhZFF}
\eea
where $C_v^f$ and $C_a^f$ are the vector and axial-vector couplings of the $Z$ gauge boson
to the fermion ($f$). Their expressions are
\bea
C_v^f = I^f_3 - 2 s_w^2 \, Q^f \qquad \qquad C_a^f = I^f_3. \nn
\eea
$I^f_3$ denotes the 3rd component of the isospin.
\item{graviton - four gauge bosons vertex}
\\ \\
\begin{minipage}{95pt}
\includegraphics[scale=1.0]{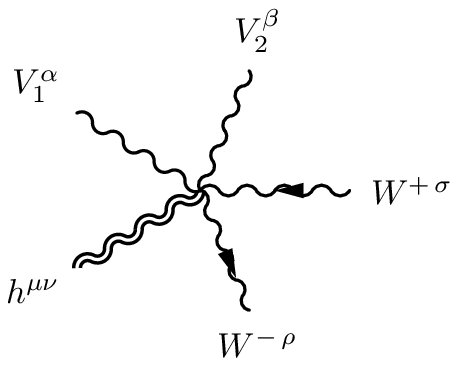}
\end{minipage}
\begin{minipage}{70pt}
\bea
\qquad \qquad = i \, e^2 \, \mathcal C_{V_1 V_2 }\frac{\kappa}{4}
\bigg\{G^{\mu\nu\alpha\beta\si\rho} + G^{\mu\nu\beta\alpha\si\rho} + G^{\mu\nu\alpha\beta\rho\si} + G^{\mu\nu\beta\alpha\rho\si}\bigg\}
\nn
\eea
\end{minipage}
\bea
\label{FRhVVWW}
\eea
where $V_1$ e $V_2$ denote $A$ or $Z$. The coefficients $\mathcal C$ are defined as
\bea
C_{A A} = 1 \qquad C_{A Z} = \frac{c_w}{s_w} \qquad C_{Z Z} = \frac{c_w^2}{s_w^2}. \nn
\eea
\item{graviton - gauge boson - gauge boson - scalar - scalar vertex}
\\ \\
\begin{minipage}{95pt}
\includegraphics[scale=1.0]{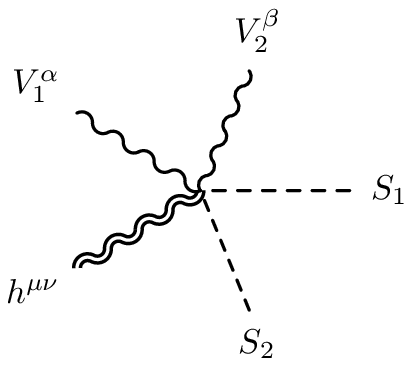}
\end{minipage}
\begin{minipage}{70pt}
\bea
 \qquad = - i \, e^2 \, \mathcal C_{V_1 V_2 S_1 S_2} \frac{\kappa}{2} \, C^{\mu\nu\alpha\beta}
\nn
\eea
\end{minipage}
\bea
\label{FRhVVSS}
\eea
where $V_1$ and $V_2$ denote the neutral gauge bosons $A$ and $Z$, while the possible scalars are
$\phi$, $\phi^{\pm}$ and $H$. The coefficients are
\bea
\mathcal C_{A A \phi^+ \phi^-} = 2
\qquad\mathcal C_{A Z \phi^+ \phi^-} = \frac{c_w^2 - s_w^2}{s_w \, c_w}\qquad \mathcal C_{Z Z \phi^+ \phi^-} = \frac{\left( c_w^2 - s_w^2 \right)^2}{2 s_w^2 \,c_w^2}
\qquad\mathcal C_{Z Z \phi \phi} =  \mathcal C_{Z Z H H} = \frac{1}{2 s_w^2 \, c_w^2} . \nn
\eea
\end{itemize}
The tensor structures $C$, $D$, $E$ and $F$ which appear in the Feynman rules defined above are given by
\bea
&& C_{\mu\nu\rho\sigma} = g_{\mu\rho}\, g_{\nu\sigma}
+g_{\mu\sigma} \, g_{\nu\rho}
-g_{\mu\nu} \, g_{\rho\sigma}\,, \nn
\\
&& D_{\mu\nu\rho\sigma} (k_1, k_2) =
g_{\mu\nu} \, k_{1 \, \sigma}\, k_{2 \, \rho}
- \biggl[g^{\mu\sigma} k_1^{\nu} k_2^{\rho}
  + g_{\mu\rho} \, k_{1 \, \sigma} \, k_{2 \, \nu}
  - g_{\rho\sigma} \, k_{1 \, \mu} \, k_{2 \, \nu}
  + (\mu\leftrightarrow\nu)\biggr], \nn \\
&& E_{\mu\nu\rho\sigma} (k_1, k_2) = g_{\mu\nu} \, (k_{1 \, \rho} \, k_{1 \, \sigma}
+k_{2 \, \rho} \, k_{2 \, \sigma} +k_{1 \, \rho} \, k_{2 \, \sigma})
-\biggl[g_{\nu\sigma} \, k_{1 \, \mu} \, k_{1 \, \rho}
+g_{\nu\rho} \, k_{2 \, \mu} \, k_{2 \, \sigma}
+(\mu\leftrightarrow\nu)\biggr],  \nn \\
&& F_{\mu\nu\rho\sigma\lambda} (k_1,k_2,k_3) =
g_{\mu\rho} \,  g_{\sigma\lambda} \, (k_2-k_3)_{\nu}
+g_{\mu\sigma} \, g_{\rho\lambda} \, (k_3-k_1)_{\nu}
+g_{\mu\lambda} \, g_{\rho\sigma}(k_1-k_2)_{\nu}
+ (\mu\leftrightarrow\nu) \,.
\eea
\section{Appendix. The Scalar integrals}
We collect in this appendix the definition of the scalar integrals appearing in the computation of the correlators.
One-, two- and three-point functions are denoted, respectively as $\mathcal A_0$, $\mathcal B_0$ and $\mathcal C_0$, with
\bea
\mathcal A_0 (m_0^2) &=& \frac{1}{i \pi^2}\int d^n l \, \frac{1}{l^2 - m_0^2} \,, \nn\\
\mathcal B_0 (k^2, m_0^2,m_1^2) &=&  \frac{1}{i \pi^2} \int d^n l \, \frac{1}{(l^2 - m_0^2) \, ((l + k )^2 - m_1^2 )} \,, \nn \\
\mathcal C_0 ((p+q)^2, p^2, q^2, m_0^2,m_1^2,m_2^2) &=& \frac{1}{i \pi^2} \int d^n l \, \frac{1}{(l^2 - m_0^2) \, ((l + p )^2 - m_1^2 ) \, ((l -q )^2 - m_2^2 ) } \,.
\eea
We have also used the finite combination of two-point scalar integrals
\bea
\mathcal D_0 (p^2, q^2, m_0^2,m_1^2) = \mathcal B_0 (p^2, m_0^2,m_1^2) - \mathcal B_0 (q^2, m_0^2,m_1^2) \,.
\eea
The explicit expressions of $A_0$, $B_0$ and $C_0$ can be found in \cite{Denner:1991kt}.
\section{Appendix. Propagators at 1 loop}
\label{propagators}
We report here the expressions of the self-energies appearing in Section \ref{BRSTsection}. They refer to the case of two vector
bosons ($V_1,V_2$), one vector boson and a scalar ($V S$) and two scalars ($S S$). The self-energies carrying Lorentz indices
are decomposed as
\bea
\Sigma^{V_1 V_2}_{\alpha\beta}(p) &=& -i \left( \eta_{\alpha\beta} - \frac{p_\alpha p_\beta}{p^2}
\right) \Sigma^{V_1 V_1}_T(p^2) - i \frac{p_\alpha p_\beta}{p^2} \Sigma^{V_1 V_1}_L(p^2) \, , \\
\Sigma^{V S}_{\alpha}(p) &=& p_\alpha \, \Sigma^{V S}_L(p^2) \,.
\eea
We denote with $\lambda$ the infrared regulator of the photon mass. We denote with $m_{l,i}$, $m_{u,i}$ and $m_{d,i}$ the masses of the lepton, u-type and d-type quarks of generation $i$ respectively.

The self-energies are then given by
\bea
\Sigma^{AA}_T(p^2)
&=& - \frac{\alpha}{4 \pi} \bigg\{ \frac{2}{3} \sum_f \, N_C^f 2 Q_f^2 \bigg[
-(p^2 + 2 m_f^2)B_0(p^2, m_f^2, m_f^2)  +  2 m_f^2 B_0(0, m_f^2, m_f^2) + \frac{1}{3}p^2 \bigg] \nn \\
&+&  \bigg[ (3 p^2 + 4 M_W^2 ) B_0(p^2, M_W^2, M_W^2) - 4 M_W^2 B_0(0, M_W^2, M_W^2)\bigg]\bigg\}\, , \\
\Sigma^{AA}_L(p^2) &=& 0\, , \\
\Sigma^{ZZ}_T(p^2)
&=& -\frac{\alpha}{4 \pi} \bigg\{ \frac{2}{3} \sum_f \,
N_C^f \bigg[ \frac{C_V^{f \, 2} + C_A^{f \, 2}}{2 s_w^2 c_w^2}
\bigg( -(p^2 + 2m_f^2) B_0(p^2, m_f^2, m_f^2)  +  2 m_f^2 B_0(0, m_f^2, m_f^2) + \frac{1}{3}p^2 \bigg) \nn\\
&+& \frac{3}{4 s_w^2 c_w^2} m_f^2 B_0(p^2,m_f^2, m_f^2) \bigg]
+  \frac{1}{6 s_w^2 c_w^2}\bigg[ \bigg( (18 c_w^4 + 2 c_w^2 -\frac{1}{2})p^2
+ (24 c_w^4 + 16 c_w^2 -10)M_W^2 \bigg)\nn\\
&\times& B_0(p^2, M_W^2, M_W^2)
-  (24 c_w^4 - 8 c_w^2 + 2)M_W^2 B_0(0, M_W^2, M_W^2) + (4 c_w^2-1)\frac{p^2}{3}  \bigg] \nn \\
&+&   \frac{1}{12 s_w^2 c_w^2} \bigg[ (2 M_H^2 -10 M_Z^2 - p^2) B_0(p^2, M_Z^2, M_H^2)
-  2 M_Z^2 B_0(0, M_Z^2, M_Z^2) - 2 M_H^2 B_0(0, M_H^2, M_H^2)  \nn \\
&-&  \frac{(M_Z^2 - M_H^2)^2}{p^2}\left( B_0(p^2, M_Z^2, M_H^2)  - B_0(0, M_Z^2, M_H^2) \right) -
\frac{2}{3} p^2   \bigg] \bigg\}\, , \\
\Sigma^{ZZ}_L(p^2)
&=& - \frac{\alpha}{2 \pi s_w^2 c_w^2} \bigg\{
\sum_f \, N_C^2 \, C_A^{f \, 2} \, m_f^2 B_0(p^2, m_f^2, m_f^2)
+ M_W^2 (c_w^4 -s_w^4) B_0(p^2, M_W^2, M_W^2) \nn \\
&& - \frac{1}{4 p^2} \bigg[ \left( (M_Z^2-M_H^2)^2 - 4 M_Z^2 p^2\right) B_0(p^2, M_Z^2, M_H^2)
 + (M_Z^2 -M_H^2)\left( A_0(M_H^2 - A_0(M_Z^2))\right)\bigg] \bigg\}\, , \\
\Sigma^{AZ}_T(p^2)
&=& \frac{\alpha}{4 \pi \, s_w \, c_w} \bigg\{ \frac{2}{3} \sum_f \, N_C^f \,
Q_f \, C_V^f \bigg[ (p^2 + 2m_f^2) B_0(p^2, m_f^2, m_f^2)
 - 2 m_f^2 B_0(0, m_f^2, m_f^2) -\frac{1}{3}p^2 \bigg] \nn \\
&& - \frac{1}{3} \bigg[ \left( (9 c_w^2 + \frac{1}{2})p^2
+ (12 c_w^2 + 4)M_W^2 \right) B_0(p^2, M_W^2, M_W^2)
-  (12 c_w^2 -2)M_W^2 B_0(0, M_W^2, M_W^2) + \frac{1}{3}p^2 \bigg]\bigg\}\, ,\nn\\
\eea
\newpage
\bea
\Sigma^{AZ}_L(p^2)
&=& -\frac{\alpha}{2 \pi \, s_w \, c_w} M_W^2 \, B_0(p^2, M_W^2, M_W^2)\, ,\\
\Sigma^{A\phi}_L(p^2)
&=& - \frac{\alpha}{2 \pi \, s_w} M_W^2 \, B_0(p^2, M_W^2, M_W^2)\, ,\\
\Sigma^{Z\phi}_L(p^2) &=& - \frac{\alpha}{2 \pi \, s_w^2 \, c_w^2} \bigg\{
C_A^{f \, 2} \frac{m_f^2}{M_Z} B_0(p^2, m_f^2, m_f^2)
+  \frac{M_W}{4} c_w (4 c_w^2 -3) B_0(p^2,M_W^2, M_W^2) \nn \\
&+&   \frac{1}{8 M_Z \, p^2} \bigg[\left( (M_H^2 -M_Z^2)^2 - 3 M_Z^2 p^2 \right) B_0(p^2, M_Z^2, M_H^2)
+  (M_Z^2 -M_H^2)\left( A_0(M_H^2) \right)\bigg]\bigg\}\, ,\\
\Sigma^{\phi\phi}(p^2) &=&  i \frac{\alpha}{4 \pi \, s_w^2 \, c_w^2 \, M_Z^2} \bigg\{\sum_f \, N_C^f \,
C_A^f \, m_f^2 \bigg[  p^2 B_0(p^2, m_f^2,m_f^2) - 2 A_0(m_f^2)\bigg]  \nn \\
&+&  \frac{1}{8} \bigg[ \left(6 M_W^2 + M_H^2 \right) A0(M_W^2)
-  4 M_W^2 \left( p^2 B_0(p^2, M_W^2, M_W^2) + M_W^2 \right) \bigg] \nn \\
&+&     \frac{1}{16} \bigg[ 2 \left( (M_H^2 - M_Z^2)^2
- 2 M_Z^2 p^2 \right)  B_0(p^2, M_Z^2, M_H^2)\nn\\
&+&  (M_H^2 + 2M_Z^2)A_0(M_H^2) + (3 M_H^2 + 4 M_Z^2)A_0(M_Z^2)   - 4 M_Z^4 \bigg]\bigg\}\, ,\nn\\
\Sigma_{HH}(p^2) &=& - \frac{\alpha}{4 \pi} \bigg\{ \sum_{f} N_C^f \frac{m_f^2}{2 s_w^2 M_W^2} \bigg[ 2 \mathcal A_0\left( m_f^2 \right) + (4 m_f^2 - p^2) \mathcal B_0 \left( p^2, m_f^2,m_f^2\right) \bigg] \nn \\
&-& \frac{1}{2 s_w^2} \bigg[ \left(6 M_W^2 - 2p^2 + \frac{M_H^4}{2 M_W^2} \right) \mathcal B_0 \left( p^2, M_W^2, M_W^2 \right) + \left( 3 + \frac{M_H^2}{2 M_W^2} \right) \mathcal A_0 \left( M_W^2 \right) - 6 M_W^2 \bigg] \nn \\
&-& \frac{1}{4 s_w^2 \, c_w^2} \bigg[ \left(6 M_Z^2 - 2p^2 + \frac{M_H^4}{2 M_Z^2} \right) \mathcal B_0 \left( p^2, M_Z^2, M_Z^2 \right) + \left( 3 + \frac{M_H^2}{2 M_Z^2} \right) \mathcal A_0 \left( M_Z^2 \right) - 6 M_Z^2 \bigg] \nn \\
&-& \frac{3}{8 s_w^2} \bigg[ 3 \frac{M_H^4}{M_W^2} \mathcal B_0 \left( p^2, M_H^2,M_H^2 \right) + \frac{M_H^2}{M_W^2} \mathcal A_0 \left( M_H^2 \right) \bigg] \bigg\}\, ,
\eea
\bea
\Sigma^{WW}_T(p^2) &=& -\frac{\alpha}{4 \pi} \bigg\{ \frac{1}{3 s_w^2} \sum_i \bigg[ \left( \frac{m_{l,i}^2}{2} - p^2 \right)
\mathcal B_0\left(p^2, 0, m_{l,i}^2 \right) + \frac{p^2}{3} + m_{l,i}^2 \mathcal B_0 \left( 0, m_{l,i}^2,m_{l,i}^2 \right)  \nn
\\
&+&  \frac{m_{l,i}^4}{2 p^2} \left( \mathcal B_0\left(p^2, 0, m_{l,i}^2 \right)
- \mathcal B_0\left(0, 0, m_{l,i}^2 \right)\right) \bigg]
+ \frac{1}{s_w^2} \sum_{i,j}|V_{ij}|^2 \bigg[\left( \frac{m_{u,i}^2 + m_{d,j}^2}{2} - p^2\right)\nn\\
&\times&\mathcal B_0 \left( p^2, m_{u,i}^2, m_{d,j}^2\right)
+ \frac{p^2}{3} + m_{u,i}^2 \mathcal B_0 \left( 0, m_{u,i}^2, m_{u,i}^2\right) + m_{d,j}^2 \mathcal B_0 \left( 0, m_{d,j}^2,
m_{d,j}^2\right)\nn\\
&+& \frac{(m_{u,i}^2 - m_{d,j}^2)^2}{2 p^2} \big( \mathcal B_0 \left( p^2, m_{u,i}^2, m_{d,j}^2\right)
- \mathcal B_0 \left( 0, m_{u,i}^2, m_{d,j}^2\right)\big) \bigg]\nn\\
&+& \frac{2}{3} \bigg[ (2 M_W^2 + 5 p^2) \mathcal B_0 \left( p^2, M_W^2, \lambda^2 \right)
- 2 M_W^2 \mathcal B_0 \left( 0, M_W^2, M_W^2\right) \nn \\
&-& \frac{M_W^4}{p^2} \big( \mathcal B_0\left( p^2, M_W^2, \lambda^2 \right) - \mathcal B_0 \left( 0,M_W^2, \lambda^2 \right)
\big) + \frac{p^2}{3}  \bigg] + \frac{1}{12 s_w^2} \bigg[
\big( (40 c_w^2 -1)p^2 \nn \\
&+& (16 c_w^2 + 54 - 10 c_w^{-2}) M_W^2 \big) \mathcal B_0 \left(p^2, M_W^2, M_Z^2 \right) - (16 c_w^2 + 2) \big( M_W^2 \mathcal
B_0 \left( 0,M_W^2,M_W^2\right) \nn \\
&+& M_Z^2 \mathcal B_0 \left( 0, M_Z^2, M_Z^2\right) \big) + (4 c_w^2 -1) \frac{2 p^2}{3} -  (8 c_w^2 +1) \frac{(M_W^2 -
M_Z^2)^2}{p^2} \big( \mathcal B_0 \left( p^2, M_W^2,M_Z^2\right) \nn \\
&-& \mathcal B_0 \left(0, M_W^2,M_Z^2 \right) \big) \bigg] + \frac{1}{12 s_w^2} \bigg[ (2 M_H^2 - 10 M_W^2 - p^2) \mathcal B_0
\left(p^2, M_W^2,M_H^2 \right)- 2 M_W^2 \mathcal B_0 \left(0,M_W^2,M_W^2 \right) \nn\\
&-& 2 M_H^2 \mathcal B_0 \left( 0, M_H^2,M_H^2\right) - \frac{(M_W^2
-M_H^2)^2}{p^2} \big( \mathcal B_0 \left( p^2, M_W^2, M_H^2\right)
- \mathcal B_0 \left( 0,M_W^2,M_H^2\right) \big) - \frac{2 p^2}{3}\bigg] \bigg\}\, . \nn\\
\eea
%


\section{Appendix. Contributions to the form factors}
\label{formfactors}
We give here the remaining coefficients appearing in the form factors of the $TAZ$ and $TZZ$ correlators.
\subsection{Form factors for the $TAZ$ vertex}
\bea
%
%
\Phi_2^{(F)}(s,0,M_Z^2,m_f^2) &=& - i \frac{\kappa}{2} \frac{\alpha}{3 \pi s_w \, c_w } \frac{Q_f \, C_v^f }{s (s-M_Z^2)^3} \bigg\{\frac{1}{6}\big(12 m_f^2 \left(s M_Z^2+M_Z^4+s^2\right) \nn \\
&&\hspace{-3cm} -9 s^2 M_Z^2+12 s M_Z^4+2 M_Z^6+s^3\big)
+ 2 m_f^2 \left(s M_Z^2+M_Z^4+3 s^2\right) \mathcal D_0\left(s,0,m_f^2,m_f^2\right) \nn \\
&&\hspace{-3cm}   -\frac{M_Z^2[4 m_f^2(3 s M_Z^2+M_Z^4+9 s^2)+s(4 s M_Z^2+M_Z^4-3 s^2)]}{2(M_Z^2-s)} \mathcal D_0\left(s,M_Z^2,m_f^2,m_f^2\right) \nn \\
&&\hspace{-3cm}  +m_f^2 \left(4 m_f^2 \left(s M_Z^2+M_Z^4+s^2\right)+3 s^2 M_Z^2+6 s
   M_Z^4+M_Z^6+2 s^3\right) \mathcal C_0\left(s,0,M_Z^2,m_f^2,m_f^2,m_f^2\right)
\bigg\}\, , \nn \\
\Phi_3^{(F)}(s,0,M_Z^2,m_f^2) &=& - i \frac{\kappa}{2} \frac{\alpha}{12 \pi s_w \, c_w } \frac{Q_f \, C_v^f }{s (s-M_Z^2)} \bigg\{
- \frac{1}{6} \left(12 m_f^2+2 M_Z^2+s\right)
-2 m_f^2 \mathcal D_0\left(s,0,m_f^2,m_f^2\right) \nn \\
&& \hspace{-3cm} + \, \frac{[ 4 m_f^2(M_Z^2+2 s)+s M_Z^2 ]}{2 \left(M_Z^2-s\right)} \mathcal D_0\left(s,M_Z^2,m_f^2,m_f^2\right)
- m_f^2 \left(4 m_f^2+M_Z^2+2 s\right) \mathcal C_0\left(s,0,M_Z^2,m_f^2,m_f^2,m_f^2\right)
\bigg\}\, , \nn \\
\Phi_4^{(F)}(s,0,M_Z^2,m_f^2) &=& - \frac{2 (2s + M_Z^2)}{s-M_Z^2} \Phi_3^{(F)}(s,0,M_Z^2,m_f^2)\, , \nn \\
\Phi_5^{(F)}(s,0,M_Z^2,m_f^2) &=& - i \frac{\kappa}{2} \frac{\alpha}{6 \pi s_w \, c_w } \frac{Q_f \, C_v^f }{(s-M_Z^2)^2} \bigg\{
M_Z^2
-8 m_f^2 \mathcal D_0\left(s,0,m_f^2,m_f^2\right) \nn \\
&& \hspace{-3cm} + \, \frac{[s M_Z^2 - 8 m_f^2 (3 M_Z^2-s )]}{s-M_Z^2} \mathcal D_0\left(s,M_Z^2,m_f^2,m_f^2\right)
-6 m_f^2 \, M_Z^2 \, \mathcal C_0\left(s,0,M_Z^2,m_f^2,m_f^2,m_f^2\right)
 \bigg\}\, , \nn \\
\Phi_6^{(F)}(s,0,M_Z^2,m_f^2) &=& - i \frac{\kappa}{2} \frac{\alpha}{3 \pi s_w \, c_w } \frac{Q_f \, C_v^f }{(s-M_Z^2)^3} \bigg\{
M_Z^2 \left(6 m_f^2-M_Z^2+2 s\right)
+10 m_f^2 \, M_Z^2 \, \mathcal D_0\left(s,0,m_f^2,m_f^2\right) \nn \\
&&\hspace{-3cm} +  \, \frac{M_Z^2[ s(M_Z^2-3 s)-4 m_f^2(9 M_Z^2+4 s)]}{2 \left(M_Z^2-s\right)} \mathcal D_0\left(s,M_Z^2,m_f^2,m_f^2\right) \nn \\
&& \hspace{-3cm}+  \, 3 m_f^2 M_Z^2 \left(4 m_f^2+M_Z^2+3 s\right) \mathcal C_0\left(s,0,M_Z^2,m_f^2,m_f^2,m_f^2\right)
 \bigg\}\, , \nn \\
\Phi_7^{(F)}(s,0,M_Z^2,m_f^2) &=& - i \frac{\kappa}{2} \frac{\alpha}{12 \pi s_w \, c_w } \frac{Q_f \, C_v^f }{ (s-M_Z^2)} \bigg\{
\frac{1}{6} \left(36 m_f^2+M_Z^2+11 s\right)
+ \left(2 s-2 M_Z^2\right) \mathcal B_0\left(s,m_f^2,m_f^2\right) \nn \\
&& \hspace{-3cm} +2 m_f^2 \mathcal D_0\left(s,0,m_f^2,m_f^2\right)
+\left(4 M_Z^2 - \frac{2(m_f^2(M_Z^2+4 s)+M_Z^4)}{M_Z^2-s}\right) \mathcal D_0\left(s,M_Z^2,m_f^2,m_f^2\right) \nn \\
&& \hspace{-3cm} +6 m_f^2 \left(2 m_f^2+s\right) \mathcal C_0\left(s,0,M_Z^2,m_f^2,m_f^2,m_f^2\right)
\bigg\}\, ,
\eea
\bea
\Phi_2^{(B)}(s,0,M_Z^2,M_W^2) &=& - i \frac{\kappa}{2} \frac{\alpha}{3 \pi s_w \, c_w } \frac{1}{ s(s-M_Z^2)^3} \bigg\{
\frac{1}{12} \big(-6 s^2 M_Z^2 \left(12 \left(s_w^4+s_w^2\right)-25\right) \nn\\
&& \hspace{-3cm} + \, M_Z^6 \left(-72 s_w^4+174
   s_w^2-103\right)-36 s M_Z^4 \left(2 s_w^4-9 s_w^2+7\right)+s^3 \left(6
   s_w^2-5\right)\big) \nn \\
&& \hspace{-3cm} - \, 2 c_w^2 M_Z^2 \left(M_Z^2-s\right)^2 \mathcal B_0\left(s,M_W^2,M_W^2\right)
+ M_Z^2 c_w^2\left(6 s_w^2 \left(s M_Z^2+M_Z^4+3 s^2\right)-9 s M_Z^2-3 M_Z^4-13 s^2\right) \times \nn \\
&& \hspace{-3cm} \times \, \mathcal D_0\left(s,0,M_W^2,M_W^2\right)
+\frac{M_Z^2}{2
   \left(M_Z^2-s\right)} \big(2 s^2 M_Z^2 \left(54 s_w^4-115 s_w^2+61\right)+2 M_Z^6 \left(6 s_w^4-11
   s_w^2+5\right) \nn \\
&& \hspace{-3cm}   +2 s M_Z^4 \left(18 s_w^4-37 s_w^2+19\right)+s^3 \left(34
   s_w^2-35\right)\big) \mathcal D_0\left(s,M_Z^2,M_W^2,M_W^2\right) \nn \\
&& \hspace{-3cm}  - M_Z^2 c_w^2\big(2 s^2 M_Z^2 \left(6 s_w^4-3 s_w^2+4\right)+s M_Z^4 \left(12 \left(s_w^2-5\right) s_w^2+41\right)+2 M_Z^6 \left(6 s_w^4-15
   s_w^2+8\right) \nn \\
&& \hspace{-3cm}    -s^3 \left(6 s_w^2+5\right)\big) \mathcal C_0\left(s,0,M_Z^2,M_W^2,M_W^2,M_W^2\right)
\bigg\}\, , \nn \\
\Phi_3^{(B)}(s,0,M_Z^2,M_W^2) &=& - i \frac{\kappa}{2} \frac{\alpha}{12 \pi s_w \, c_w } \frac{1}{ s(s-M_Z^2)} \bigg\{
\frac{1}{12} \left(M_Z^2 \left(72 s_w^4-174 s_w^2+103\right)+s \left(5-6 s_w^2\right)\right) \nn \\
&& \hspace{-3cm} +2 M_Z^2 c_w^2 \mathcal B_0\left(s,M_W^2,M_W^2\right)
+3 M_Z^2 \left(2 s_w^4-3 s_w^2+1\right) \mathcal D_0\left(s,0,M_W^2,M_W^2\right) \nn \\
&& \hspace{-3cm} -\frac{M_Z^2 c_w^2}{s-M_Z^2} \left(M_Z^2 \left(6 s_w^2-5\right)+2 s \left(6 s_w^2-7\right)\right)\mathcal D_0\left(s,M_Z^2,M_W^2,M_W^2\right)\nn\\
&& \hspace{-3cm}+M_Z^2 c_w^2 \left(2 M_Z^2 \left(6 s_w^4-15 s_w^2+8\right)+s \left(7-6 s_w^2\right)\right) \mathcal C_0\left(s,0,M_Z^2,M_W^2,M_W^2,M_W^2\right)
\bigg\}\, , \nn \\
\Phi_4^{(B)}(s,0,M_Z^2,M_W^2) &=& - i \frac{\kappa}{2} \frac{\alpha}{6 \pi s_w \, c_w } \frac{1}{ s(s-M_Z^2)^2} \bigg\{
-\frac{1}{12} \left(M_Z^2+2 s\right) \big(M_Z^2 \left(72 s_w^4-174 s_w^2+103\right) \nn \\
&& \hspace{-3cm} +s \left(5-6 s_w^2\right)\big)
- 2 M_Z^2 c_w^2 \left(M_Z^2-4 s\right) \mathcal B_0\left(s,M_W^2,M_W^2\right)
+ 3 M_Z^2 c_w^2 \big(2 s_w^2 \left(M_Z^2+2 s\right) \nn \\
&& \hspace{-3cm}  -M_Z^2-6 s \big) \mathcal D_0\left(s,0,M_W^2,M_W^2\right)
+\frac{M_Z^2 \, c_w^2 }{s-M_Z^2} \left(M_Z^2+2 s\right) \left(M_Z^2 \left(6 s_w^2-5\right)+2 s \left(6 s_w^2-7\right)\right) \times \nn \\
&& \hspace{-3cm}  \times \, \mathcal D_0\left(s,M_Z^2,M_W^2,M_W^2\right)
- M_Z^2 c_w^2 \left(M_Z^2+2 s\right) \big(2 M_Z^2 \left(6 s_w^4-15 s_w^2+8\right) \nn \\
&& \hspace{-3cm} +s \left(7-6 s_w^2\right)\big) \mathcal C_0\left(s,0,M_Z^2,M_W^2,M_W^2,M_W^2\right)
\bigg\}\, , \nn \\
\Phi_5^{(B)}(s,0,M_Z^2,M_W^2) &=& - i \frac{\kappa}{2} \frac{\alpha}{6 \pi s_w \, c_w } \frac{1}{(s-M_Z^2)^2} \bigg\{
M_Z^2 \left(18 s_w^2-19\right)
+12 M_Z^2 c_w^2 \mathcal B_0\left(s,M_W^2,M_W^2\right) \nn \\
&& \hspace{-3cm} +8 M_Z^2 \left(3 s_w^4-4 s_w^2+1\right) \mathcal D_0\left(s,0,M_W^2,M_W^2\right)
- \frac{M_Z^2}{s-M_Z^2} \big(s \left(24 s_w^4-62 s_w^2+39\right) \nn \\
&& \hspace{-3cm}  -12 M_Z^2 \left(6 s_w^4-11 s_w^2+5\right)\big) \mathcal D_0\left(s,M_Z^2,M_W^2,M_W^2\right) \nn \\
&& \hspace{-3cm}  +6 M_Z^2 c_w^2 \left(M_Z^2 \left(2 s_w^2-1\right)-2 s\right) \mathcal C_0\left(s,0,M_Z^2,M_W^2,M_W^2,M_W^2\right)
\bigg\}\, , \nn \\
\Phi_6^{(B)}(s,0,M_Z^2,M_W^2) &=& - i \frac{\kappa}{2} \frac{\alpha}{3 \pi s_w \, c_w } \frac{1}{(s-M_Z^2)^3} \bigg\{
-\frac{1}{4} M_Z^2 \left(M_Z^2 \left(72 s_w^4-90 s_w^2+17\right)+s \left(53-54 s_w^2\right)\right) \nn \\
&& \hspace{-3cm} -5 M_Z^4 \left(6 s_w^4-11 s_w^2+5\right) \mathcal D_0\left(s,0,M_W^2,M_W^2\right)
-\frac{M_Z^2}{2 \left(s-M_Z^2\right)} \big(18 M_Z^4 \left(6 s_w^4-11 s_w^2+5\right) \nn \\
&& \hspace{-3cm} +s M_Z^2 \left(48 s_w^4-70 s_w^2+21\right)+24 s^2 c_w^2\big) \mathcal D_0\left(s,M_Z^2,M_W^2,M_W^2\right)
- 3 M_Z^2 c_w^2 \big(M_Z^4 \left(12 s_w^4-20 s_w^2+9\right) \nn \\
&& \hspace{-3cm} +s M_Z^2 \left(9-14 s_w^2\right)+2 s^2\big) \mathcal C_0\left(s,0,M_Z^2,M_W^2,M_W^2,M_W^2\right)
\bigg\}\, , \nn\eea
\bea
\Phi_7^{(B)}(s,0,M_Z^2,M_W^2) &=& - i \frac{\kappa}{2} \frac{\alpha}{6 \pi s_w \, c_w } \frac{1}{(s-M_Z^2)} \bigg\{
\frac{1}{24} \left(M_Z^2 \left(54 s_w^2 \left(7-4 s_w^2\right)-161\right)+s \left(270 s_w^2-277\right)\right) \nn \\
&& \hspace{-3cm} + \frac{1}{4} \left(M_Z^2 \left(43-42 s_w^2\right)+s \left(18 s_w^2-19\right)\right) \mathcal B_0\left(s,M_W^2,M_W^2\right)
+\frac{1}{2} c_w^2 \left(M_Z^2 \left(6 s_w^2-11\right)-6 s\right) \times \nn \\
&& \hspace{-3cm} \times \, \mathcal D_0\left(s,0,M_W^2,M_W^2\right)
- \frac{1}{4\left(s-M_Z^2\right)}\big(M_Z^4 \left(12 s_w^4+8 s_w^2-21\right)+2 s M_Z^2 \left(24 s_w^4-74 s_w^2+51\right) \nn \\
&& \hspace{-3cm}  +12 s^2c_w^2\big) \mathcal D_0\left(s,M_Z^2,M_W^2,M_W^2\right)
- 3 c_w^2  \big(M_Z^4 \left(6 s_w^4-11 s_w^2+5\right) \nn \\
&& \hspace{-3cm} +s M_Z^2 \left(6-8 s_w^2\right)+2 s^2\big) \mathcal C_0\left(s,0,M_Z^2,M_W^2,M_W^2,M_W^2\right)
\bigg\}\, , \nn \\
\Phi_8^{(B)}(s,0,M_Z^2,M_W^2) &=& \frac{i \alpha  \kappa  c_w M_Z^2 }{6 \pi  s s_w} \mathcal B_0(0,M_W^2,M_W^2)\, , \nn \\
\Phi_9^{(B)}(s,0,M_Z^2,M_W^2) &=& -\frac{i \alpha  \kappa  c_w M_Z^2 }{6 \pi  s s_w \left(s-M_Z^2\right)} \mathcal B_0(0,
M_W^2,M_W^2)\, .\eea

\subsection{Form factors for the $TZZ$ vertex in the fermionic sector}
\label{fermionicFF}
The coefficients of Eq. (\ref{fermionhZZ}) are given by
\small
\bea
{C_{(F)}}_0^2 &=& \frac{i \kappa\, \alpha, m_f^2}{6 \pi  s^2 c_w^2 \left(s-4 M_Z^2\right) s_w^2}
\left(\left(2 M_Z^4-4 s M_Z^2+s^2\right) C_a^{f \, 2}+2 C_v^{f \, 2} M_Z^4\right)\, ,  \nn \\
{C_{(F)}}_1^2 &=&   0  \, ,\nn\\
{C_{(F)}}_2^2 &=&     \frac{i \kappa \, \alpha\,m_f^2}{6 \pi  s c_w^2 s_w^2}\,C_a^{f \, 2}  \, ,\nn\\
{C_{(F)}}_3^2 &=&   \frac{i \kappa \, \alpha\,  m_f^2 \,M_Z^2}{3 \pi  s^2 c_w^2 \left(s-4 M_Z^2\right){}^2 s_w^2}
\left(s^2 C_a^{f \, 2}-2 \left(C_a^{f \, 2}+C_v^{f \, 2}\right) M_Z^4
+2 s \left(C_v^{f \, 2}-C_a^{f \, 2}\right) M_Z^2\right) \, ,\nn\\
{C_{(F)}}_4^2 &=&   \frac{i \kappa \, \alpha \, m_f^2}{6 \pi  s^2 c_w^2 \left(s-4 M_Z^2\right){}^2 s_w^2}
\bigg(\bigg(4 M_Z^8-2 \left(8 m_f^2+5 s \right) M_Z^6+3 s \left(12 m_f^2+s \right) M_Z^4\nn\\
&-&16 s^2 m_f^2 M_Z^2+2 s^3 m_f^2\bigg) C_a^{f \,2}+C_v^{f \, 2} M_Z^4 \left(4 M_Z^4-2 \left(8 m_f^2+s \right) M_Z^2
+s \left(4 m_f^2+s \right)\right)\bigg)\, ,
\eea
\bea
{C_{(F)}}_0^3 &=& \frac{i \kappa \, \alpha}{192 \pi  c_w^2 \left(s-4 M_Z^2\right) s_w^2}
\bigg(4 \left(C_a^{f \, 2}+C_v^{f \, 2}\right) M_Z^4-2 \left(32 m_f^2 C_a^{f \, 2}+7 s \left(C_a^{f \, 2}+C_v^{f \,
2}\right)\right) M_Z^2\nn\\
&+& s\left(16 m_f^2 C_a^{f \, 2}+3 s \left(C_a^{f \, 2}+C_v^{f \, 2}\right)\right)\bigg)\, ,\nn\\
{C_{(F)}}_1^3 &=&     \frac{i \kappa \, \alpha}{48 \pi  c_w^2 s_w^2}  \left(C_a^{f \, 2}+C_v^{f \, 2}\right)\, , \nn\\
{C_{(F)}}_2^3 &=&     \frac{i \kappa \, \alpha}{48 \pi  c_w^2 s_w^2}\,\left(\left(3 m_f^2+s \right) C_a^{f \, 2}+C_v^{f \, 2} \left(s-m_f^2\right)\right)-\frac{i \alpha  \kappa}{24 \pi  c_w^2 s_w^2}\,\left(C_a^{f \, 2}+C_v^{f \, 2}\right) M_Z^2\, , \nn\\
{C_{(F)}}_3^3 &=&     \frac{i \kappa \, \alpha}{48 \pi  c_w^2 \left(s-4 M_Z^2\right){}^2 s_w^2}
\bigg(8 s^2 m_f^2 C_a^{f \, 2}+14 \left(C_a^{f \, 2}+C_v^{f \, 2}\right) M_Z^6
+\bigg(8 \left(5 C_a^{f \, 2}+C_v^{f \, 2}\right) m_f^2\nn\\
&-&17 s \left(C_a^{f \, 2}+C_v^{f \, 2}\right)\bigg) M_Z^4
+s \left(3 s \left(C_a^{f \, 2}+C_v^{f \, 2}\right)-2 \left(21 C_a^{f \, 2}+C_v^{f \, 2}\right) m_f^2\right) M_Z^2\bigg)\, ,
\nn\\
{C_{(F)}}_4^3 &=& \frac{i \kappa \, \alpha}{48 \pi  c_w^2 \left(s-4 M_Z^2\right){}^2 s_w^2}
\bigg(18 \left(C_a^{f \, 2}+C_v^{f \, 2}\right) M_Z^8-2 \left(8 \left(7 C_a^{f \, 2}+C_v^{f \, 2}\right) m_f^2+9 s \left(C_a^{f \, 2}+C_v^{f
\, 2}\right)\right) M_Z^6\nn\\
&+&\left(\left(160 m_f^4+116 s m_f^2+3 s^2\right) C_a^{f \, 2}+C_v^{f \, 2} \left(4 m_f^2+3 s \right) \left(8 m_f^2+s
\right)\right) M_Z^4\nn\\
&-&2 s m_f^2 \left(\left(40 m_f^2+21 s \right) C_a^{f \, 2} + C_v^{f \, 2} \left(8 m_f^2+5 s \right)\right)
M_Z^2+s^2 \left(5 C_a^{f \, 2}+C_v^{f \, 2}\right) m_f^2 \left(2 m_f^2+s \right)\bigg)\, ,
\eea
\bea
{C_{(F)}}_0^4 &=& -\frac{i \kappa \, \alpha}{144 \pi  s c_w^2 \left(s-4 M_Z^2\right){}^2 s_w^2}
\bigg(-44 \left(C_a^{f \, 2}+C_v^{f \, 2}\right) M_Z^6
+2 \bigg(96 \left(2 C_a^{f \, 2}+C_v^{f \, 2}\right) m_f^2
+ 31 s \left(C_a^{f \, 2}+C_v^{f \, 2}\right)\bigg) M_Z^4\nn\\
&-&s \left(48 \left(2 C_a^{f \, 2}+3 C_v^{f \, 2}\right) m_f^2+13 s
\left(C_a^{f \, 2}+C_v^{f \, 2}\right)\right) M_Z^2+s^2 \left(s C_a^{f \, 2}+C_v^{f \, 2} \left(24 m_f^2+s \right)\right)\bigg)
\, ,  \nn\\
{C_{(F)}}_1^4 &=&    \frac{i \kappa \, \alpha}{12 \pi  s c_w^2 \left(s-4
M_Z^2\right) s_w^2} \left(C_a^{f \, 2}+C_v^{f \, 2}\right) \left(s-3 M_Z^2\right) \, , \nn\\
{C_{(F)}}_2^4 &=& \frac{i \kappa \, \alpha \, m_f^2}{12 \pi  s c_w^2 \left(s-4 M_Z^2\right) s_w^2}\left(C_a^{f \, 2} \left(s-5 M_Z^2\right)-C_v^{f \, 2} \left(s-3 M_Z^2\right)\right)\, , \nn\\
{C_{(F)}}_3^4 &=&   \frac{i \kappa \, \alpha}{24 \pi  s c_w^2 \left(s-4 M_Z^2\right){}^3 s_w^2}
\bigg(-36 \left(C_a^{f \, 2}+C_v^{f \, 2}\right) M_Z^8+2 \bigg(56 \left(3 C_a^{f \, 2}-C_v^{f \, 2}\right) m_f^2\nn\\
&+&9 s \left(C_a^{f \, 2}+C_v^{f \, 2}\right)\bigg) M_Z^6
-2 s \left(\left(98 m_f^2+5 s \right) C_a^{f \, 2}+C_v^{f \, 2} \left(5 s-22 m_f^2\right)\right) M_Z^4\nn\\
&+&s^2 \left(C_a^{f \, 2}
+C_v^{f \, 2}\right) \left(12 m_f^2+s \right) M_Z^2+4 s^3 \left(C_a^f-C_v^f\right) \left(C_a^f+C_v^f\right) m_f^2\bigg)\, ,\nn\\
{C_{(F)}}_4^4 &=&    \frac{i \kappa \, \alpha}{12 \pi  s c_w^2 \left(s-4 M_Z^2\right){}^3 s_w^2}
\bigg(18 \left(C_a^{f \, 2}+C_v^{f \, 2}\right) M_Z^{10}-4 \bigg(4 \left(7 C_a^{f \, 2}+C_v^{f \, 2}\right) m_f^2\nn\\
&+&3 s \left(C_a^{f \, 2}+C_v^{f \, 2}\right)\bigg) M_Z^8 + 4 m_f^2 \left(\left(40 m_f^2+21 s \right) C_a^{f \, 2}+C_v^{f \, 2}
\left(8 m_f^2+11 s \right)\right) M_Z^6\nn\\
&-&2 s m_f^2 \left(24 \left(C_a^{f \,2}+C_v^{f \, 2}\right) m_f^2+s \left(7 C_a^{f \, 2}+13 C_v^{f \, 2}\right)\right) M_Z^4
+2 s^2 m_f^2 \bigg(C_v^{f \, 2} \left(9 m_f^2+4 s \right)\nn\\
&-&C_a^{f \, 2} \left(3 m_f^2+2 s \right)\bigg) M_Z^2+s^3 \left(C_a^f-C_v^f\right) \left(C_a^f+C_v^f\right) m_f^2 \left(2 m_f^2+s
\right)\bigg)\, ,
\eea
\bea
{C_{(F)}}_0^5 &=& -\frac{i \kappa \, \alpha}{288 \pi  s c_w^2 \left(s-4 M_Z^2\right){}^2 s_w^2}
\bigg(-88 \left(C_a^{f \, 2}+C_v^{f \, 2}\right) M_Z^6
+24 \bigg(16 \left(2 C_a^{f \, 2}+C_v^{f \, 2}\right) m_f^2
+5 s \left(C_a^{f \,2}+C_v^{f \, 2}\right)\bigg) M_Z^4\nn\\
&-&12 s \left(8 \left(6 C_a^{f \, 2}+C_v^{f \, 2}\right) m_f^2+5 s \left(C_a^{f \,
2}+C_v^{f \,2}\right)\right) M_Z^2+s^2 \left(96 m_f^2 C_a^{f \, 2}+7 s \left(C_a^{f \, 2}+C_v^{f \, 2}\right)\right)\bigg)\, ,
  \nn\\
{C_{(F)}}_1^5 &=&    \frac{i \kappa \, \alpha}{24 \pi  s c_w^2 \left(s-4
M_Z^2\right) s_w^2}\left(C_a^{f \, 2}+C_v^{f \, 2}\right) \left(s-6 M_Z^2\right) \, , \nn\\
{C_{(F)}}_2^5 &=&    -\frac{i \kappa \, \alpha}{24 \pi  s c_w^2 \left(s-4 M_Z^2\right) s_w^2}
\bigg(s \left(\left(s-3 m_f^2\right) C_a^{f \, 2}+C_v^{f \, 2} \left(m_f^2+s \right)\right)\nn\\
&-&2 \bigg(\left(3 C_v^{f \, 2}-5 C_a^{f \, 2}\right) m_f^2
+2 s \left(C_a^{f \, 2}+C_v^{f \, 2}\right)\bigg) M_Z^2\bigg) \, , \nn\\
{C_{(F)}}_3^5 &=&    -\frac{i \kappa \, \alpha}{24 \pi  s c_w^2 \left(s-4 M_Z^2\right){}^3 s_w^2}
\bigg(36 \left(C_a^{f \, 2}+C_v^{f \, 2}\right) M_Z^8+4 \bigg(\left(s-84 m_f^2\right) C_a^{f \, 2}\nn\\
&+&C_v^{f \, 2} \left(28 m_f^2+s\right)\bigg) M_Z^6
-3 s \left(44 \left(C_v^{f \, 2}-3 C_a^{f \, 2}\right) m_f^2+5 s \left(C_a^{f \, 2}+C_v^{f \, 2}\right)\right)
M_Z^4\nn\\
&+&2 s^2 \bigg(\left(s-79 m_f^2\right) C_a^{f \, 2}
+C_v^{f \, 2} \left(29 m_f^2+s \right)\bigg) M_Z^2+4 s^3 \left(5 C_a^{f \, 2}-2 C_v^{f \, 2}\right) m_f^2\bigg)\, , \nn\\
{C_{(F)}}_4^5 &=& -\frac{i \kappa \, \alpha}{24 \pi  s c_w^2 \left(s-4 M_Z^2\right){}^3 s_w^2}
\bigg(-36 \left(C_a^{f \, 2}+C_v^{f \, 2}\right) M_Z^{10}\nn\\
&+&2 \bigg(16 \left(7 C_a^{f \, 2}+C_v^{f \, 2}\right) m_f^2
+33 s \left(C_a^{f \,2}+C_v^{f \, 2}\right)\bigg) M_Z^8\nn\\
&-&2 \bigg(32 \left(5 C_a^{f \, 2}+C_v^{f \, 2}\right) m_f^4+4 s \left(51 C_a^{f \, 2}+5 C_v^{f \, 2}\right)m_f^2
+15 s^2 \left(C_a^{f \, 2}+C_v^{f \, 2}\right)\bigg) M_Z^6\nn\\
&+&s \bigg(384 m_f^4 C_a^{f \, 2}+16 s \left(16 C_a^{f \, 2}+C_v^{f \,2}\right)m_f^2
+3 s^2 \left(C_a^{f \, 2}+C_v^{f \, 2}\right)\bigg) M_Z^4\nn\\
&+&2 s^2 m_f^2 \bigg(C_v^{f \, 2} \left(6 m_f^2+s \right)
-C_a^{f \, 2} \left(66 m_f^2+35 s \right)\bigg) M_Z^2
+s^3 \left(7 C_a^{f \, 2}-C_v^{f \, 2}\right) m_f^2 \left(2 m_f^2+s \right)\bigg)\, ,
\eea
\bea
{C_{(F)}}_0^6 &=& \frac{i \kappa \, \alpha}{288 \pi  s c_w^2 \left(s-4 M_Z^2\right){}^2 s_w^2}
\left(C_a^{f \, 2}+C_v^{f \, 2}\right) \bigg(-72 M_Z^6+8 \left(48 m_f^2+19 s \right) M_Z^4\nn\\
&-&2 s \left(144 m_f^2+35 s \right) M_Z^2 +s^2 \left(48 m_f^2+11 s \right)\bigg)\, ,\nn\\
{C_{(F)}}_1^6 &=&    -\frac{i \kappa \, \alpha}{24 \pi  s c_w^2 \left(s-4 M_Z^2\right) s_w^2}
\left(C_a^{f \, 2}+C_v^{f \, 2}\right) \left(s-2 M_Z^2\right)\, , \nn\\
{C_{(F)}}_2^6 &=&   \frac{i \kappa\, \alpha}{24 \pi  s c_w^2 \left(s-4 M_Z^2\right) s_w^2}\,
\left(C_a^{f \, 2}+C_v^{f \, 2}\right) \left(s \left(m_f^2+s \right)-2 \left(m_f^2+2 s \right) M_Z^2\right) \, , \nn\\
{C_{(F)}}_3^6 &=&    \frac{i \kappa \, \alpha}{48 \pi  s c_w^2 \left(s-4 M_Z^2\right){}^3 s_w^2}
\bigg(24 \left(C_a^{f \, 2}+C_v^{f \, 2}\right) M_Z^8+24 \left(C_a^{f \, 2}+C_v^{f \, 2}\right) \left(s-4 m_f^2\right) M_Z^6\nn\\
&+&4 s \left(6\left(11 C_a^{f \, 2}+3 C_v^{f \, 2}\right) m_f^2-7 s \left(C_a^{f \, 2}+C_v^{f \, 2}\right)\right) M_Z^4\nn\\
&+&s^2\left(7 \left(s-20 m_f^2\right) C_a^{f \, 2}+C_v^{f \, 2} \left(7 s-44 m_f^2\right)\right) M_Z^2
+4 s^3 \left(5 C_a^{f \, 2}+2 C_v^{f \, 2}\right) m_f^2\bigg)\, ,\nn
\eea
\bea
{C_{(F)}}_4^6 &=&    \frac{i \kappa \, \alpha}{8 \pi  s c_w^2 \left(s-4 M_Z^2\right){}^3 s_w^2}
\bigg(\left(M_Z^4-4 m_f^2 M_Z^2+s m_f^2\right) \bigg(-4 M_Z^6+2 \left(8 m_f^2+9 s \right) M_Z^4\nn\\
&-&s \left(12 m_f^2+11 s \right) M_Z^2 +2 s^2\left(m_f^2+s \right)\bigg) C_a^{f \, 2}\nn\\
&+&C_v^{f \, 2} \bigg(-4 M_Z^{10}+2 \left(16 m_f^2+9 s \right) M_Z^8-\left(64 m_f^4+56 s m_f^2+11 s^2\right) M_Z^6\nn\\
&+& 2 s \left(32 m_f^4+16 s m_f^2+s^2\right) M_Z^4-s^2 m_f^2 \left(20 m_f^2+9 s \right) M_Z^2+s^3 m_f^2 \left(2 m_f^2+s
\right)\bigg)\bigg)\, ,
\eea
\bea
{C_{(F)}}_0^7 &=&      -\frac{i \kappa\, \alpha}{24 \pi  s c_w^2 \left(s-4 M_Z^2\right){}^2 s_w^2}
\left(C_a^{f \, 2}+C_v^{f \, 2}\right) M_Z^2 \left(6 M_Z^4-\left(32 m_f^2+7 s \right) M_Z^2+2 s \left(4 m_f^2+s \right)\right)
\, ,\nn\\
{C_{(F)}}_1^7 &=&    \frac{i \kappa \, \alpha}{12 \pi  s c_w^2 \left(s-4 M_Z^2\right) s_w^2}
\left(C_a^{f \, 2}+C_v^{f \, 2}\right) M_Z^2  \, ,\nn\\
{C_{(F)}}_2^7 &=&     -\frac{i \kappa \, \alpha}{12 \pi  s c_w^2 \left(s-4 M_Z^2\right) s_w^2}
\left(C_a^{f \, 2}+C_v^{f \, 2}\right) m_f^2 M_Z^2 \, ,\nn\\
{C_{(F)}}_3^7 &=&     -\frac{i \kappa \, \alpha}{48 \pi  s c_w^2 \left(s-4 M_Z^2\right){}^3 s_w^2}
\bigg(12 s^3 m_f^2 C_a^{f \, 2}-24 \left(C_a^{f \, 2}+C_v^{f \, 2}\right) M_Z^8\nn\\
&-& 4 \left(C_a^{f \, 2}+C_v^{f \, 2}\right) \left(s-24 m_f^2\right) M_Z^6
- 2 s \left(\left(s-20 m_f^2\right) C_a^{f \, 2}+C_v^{f \, 2} \left(76
m_f^2+s \right)\right) M_Z^4\nn\\
&+&s^2 \left(32 \left(C_v^{f \, 2}-2 C_a^{f \, 2}\right) m_f^2+3 s \left(C_a^{f \, 2}+C_v^{f \, 2}\right)\right)M_Z^2\bigg)
\, ,\nn\\
{C_{(F)}}_4^7 &=&    -\frac{i \kappa \, \alpha}{8 \pi  s c_w^2 \left(s-4 M_Z^2\right){}^3 s_w^2}
\bigg(\left(M_Z^4-4 m_f^2 M_Z^2+s m_f^2\right) \bigg(4 M_Z^6-16 m_f^2 M_Z^4\nn\\
&+&s \left(4 m_f^2-3s \right) M_Z^2+s^3\bigg) C_a^{f \, 2}
+C_v^{f \, 2} M_Z^2 \bigg(4 \left(M_Z^2-4 m_f^2\right){}^2 M_Z^4+8 s m_f^2
\left(5 M_Z^2-4 m_f^2\right) M_Z^2\nn\\
&+&s^3 \left(3 m_f^2+M_Z^2\right)+s^2 \left(4 m_f^4-20 M_Z^2 m_f^2-3 M_Z^4\right)\bigg)\bigg)\, ,
\eea
\bea
{C_{(F)}}_0^8 &=& \frac{i \kappa \, \alpha}{144 \pi  s^2 c_w^2 \left(s-4 M_Z^2\right){}^3 s_w^2}
\left(C_a^{f \, 2}+C_v^{f \, 2}\right) \bigg(-216 M_Z^8+8 \left(96 m_f^2+19 s \right) M_Z^6\nn\\
&-&24 s \left(16 m_f^2+s \right) M_Z^4+3 s^2 \left(3 s-16 m_f^2\right) M_Z^2+s^3 \left(24 m_f^2+s \right)\bigg) \, , \nn\\
{C_{(F)}}_1^8 &=&      -\frac{i \kappa \, \alpha}{12 \pi  s^2 c_w^2 \left(s-4 M_Z^2\right){}^2 s_w^2}
\left(C_a^{f \, 2}+C_v^{f \, 2}\right) \left(-6 M_Z^4+s M_Z^2+s^2\right)\, ,  \nn\\
{C_{(F)}}_2^8 &=&     \frac{i \kappa \, \alpha}{12 \pi  s^2 c_w^2 \left(s-4 M_Z^2\right){}^2 s_w^2}
\left(C_a^{f \, 2}+C_v^{f \, 2}\right) m_f^2 \left(-6 M_Z^4+s M_Z^2+s^2\right) \, , \nn \\
{C_{(F)}}_3^8 &=&     \frac{i \kappa \, \alpha}{12 \pi  s^2 c_w^2 \left(s-4 M_Z^2\right){}^4 s_w^2}
\bigg(36 \left(C_a^{f \, 2}+C_v^{f \, 2}\right) M_Z^{10}-4 \left(C_a^{f \, 2}+C_v^{f \, 2}\right) \left(20 m_f^2+3 s \right)
M_Z^8 \nn\\
&-&s \left(\left(20 m_f^2+s \right) C_a^{f \, 2}+C_v^{f \, 2} \left(s-108 m_f^2\right)\right) M_Z^6+s^2 \left(\left(34 m_f^2+3 s
\right)
C_a^{f \, 2}+C_v^{f \, 2} \left(3 s-62 m_f^2\right)\right) M_Z^4\nn\\
&+&s^3 \left(\left(s-22 m_f^2\right) C_a^{f \, 2}+C_v^{f \, 2} \left(2 m_f^2+s \right)\right) M_Z^2+2 s^4 \left(2 C_a^{f \, 2}+C_v^{f \, 2}\right) m_f^2\bigg) \, ,\nn\eea
\bea
{C_{(F)}}_4^8 &=&      \frac{i \kappa \, \alpha}{12 \pi  s^2 c_w^2 \left(s-4 M_Z^2\right){}^4 s_w^2}
\bigg(-36 \left(C_a^{f \, 2}+C_v^{f \, 2}\right) M_Z^{12}+2 \left(C_a^{f \, 2}+C_v^{f \, 2}\right) \left(112 m_f^2+9 s \right) M_Z^{10}\nn\\
&+&4\left(-80 \left(C_a^{f \, 2}+C_v^{f \, 2}\right) m_f^4-2 s \left(15 C_a^{f \, 2}+31 C_v^{f \, 2}\right) m_f^2+3 s^2 \left(C_a^{f \, 2}
+C_v^{f \, 2}\right)\right) M_Z^8\nn\\
&+&s \left(256 \left(C_a^{f \, 2}+C_v^{f \, 2}\right) m_f^4+8 s \left(13 C_v^{f \, 2}-7 C_a^{f \, 2}\right)
m_f^2-9 s^2 \left(C_a^{f \, 2}+C_v^{f \, 2}\right)\right) M_Z^6\nn\\
&+&s^2 \bigg(-36 \left(C_a^{f \, 2}+C_v^{f \, 2}\right) m_f^4
+ 2s \left(23 C_a^{f \, 2}-13 C_v^{f \, 2}\right) m_f^2+3 s^2 \left(C_a^{f \, 2}+C_v^{f \, 2}\right)\bigg) M_Z^4\nn\\
&-&s^3 m_f^2 \left(10 \left(C_a^{f \, 2}+C_v^{f \, 2}\right) m_f^2+s \left(15 C_a^{f \, 2}+C_v^{f \, 2}\right)\right) M_Z^2\nn\\
&+&s^4 m_f^2 \left(2 \left(m_f^2+s \right) C_a^{f \, 2}+C_v^{f \, 2} \left(2 m_f^2+s \right)\right)\bigg)\, ,\eea
\bea
{C_{(F)}}_0^9 &=&    -\frac{i \kappa \, \alpha}{72 \pi  s^2 c_w^2 \left(s-4 M_Z^2\right){}^3 s_w^2}
\left(C_a^{f \, 2}+C_v^{f \, 2}\right) \bigg(108 M_Z^8-2 \left(192 m_f^2+83 s \right) M_Z^6\nn\\
&+&3 s \left(128 m_f^2+23 s \right) M_Z^4-6 s^2\left(28 m_f^2+s \right) M_Z^2+s^3 \left(24 m_f^2+s \right)\bigg)\, ,\nn\\
{C_{(F)}}_1^9 &=&    \frac{i \kappa \, \alpha}{6 \pi  s^2 c_w^2 \left(s-4 M_Z^2\right){}^2 s_w^2}
\left(C_a^{f \, 2}+C_v^{f \, 2}\right) \left(3 M_Z^4-3 s M_Z^2+s^2\right)\, ,\nn\\
{C_{(F)}}_2^9 &=&   -\frac{i \kappa \, \alpha}{6 \pi  s^2 c_w^2 \left(s-4 M_Z^2\right){}^2 s_w^2}
\left(C_a^{f \, 2}+C_v^{f \, 2}\right) m_f^2 \left(3 M_Z^4-3 s M_Z^2+s^2\right)  \, ,\nn\\
{C_{(F)}}_3^9 &=&     -\frac{i \kappa \, \alpha}{24 \pi  s^2 c_w^2 \left(s-4 M_Z^2\right){}^4 s_w^2}
\bigg(-72 \left(C_a^{f \, 2}+C_v^{f \, 2}\right) M_Z^{10}+4 \left(C_a^{f \, 2}+C_v^{f \, 2}\right)\left(40 m_f^2+21 s \right) M_Z^8\nn\\
&-& 4 s\left(2 \left(C_a^{f \, 2}+33 C_v^{f \, 2}\right) m_f^2+15 s \left(C_a^{f \, 2}+C_v^{f \, 2}\right)\right) M_Z^6 +4 s^2 \bigg(\left(5
s-18 m_f^2\right) C_a^{f \, 2}\nn\\
&+&5 C_v^{f \, 2} \left(6 m_f^2+s \right)\bigg) M_Z^4
+s^3 \left(s C_a^{f \, 2}+C_v^{f \, 2} \left(s-48 m_f^2\right)\right) M_Z^2+4 s^4 \left(C_a^{f \, 2}+2 C_v^{f \, 2}\right) m_f^2\bigg) \, ,\nn\\
{C_{(F)}}_4^9 &=&    -\frac{i \kappa\, \alpha}{12 \pi  s^2 c_w^2 \left(s-4 M_Z^2\right){}^4 s_w^2}
\bigg(36 \left(C_a^{f \, 2}+C_v^{f \, 2}\right) M_Z^{12}-16 \left(C_a^{f \, 2}+C_v^{f \,
2}\right) \left(14 m_f^2+3 s \right) M_Z^{10}\nn\\
&+&8 \left(40 \left(C_a^{f \, 2}+C_v^{f \, 2}\right) m_f^4+s \left(33 C_a^{f \, 2}+49 C_v^{f \,
2}\right) m_f^2+3 s^2 \left(C_a^{f \, 2}+C_v^{f \, 2}\right)\right) M_Z^8\nn\\
&-&s \bigg(352 \left(C_a^{f \, 2}+C_v^{f \, 2}\right) m_f^4
+20 s\left(5 C_a^{f \, 2}+13 C_v^{f \, 2}\right) m_f^2+9 s^2 \left(C_a^{f \, 2}+C_v^{f \, 2}\right)\bigg) M_Z^6\nn\\
&+&s^2 \left(180 \left(C_a^{f \,2}+C_v^{f \, 2}\right) m_f^4+8 s \left(C_a^{f \, 2}+10 C_v^{f \, 2}\right) m_f^2+3 s^2 \left(C_a^{f \,
2}+C_v^{f \, 2}\right)\right) M_Z^4 \nn\\
&-&s^3 m_f^2 \left(44 \left(C_a^{f \, 2}+C_v^{f \, 2}\right) m_f^2+s \left(3 C_a^{f \, 2}+17 C_v^{f \,2}\right)\right) M_Z^2\nn\\
&+&s^4 m_f^2 \left(\left(4 m_f^2+s \right) C_a^{f \, 2}+2 C_v^{f \, 2} \left(2 m_f^2+s \right)\right)\bigg)\, .
\eea

\newpage

\subsection{Form factors for the $TZZ$ vertex in the $W$ sector}
The coefficients corresponding to Eq. (\ref{boson1hZZ}) are given by

\bea
{C_{(W)}}_0^2 &=& \frac{-i \kappa \, \alpha \,  M_Z^2} {12 \, s_w^2 \, c_w^2 \, \pi \, s^2  (s-4M_Z^2)}   \bigg(  2 M_Z^4 \left(-12 s_w^6+32 s_w^4-29 s_w^2+9\right)\nn\\
&+&s M_Z^2 \left(4\left(s_w^4+s_w^2\right)-7\right)-2 s^2 \left(s_w^2-1\right) \bigg)\, , \nn \\
{C_{(W)}}_1^2 &=& 0 \, ,\nn\\
{C_{(W)}}_2^2 &=&  \frac{-i \kappa \, \alpha \, M_Z^2} {6 \, s_w^2 \, c_w^2 \, \pi \, s} (-2 s_w^4+3s_w^2-1)\, , \nn\\
{C_{(W)}}_3^2 &=& \frac{-i \kappa \, \alpha \, M_Z^2} {6\, s_w^2 \, c_w^2 \, \pi \, s^2 (s-M_Z^2)^2} \bigg( 2 M_Z^6 \left(12
s_w^6-32 s_w^4+29 s_w^2-9\right) +s M_Z^4 \left(-24 s_w^6+92 s_w^4-110 s_w^2+41\right) \nn \\
&+& s^2 M_Z^2 \left(-12 s_w^4+26 s_w^2-13\right)+ 2 s^3 \left(s_w^2-1\right)^2   \bigg)\, ,\nn\\
{C_{(W)}}_4^2 &=&  \frac{-i \kappa \, \alpha \, M_Z^2} {24 \, s_w^2 \, c_w^2 \, \pi \, s^2 (s-M_Z^2)^2} \bigg( -8 M_Z^8 \left(s_w^2 - 1
\right) \left(4 s_w^2-3\right) \left(12 s_w^4-20 s_w^2+9\right) \nn \\
&+&   4 s M_Z^6 \left(24 s_w^8-60 s_w^6+30 s_w^4+25 s_w^2-18\right)+ 2 s^2 M_Z^4 (-20 s_w^6+76 s_w^4 \nn \\
&-&   103 s_w^2+46) + s^3 M_Z^2 \left(-4 s_w^4+24 s_w^2-19\right) - 2 s^4 \left(s_w^2-1\right)   \bigg)\, ,\\
{C_{(W)}}_0^3 &=&  \frac{-i \kappa \, \alpha}{384 \, s_w^2 \, c_w^2 \, \pi (s-4 M_Z^2)}  \bigg(  4 M_Z^4 \left(124 s_w^4-228
s_w^2+101\right)\nn\\
&-&2 s M_Z^2 \left(412 s_w^4-836s_w^2+417\right)+ s^2 \left(172 s_w^4-356 s_w^2+181\right)\bigg) \, ,\nn \\
{C_{(W)}}_1^3 &=&  \frac{-i \kappa \, \alpha}{48\, s_w^2 \, c_w^2 \, \pi}   \bigg(6 s_w^4 - 10 s_w^2 + \frac{9}{2} \bigg)
\, ,\nn\\
{C_{(W)}}_2^3 &=&  \frac{-i \kappa \, \alpha}{48\, s_w^2 \, c_w^2 \, \pi}  \bigg(\frac{1}{2} M_Z^2 \left(3 s_w^2 \left(4
\left(s_w^2-7\right) s_w^2+43\right)-56\right)+s \left(9 s_w^4-19 s_w^2+\frac{39}{4}\right)  \bigg) \, ,\nn\\
{C_{(W)}}_3^3 &=&   \frac{-i \kappa \, \alpha}{96\, s_w^2 \, c_w^2 \, \pi (s-4M_Z^2)^2}  \bigg( 2 M_Z^6 \left(-48 s_w^6+196
s_w^4-304 s_w^2+151\right)+s M_Z^4 (24 s_w^6-148 s_w^4 \nn \\
&+&   342 s_w^2-209) -2 s^2 M_Z^2 \left(44 s_w^4-76 s_w^2+33\right) +24 s^3 \left(s_w^2-1\right)^2\bigg) \, ,\nn\\
{C_{(W)}}_4^3 &=&   \frac{-i \kappa \, \alpha}{48\, s_w^2 \, c_w^2 \, \pi (s-4 M_Z^2)^2}
\bigg(M_Z^8 \left(4 s_w^2-3\right) \left(48 s_w^6-36 s_w^4+16 s_w^2-31\right) \nn \\
&-& 2 s M_Z^6 \left(48 s_w^8+228 s_w^6-532 s_w^4+247 s_w^2+7\right)
+s^2 M_Z^4 (12 s_w^8+276 s_w^6  -407 s_w^4 \nn \\
&-& 10 s_w^2+128)-s^3 M_Z^2 \left(36 s_w^6+16 s_w^4-133 s_w^2+81\right)+12 s^4 \left(s_w^2-1\right)^2\bigg)\, ,
\eea
\bea
{C_{(W)}}_0^4 &=&  \frac{-i \kappa \, \alpha}{288\, s_w^2 \, c_w^2 \, \pi s(s-4M_Z^2)^2}  \bigg(   s^2 M_Z^2 \left(4 \left(72 s_w^4-93
s_w^2-35\right) s_w^2+225\right)
+4 M_Z^6 (576 s_w^6-1164 s_w^4\nn\\
&+&740 s_w^2-153)-2 s M_Z^4 \left(864 s_w^6-1332 s_w^4+100 s_w^2+369\right) +s^3 \left(-12 s_w^4+20 s_w^2-9\right)\bigg)
\, , \nn\\
{C_{(W)}}_1^4 &=&  \frac{-i \kappa \, \alpha \, (12 s_w^4-20s_w^2+9)(s-3 M_Z^2) }{24\, s_w^2 \, c_w^2 \, \pi s(s-4M_Z^2)}
\, ,\nn\\
{C_{(W)}}_2^4 &=&   \frac{-i \kappa \, \alpha \,   M_Z^2 \left(s_w^2-1\right) \left(M_Z^2 \left(-36 s_w^4+92 s_w^2-43\right)+s \left(12 s_w^4-28 s_w^2+13\right)\right)}{24\, s_w^2 \, c_w^2 \, \pi s(s-4M_Z^2)} \, , \nn\\
{C_{(W)}}_3^4 &=&    \frac{-i \kappa \, \alpha}{48 \, s_w^2 \, c_w^2 \, \pi s(s-4M_Z^2)^3 }  \bigg(
4 M_Z^8 \left(4 s_w^2 \left(84 s_w^4-375 s_w^2+452\right)-641\right)
+2 s M_Z^6 (-264 s_w^6+1396 s_w^4 - 1786 s_w^2+653)\nn\\
&-&4 s^2 M_Z^4 \left(36 s_w^6+32 s_w^4-169 s_w^2+101\right)
+ s^3 M_Z^2 \left(48 s_w^6-60 s_w^4-40 s_w^2+51\right)\bigg) \, , \nn\\
{C_{(W)}}_4^4 &=&    \frac{-i \kappa \, \alpha}{24 \, s_w^2 \, c_w^2 \, \pi s(s-4M_Z^2)^3 }  \bigg(
2 M_Z^{10} \left(4 s_w^2-3\right) \left(48 s_w^6-36 s_w^4+16 s_w^2-31\right)
+ 2 s M_Z^8 \big(-288 s_w^8+312 s_w^6+52 s_w^4\nn\\
&+& 14 s_w^2-89\big) + 2 s^2 M_Z^6 (108 s_w^8 - 60 s_w^6-103 s_w^4-2 s_w^2+58)\nn\\
&+& s^3 M_Z^4 \left(-24 s_w^8-48 s_w^6+98 s_w^4+34 s_w^2-61\right)
+  s^4 M_Z^2 \left(12 s_w^6-16 s_w^4-5 s_w^2+9\right) \bigg)\, ,\\
{C_{(W)}}_0^5 &=&   \frac{-i \kappa \, \alpha}{576\, s_w^2 \, c_w^2 \, \pi s(s-4M_Z^2)^2}  \bigg(
8 M_Z^6 \left(576 s_w^6-1164 s_w^4+740 s_w^2-153\right)
-48 s M_Z^4 \big(24 s_w^6+92 s_w^4\nn\\
&-&246 s_w^2+127\big) + 24 s^2 M_Z^2 \left(156 s_w^4-316 s_w^2+157\right)
+s^3 \left(-492 s_w^4+1028 s_w^2-525\right) \bigg)\, ,\nn\\
{C_{(W)}}_1^5 &=&  \frac{- i \kappa \, \alpha}{24\, s_w^2 \, c_w^2 \, \pi s(s-4M_Z^2)}
\bigg( (6 s_w^4 -10 s_w^2 + \frac{9}{2})(s-6 M_Z^2)  \bigg) \, , \nn\\
{C_{(W)}}_2^5 &=& \frac{- i \kappa \, \alpha}{24\, s_w^2 \, c_w^2 \, \pi s(s-4M_Z^2)}   \bigg(   M_Z^4 \left(-36 s_w^6+128 s_w^4-135
s_w^2+43\right)\nn\\
&+&\frac{1}{2} s M_Z^2\left(12 s_w^6+24 s_w^4-99 s_w^2+61\right) +s^2 \left(-9 s_w^4+19 s_w^2-\frac{39}{4}\right)\bigg)\, , \nn\\
{C_{(W)}}_3^5 &=&   \frac{-i \kappa \, \alpha}{48\, s_w^2 \, c_w^2 \, \pi s(s-4M_Z^2)^3}   \bigg(
4 M_Z^8 \left(4 s_w^2 \left(84 s_w^4-375 s_w^2+452\right)-641\right)\nn\\
&+&4 s M_Z^6 (-396 s_w^6+1600 s_w^4- 1781 s_w^2+577)+s^2 M_Z^4 \left(696 s_w^6-2620 s_w^4+2758 s_w^2-839\right) \nn \\
&+&  2 s^3 M_Z^2 \left(-48 s_w^6+228 s_w^4-284 s_w^2+105\right)-24 s^4 \left(s_w^2-1\right)^2\bigg)\, ,\nn\\
{C_{(W)}}_4^5 &=&   \frac{-i \kappa \, \alpha}{24\, s_w^2 \, c_w^2 \, \pi s(s-4M_Z^2)^3} \bigg(
2 M_Z^{10} \left(4 s_w^2-3\right) \left(48 s_w^6-36 s_w^4+16 s_w^2-31\right)
+ s M_Z^8 (29  -4 s_w^2 \left(540 s_w^4-995 s_w^2+458\right))\nn\\
&+&2 s^2 M_Z^6 (-36 s_w^8+840 s_w^6-1243 s_w^4
+ 229 s_w^2+205)+s^3 M_Z^4 \left(12 s_w^8-372 s_w^6+245 s_w^4+508 s_w^2-391\right) \nn \\
&+& 8 s^4 M_Z^2 \left(3 s_w^6+10 s_w^4-28 s_w^2+15\right)-12 s^5 \left(s_w^2-1\right)^2\bigg)\, ,
\eea
\bea
{C_{(W)}}_0^6 &=&  \frac{-i \kappa \, \alpha}{576 \, s_w^2 \, c_w^2 \, \pi s(s-4M_Z^2)^2} \bigg(
-24 M_Z^6 \left(16 s_w^2-13\right) \left(12 s_w^4-20 s_w^2+9\right)+32 s M_Z^4 (108 s_w^6-72 s_w^4\nn\\
&-&179 s_w^2+141) -8 s^2 M_Z^2 \left(72 s_w^6+276 s_w^4-790 s_w^2+435\right)+s^3 \left(540 s_w^4-1108 s_w^2+561\right)\bigg)
\, ,\nn\\
{C_{(W)}}_1^6 &=& \frac{ -i \kappa \, \alpha}{24\, s_w^2 \, c_w^2 \, \pi s(s-4M_Z^2)}  \bigg( (-6 s_w^4 +10 s_w^2 - \frac{9}{2})(s - 2 M_Z^2)\bigg)\, , \nn\\
{C_{(W)}}_2^6 &=& \frac{ -i \kappa \, \alpha}{24\, s_w^2 \, c_w^2 \, \pi s(s-4M_Z^2)}
\bigg( M_Z^4 \left(12 s_w^6-32 s_w^4+29 s_w^2-9\right)\nn\\
&-&\frac{1}{2} s M_Z^2 \left(12 s_w^6+40 s_w^4-123 s_w^2+69\right)
+s^2 \left(9 s_w^4-19 s_w^2+\frac{39}{4}\right) \bigg)\, , \nn\\
{C_{(W)}}_3^6 &=& \frac{-i \kappa \, \alpha}{48\, s_w^2 \, c_w^2 \, \pi s(s-4M_Z^2)^3} \bigg(
12 M_Z^8 \left(4 s_w^2-3\right) \left(12 s_w^4-20 s_w^2+9\right)\nn\\
&+&12 s M_Z^6 \left(-36 s_w^6+40 s_w^4+s_w^2-7\right)
+s^2 M_Z^4 \left(264 s_w^6-260 s_w^4-102 s_w^2+111\right)\nn\\
&-&s^3 M_Z^2 (48 s_w^6+28 s_w^4 -180 s_w^2+105) +24 s^4 \left(s_w^2-1\right)^2 \bigg)\, , \nn\\
{C_{(W)}}_4^6 &=& \frac{-i \kappa \, \alpha}{8\, s_w^2 \, c_w^2 \, \pi s(s-4M_Z^2)^3} \bigg(
-2 M_Z^{10} \left(3-4 s_w^2\right){}^2 \left(12 s_w^4-20 s_w^2+9\right) \nn \\
&+& s M_Z^8 \left(4 s_w^2-3\right) \left(96 s_w^6-76 s_w^4-36 s_w^2+15\right)
-s^2 M_Z^6 (120 s_w^8+152 s_w^6 - 538 s_w^4+212 s_w^2+53)\nn\\
&+& s^3 M_Z^4 \left(12 s_w^8+120 s_w^6-163 s_w^4-61 s_w^2+92\right)
- 2 s^4 M_Z^2 \left(8 s_w^6+2 s_w^4-27 s_w^2+17\right) +4 s^5 \left(s_w^2-1\right)^2\bigg)\, ,\\
{C_{(W)}}_0^7 &=&   \frac{-i \kappa \, \alpha \, M_Z^2}{384\, s_w^2 \, c_w^2 \, \pi s(s-4M_Z^2)} \bigg(
4 M_Z^2 \left(16 s_w^2-13\right) \left(12 s_w^4-20 s_w^2+9\right)\nn\\
&+&s \left(-372 s_w^4+748 s_w^2-375\right)
-\frac{5 s^2 \left(12 s_w^4-20 s_w^2+9\right)}{s-4M_Z^2} \bigg)\, ,\nn\\
{C_{(W)}}_1^7 &=&    \frac{-i \kappa \, \alpha  \, M_Z^2}{24\, s_w^2 \, c_w^2 \, \pi s(s-4M_Z^2)}( 12 s_w^4 -20 s_w^2+ 9 )
\, ,\nn\\
{C_{(W)}}_2^7 &=&     \frac{-i \kappa \, \alpha \,  M_Z^4 }{24\, s_w^2 \, c_w^2 \, \pi s(s-4M_Z^2)}   \bigg( 12 s_w^6 - 32 s_w^4 + 29 s_w^2 -9\bigg)\, ,\nn\\
{C_{(W)}}_3^7 &=&   \frac{-i \kappa \, \alpha \, M_Z^2}{24 \, s_w^2 \, c_w^2 \, \pi s(s-4M_Z^2)^3}  \bigg(
6 M_Z^6 \left(4 s_w^2-3\right) \left(12 s_w^4-20 s_w^2+9\right) +s M_Z^4 (-456 s_w^6+1348 s_w^4\nn\\
&-&1290 s_w^2+393 )+ s^2 M_Z^2 \left(96 s_w^6-220 s_w^4+108 s_w^2+15\right)
-6 s^3 \left(4 s_w^4-9 s_w^2+5\right) \bigg)\, ,\nn\\
{C_{(W)}}_4^7 &=&   \frac{-i \kappa \, \alpha \, M_Z^2}{8 \, s_w^2 \, c_w^2 \, \pi s(s-4M_Z^2)^3}  \bigg(
-2 M_Z^8 \left(3-4 s_w^2\right){}^2 \left(12 s_w^4-20 s_w^2+9\right) \nn \\
&+&   2 s M_Z^6 \left(96 s_w^8 -152 s_w^6+20 s_w^4+58 s_w^2-21\right)
-s^2 M_Z^4 (24s_w^8+112 s_w^6 \nn \\
&-& 330 s_w^4+254 s_w^2-59 )+ s^3 M_Z^2 \left(28 s_w^6-48 s_w^4+15 s_w^2+5\right)-4 s^4 \left(s_w^2-1\right)^2 \bigg)\, ,
\eea
\bea
{C_{(W)}}_0^8 &=&    \frac{-i \kappa \, \alpha}{288 \, s_w^2 \, c_w^2 \, \pi s^2(s-4M_Z^2)^3}  \bigg(
-24 M_Z^8 \left(32 s_w^2-23\right) \left(12 s_w^4-20 s_w^2+9\right)\nn\\
&+&8 s M_Z^6 (576 s_w^6-1068 s_w^4 + 484 s_w^2+15)
+24 s^2 M_Z^4 \left(24 s_w^6-116 s_w^4+166 s_w^2-73\right) \nn \\
&+&  3 s^3 M_Z^2 \left(-96 s_w^6+332 s_w^4-380 s_w^2+145\right)
+s^4 \left(12 s_w^4-20 s_w^2+9\right)\bigg)  \, , \nn\\
{C_{(W)}}_1^8 &=&   \frac{-i \kappa \, \alpha}{24\, s_w^2 \, c_w^2 \, \pi s^2(s-4M_Z^2)^2}
\bigg((-12 s_w^4 + 20 s_w^2 - 9)(s^2 + s M_Z^2 -6 M_Z^4) \bigg)\, ,\nn\\
{C_{(W)}}_2^8 &=&   \frac{-i \kappa \, \alpha \, M_Z^2}{24\, s_w^2 \, c_w^2 \, \pi s^2(s-4M_Z^2)^2}
\bigg( (-12 s_w^6 + 32 s_w^4 - 29 s_w^2 +9)(s^2 + s M_Z^2 - 6 M_Z^2) \bigg)  \, , \nn\\
{C_{(W)}}_3^8 &=&    \frac{-i \kappa \, \alpha}{48 \, s_w^2 \, c_w^2 \, \pi s^2(s-4M_Z^2)^4}  \bigg(
8 M_Z^{10} \left(20 s_w^2-11\right) \left(12 s_w^4-20 s_w^2+9\right)\nn\\
&-&8 s M_Z^8 (324 s_w^6-908 s_w^4 + 835 s_w^2-244)+2 s^2 M_Z^6 \left(744 s_w^6-1996 s_w^4+1626 s_w^2-375\right) \nn \\
&+& 4 s^3 M_Z^4 \left(-12 s_w^6+20 s_w^4+31 s_w^2-36\right)s^4 M_Z^2 \left(-48 s_w^6+172 s_w^4-208 s_w^2+85\right)\bigg)\, ,\nn\\
{C_{(W)}}_4^8 &=&   \frac{-i \kappa \, \alpha}{24 \, s_w^2 \, c_w^2 \, \pi s^2(s-4M_Z^2)^4}
\bigg(-4 M_Z^{12} \left(4 s_w^2-3\right) \left(20 s_w^2-11\right) \left(12 s_w^4-20 s_w^2+9\right) \nn \\
&+& 2 s M_Z^{10} \left(4 \left(384 s_w^6-876 s_w^4+691 s_w^2-218\right) s_w^2+93\right)  \nn \\
&-& 4 s^2 M_Z^8 \left(108 s_w^8+156 s_w^6-695 s_w^4+564 s_w^2-130\right)-2 s^3 M_Z^6 (60 s_w^8
- 496 s_w^6+865 s_w^4-513 s_w^2+84)\nn\\
&+&2 s^4 M_Z^4 \left(12 s_w^8-48 s_w^6+41 s_w^4+17 s_w^2-21\right)
+  s^5 M_Z^2 \left(-12 s_w^6+40 s_w^4-45 s_w^2+17\right)\bigg)\, ,\\
{C_{(W)}}_0^9 &=& \frac{-i \kappa \, \alpha}{144 \, s_w^2 \, c_w^2 \, \pi s^2(s-4M_Z^2)^3}
\bigg(-12 M_Z^8 \left(32 s_w^2-23\right) \left(12 s_w^4-20 s_w^2+9\right)\nn\\
&+&2 s M_Z^6 (2304 s_w^6-4668 s_w^4
+ 2852 s_w^2-429)+3 s^2 M_Z^4 \left(-672 s_w^6+1356 s_w^4-812 s_w^2+113\right) \nn \\
&+& 3 s^3 M_Z^2 \left(96 s_w^6-212 s_w^4+148 s_w^2-31\right)+s^4 \left(-12 s_w^4+20 s_w^2-9\right) \bigg) \, ,\nn\\
{C_{(W)}}_1^9 &=&   \frac{-i \kappa \, \alpha}{12\, s_w^2 \, c_w^2 \, \pi s^2(s-4M_Z^2)^2}   \bigg( (12 s_w^4 - 20 s_w^2 + 9) (s^2 - 3 s M_Z^2 + 3 M_Z^4) \bigg) \, ,  \nn\\
{C_{(W)}}_2^9 &=&   \frac{-i \kappa \, \alpha \, M_Z^2}{12\, s_w^2 \, c_w^2 \, \pi s^2(s-4M_Z^2)^2}
\bigg( (12s_w^6 - 32 s_w^4 + 29 s_w^2 -9)(s^2 - 3 s M_Z^2 + 3 M_Z^4) \bigg)\, ,   \nn\\
{C_{(W)}}_3^9 &=&   \frac{-i \kappa \, \alpha}{24\, s_w^2 \, c_w^2 \, \pi s^2(s-4M_Z^2)^4}  \bigg(
4 M_Z^{10} \left(20 s_w^2-11\right) \left(12 s_w^4-20 s_w^2+9\right)\nn\\
&+&2 s M_Z^8 (-792 s_w^6+2020 s_w^4 - 1718 s_w^2+461)+6 s^2 M_Z^6 \left(120 s_w^6-260 s_w^4+158 s_w^2-13\right) \nn \\
&+& s^3 M_Z^4 \left(-288 s_w^6+588 s_w^4-316 s_w^2+9\right)+ s^4 M_Z^2 \left(2 s_w^2-1\right) \left(4 s_w^2-5\right)
\left(6 s_w^2-5\right) \bigg)  \, ,    \nn\\
{C_{(W)}}_4^9 &=&     \frac{-i \kappa \, \alpha}{12\, s_w^2 \, c_w^2 \, \pi s^2(s-4M_Z^2)^4}  \bigg(*
-2 M_Z^{12} \left(4 s_w^2-3\right) \left(20 s_w^2-11\right) \left(12 s_w^4-20 s_w^2+9\right) \nn \\
&+&   4 s M_Z^{10} \left(528 s_w^8-1188 s_w^6+892 s_w^4-227 s_w^2+3\right)-2 s^2 M_Z^8 (540 s_w^8-960 s_w^6 \nn \\
&+& 361 s_w^4+147 s_w^2-76)+s^3 M_Z^6 \left(264 s_w^8-368 s_w^6+26 s_w^4+90 s_w^2-3\right) \nn \\
&-&   2 s^4 M_Z^4 \left(12 s_w^8+12 s_w^6-55 s_w^4+35 s_w^2-3\right)s^5 M_Z^2 \left(12 s_w^6-28 s_w^4+21 s_w^2-5\right)\bigg)\, .
\eea
\subsection{Form factors for the $TZZ$ vertex in the $(Z,H)$ sector}
The coefficients corresponding to Eq. (\ref{boson2hZZ}) are given by
\bea
{C_{(Z,H)}}_0^2  &=&   -\frac{i \kappa\, \alpha  M_Z^2}{12 \pi  s^2 c_w^2 \left(s-4 M_Z^2\right) s_w^2} \left(M_Z^4+M_H^2 M_Z^2-3 s M_Z^2+s^2\right)\, , \nn\\
{C_{(Z,H)}}_1^2  &=&    \frac{i \kappa\, \alpha}{24 \pi  s^2 c_w^2 \left(s-4 M_Z^2\right)
s_w^2}  \left(M_H^2-M_Z^2 \right)\left(s-2 M_Z^2\right)\, , \nn\\
{C_{(Z,H)}}_2^2  &=&  -{C_{(Z,H)}}_1^2 \, , \nn\\
{C_{(Z,H)}}_3^2  &=&  -\frac{i \kappa\, \alpha}{48 \pi  s^2 c_w^2 \left(s-4 M_Z^2\right){}^2 s_w^2}
\left(\left(8 M_Z^6+s^3\right) M_H^2+M_Z^2 \left(s-4 M_Z^2\right) \left(s-2 M_Z^2\right)\left(3 s-2 M_Z^2\right)\right)\, , \nn\\
{C_{(Z,H)}}_4^2  &=&    \frac{i \kappa\, \alpha}{48 \pi  s^2 c_w^2 \left(s-4 M_Z^2\right){}^2 s_w^2}
\left(2 \left(4 M_H^4-s^2\right) M_Z^4-s \left(2 M_H^2+s \right){}^2 M_Z^2+s^2 M_H^2\left(2 M_H^2+s \right)\right)\, ,\nn\\
{C_{(Z,H)}}_5^2  &=& \frac{i \kappa\, \alpha  M_Z^2}{6 \pi  s^2 c_w^2 \left(s-4 M_Z^2\right){}^2 s_w^2}
\left(-s M_H^4-\left(3 M_H^2+5 s \right) M_Z^4+\left(M_H^2+s \right)\left(M_H^2+2 s \right) M_Z^2\right) \, ,\nn\\
{C_{(Z,H)}}_6^2  &=&  -\frac{i \kappa\, \alpha  \left(2 M_H^2+s \right)}{48 \pi  s^2 c_w^2 \left(s-4 M_Z^2\right)^2 s_w^2}
\bigg(4 \left(7 s-4 M_H^2\right) M_Z^6+4 \left(M_H^2-s \right) \left(M_H^2+3 s \right) M_Z^4 \nn \\
&+& 2 s \left(-M_H^4-2 s M_H^2+s^2\right) M_Z^2+s^2 M_H^4\bigg)\, , \nn\\
{C_{(Z,H)}}_7^2  &=& -\frac{i \kappa\, \alpha}{48 \pi  s^2 c_w^2 \left(s-4 M_Z^2\right){}^2 s_w^2}
\bigg(\left(8 M_Z^6+s^3\right) M_H^4\nn\\
&+&4 M_Z^2 \left(s-4 M_Z^2\right) \left(2 M_Z^4-s M_Z^2+s^2\right) M_H^2+4 s M_Z^4 \left(s-4 M_Z^2\right){}^2\bigg)\, ,\\
{C_{(Z,H)}}_0^3  &=&  \frac{i \kappa\, \alpha}{384 \pi  c_w^2\left(s-4 M_Z^2\right) s_w^2}
\left(4 M_H^4+80 M_Z^4+3 s^2-2 \left(4 M_H^2+15 s \right) M_Z^2\right)\, ,\nn\\
{C_{(Z,H)}}_1^3  &=&  \frac{i \kappa\, \alpha}{192 \pi  c_w^2 \left(s-4 M_Z^2\right) s_w^2}  \left(4 M_H^2-s \right) \, ,\nn\\
{C_{(Z,H)}}_2^3  &=&   -\frac{i \kappa\, \alpha}{192 \pi  c_w^2 \left(s-4 M_Z^2\right) s_w^2}  \left(4 M_H^2-8 M_Z^2+s \right)
\, ,\nn\\
{C_{(Z,H)}}_3^3  &=&   \frac{i \kappa\, \alpha}{384 \pi  c_w^2 \left(s-4 M_Z^2\right){}^2 s_w^2}
\left(2 M_H^2-4 M_Z^2+s \right) \left(6 M_H^4+6 \left(s-4 M_Z^2\right) M_H^2+\left(s-28 M_Z^2\right) \left(s-4
M_Z^2\right)\right)\, ,\nn\\
{C_{(Z,H)}}_4^3  &=&    -\frac{i \kappa\, \alpha}{384 \pi  c_w^2 \left(s-4 M_Z^2\right){}^2 s_w^2}\left(2 M_H^2-s \right)
\left(6 M_H^4-4 \left(2 M_Z^2+s \right) M_H^2+\left(s-16 M_Z^2\right) \left(s-6 M_Z^2\right)\right)\, ,\nn\\
{C_{(Z,H)}}_5^3  &=&    \frac{i \kappa\, \alpha}{96 \pi  c_w^2 \left(s-4 M_Z^2\right){}^2 s_w^2}
\left(32 M_Z^6-2 \left(4 M_H^2+19 s \right) M_Z^4+\left(6 M_H^4+11 s M_H^2+6 s^2\right) M_Z^2-6 s M_H^4\right)\, , \nn\\
{C_{(Z,H)}}_6^3  &=&   \frac{i \kappa\, \alpha} {96 \pi  c_w^2 \left(s-4 M_Z^2\right)^2 s_w^2} \bigg(3 M_H^8-4 \left(2 M_Z^2+s
\right) M_H^6+\left(32 M_Z^4+6 s M_Z^2+s^2\right) M_H^4 \nn \\
&-& 8 M_Z^4 \left(8 M_Z^2+s \right) M_H^2+s M_Z^2 \left(16 M_Z^4+3 s M_Z^2-s^2\right)\bigg) \, ,\nn\\
{C_{(Z,H)}}_7^3  &=&   \frac{i \kappa\, \alpha } {96 \pi  c_w^2 \left(s-4 M_Z^2\right){}^2 s_w^2} \bigg(3 M_H^8+6 \left(s-4
M_Z^2\right) M_H^6+4 \left(s-7 M_Z^2\right) \left(s-4 M_Z^2\right) M_H^4 \nn \\
&+& \left(s-16 M_Z^2\right) \left(s-4 M_Z^2\right)^2 M_H^2-2 M_Z^2 \left(s-4 M_Z^2\right){}^2 \left(3 s-4 M_Z^2\right)\bigg)\, ,
\eea
\bea
{C_{(Z,H)}}_0^4  &=&       \frac{i \kappa\, \alpha}{288 \pi  s c_w^2 \left(s-4 M_Z^2\right)^2 s_w^2}
\bigg(304 M_Z^6-10 s M_Z^4-13 s^2 M_Z^2+s^3\nn\\
&+&12 M_H^4 \left(M_Z^2+s \right)+12 M_H^2 \left(6 M_Z^4-8 s M_Z^2+s^2\right)\bigg)\, , \nn\\
{C_{(Z,H)}}_1^4  &=& \frac{i \kappa\, \alpha}{48\pi  s c_w^2 M_Z^2 \left(s-4 M_Z^2\right){}^2 s_w^2}
\left(-8 M_Z^6+5 s M_Z^4-2 s^2 M_Z^2+M_H^2 \left(-4 M_Z^4+2 s M_Z^2+s^2\right)\right)\, ,\nn\\
{C_{(Z,H)}}_2^4  &=&     -\frac{i \kappa\, \alpha}{48 \pi  s c_w^2 M_Z^2 \left(s-4 M_Z^2\right){}^2 s_w^2}
\left(16 M_Z^6-9 s M_Z^4+M_H^2 \left(-4 M_Z^4+2 s M_Z^2+s^2\right)\right)\, , \nn\\
{C_{(Z,H)}}_3^4  &=&     \frac{i \kappa\, \alpha  \left(2 M_H^2-4 M_Z^2+s \right)}{96 \pi  s c_w^2
\left(s-4 M_Z^2\right)^3 s_w^2} \bigg(6\left(M_Z^2+s \right) M_H^4+6 \left(s-4 M_Z^2\right) \left(M_Z^2+s \right) M_H^2 \nn \\
&+& M_Z^2 \left(4 M_Z^2-s \right) \left(28 M_Z^2+11 s \right)\bigg)\, ,\nn\\
{C_{(Z,H)}}_4^4  &=&    -\frac{i \kappa\, \alpha}{96 \pi  s c_w^2 \left(s-4 M_Z^2\right){}^3 s_w^2}
\left(2 M_H^2-s \right)\bigg(96 M_Z^6+26 s M_Z^4-5 s^2 M_Z^2\nn\\
&+&6 M_H^4 \left(M_Z^2+s \right)-4 M_H^2 \left(2 M_Z^4+7 s M_Z^2\right)\bigg)\, ,\nn\\
{C_{(Z,H)}}_5^4  &=&      \frac{i \kappa\, \alpha }{48 \pi  s c_w^2 M_Z^2 \left(s-4 M_Z^2\right)^3 s_w^2}
\bigg(448 M_Z^{10}-236 s M_Z^8-8 s^2 M_Z^6+6 s^3 M_Z^4 \nn \nn\\
&-& 2 M_H^2 \left(2 s-3 M_Z^2\right) \left(-24 M_Z^4-7 s M_Z^2+s^2\right) M_Z^2
+ M_H^4 \left(44 M_Z^6-16 s M_Z^4-14 s^2 M_Z^2+s^3\right)\bigg) \, ,\nn\\
{C_{(Z,H)}}_6^4  &=&     \frac{i \kappa\, \alpha}{48 \pi  s c_w^2 \left(s-4 M_Z^2\right)^3 s_w^2}  \bigg(6 \left(M_Z^2+s \right) M_H^8-4
\left(4 M_Z^4+10 s M_Z^2+s^2\right) M_H^6\nn\\
&+&(64 M_Z^6+92 s M_Z + 14 s^2 M_Z^2+s^3) M_H^4-16 \left(8 M_Z^8+5 s M_Z^6+2 s^2 M_Z^4\right) M_H^2\nn\\
&+& s M_Z^2 \left(32 M_Z^6+22 s M_Z^4+4 s^2 M_Z^2-s^3\right)\bigg)\, ,\nn \\
{C_{(Z,H)}}_7^4  &=&     \frac{i \kappa\, \alpha}{48 \pi  s c_w^2 \left(s-4 M_Z^2\right){}^3 s_w^2}
\bigg(6 \left(M_Z^2+s \right) M_H^8+12 \left(s-4 M_Z^2\right) \left(M_Z^2+s \right) M_H^6\nn\\
&+&\left(s-4 M_Z^2\right)\left(-56 M_Z^4 - 32 s M_Z^2+7 s^2\right) M_H^4\nn\\
&+&\left(s-16 M_Z^2\right)\left(s-4 M_Z^2\right){}^2 \left(2 M_Z^2+s\right) M_H^2 - 4 M_Z^4 \left(s-4M_Z^2\right){}^3\bigg)\, ,\\
{C_{(Z,H)}}_0^5  &=&    -\frac{i \kappa\, \alpha}{576 \pi  s c_w^2 \left(s-4 M_Z^2\right){}^2 s_w^2}
\bigg(-608 M_Z^6+12 \left(59 s-12 M_H^2\right) M_Z^4\nn\\
&-&24 \left(M_H^4+s M_H^2+7 s^2\right) M_Z^2+36 s M_H^4+11 s^3\bigg)\, ,\nn\\
{C_{(Z,H)}}_1^5  &=&   \frac{i \kappa\, \alpha}{96 \pi  s c_w^2 M_Z^2 \left(s-4 M_Z^2\right){}^2 s_w^2}
\left(M_Z^2 \left(s-2 M_Z^2\right) \left(8 M_Z^2+3 s \right)-4 M_H^2 \left(2 M_Z^4-2 s M_Z^2+s^2\right)\right)\, , \nn\\
{C_{(Z,H)}}_2^5  &=&   \frac{i \kappa\, \alpha}{96 \pi  s c_w^2 M_Z^2 \left(s-4 M_Z^2\right){}^2 s_w^2}
\left(-32 M_Z^6+18 s M_Z^4-5 s^2 M_Z^2+4 M_H^2 \left(2 M_Z^4-2 s M_Z^2+s^2\right)\right)\, , \nn\\
{C_{(Z,H)}}_3^5  &=&   -\frac{i \kappa\, \alpha  \left(2 M_H^2-4 M_Z^2+s \right)}
{192 \pi  s c_w^2 \left(s-4 M_Z^2\right){}^3 s_w^2} \bigg(6\left(3 s-2 M_Z^2\right) M_H^4 \nn \\
&+& 6 \left(s-4 M_Z^2\right) \left(3 s-2 M_Z^2\right) M_H^2+\left(s-4 M_Z^2\right) \left(56 M_Z^4-54 s M_Z^2+s^2\right)\bigg)
\, ,\nn \eea
\bea
{C_{(Z,H)}}_4^5  &=& \frac{i \kappa\, \alpha\left(2 M_H^2-s \right)} {192 \pi  s c_w^2 \left(s-4 M_Z^2\right)^3 s_w^2}
\bigg(-192 M_Z^6+4 \left(4 M_H^2+59 s \right) M_Z^4\nn\\
&-&12 \left(M_H^2+s \right) \left(M_H^2+3 s \right) M_Z^2 + s \left(18 M_H^4-4 s M_H^2+s^2\right)\bigg)\, ,\nn\\
{C_{(Z,H)}}_5^5  &=&   -\frac{i \kappa\, \alpha }{48 \pi  s c_w^2 M_Z^2 \left(s-4 M_Z^2\right){}^3 s_w^2} \bigg(-448 M_Z^{10}+4
\left(36 M_H^2+131 s \right) M_Z^8 -2(22 M_H^4 + 95 s M_H^2\nn\\
&+&107 s^2) M_Z^6 + s \left(62 M_H^4+93 s M_H^2+24 s^2\right) M_Z^4
-8 s^2 M_H^2 \left(4 M_H^2+s \right) M_Z^2+2 s^3 M_H^4\bigg)\, ,\nn\\
{C_{(Z,H)}}_6^5  &=&    \frac{i \kappa\, \alpha} {48 \pi  s c_w^2 \left(s-4 M_Z^2\right){}^3 s_w^2} \bigg(\left(6 M_Z^2-9 s
\right) M_H^8+8 \left(-2 M_Z^4+4 s M_Z^2+s^2\right) M_H^6\nn\\
&-&2(-32 M_Z^6+ 50 s M_Z^4+8 s^2 M_Z^2+s^3) M_H^4+8 \left(-16 M_Z^8+14 s M_Z^6+5 s^2 M_Z^4\right) M_H^2  \nn \\
&+& s M_Z^2 \left(32 M_Z^6-26 s M_Z^4-11 s^2 M_Z^2+2 s^3\right)\bigg)\, ,\nn\\
{C_{(Z,H)}}_7^5  &=&   -\frac{i \kappa\, \alpha}{48 \pi  s c_w^2 \left(s-4 M_Z^2\right){}^3 s_w^2}
\bigg(\left(9 s-6 M_Z^2\right) M_H^8+6 \left(s-4 M_Z^2\right) \left(3 s-2 M_Z^2\right) M_H^6\nn\\
&+&\left(s-4 M_Z^2\right)(56 M_Z^4 - 76 s M_Z^2+11 s^2) M_H^4+2 \left(s-16 M_Z^2\right)
\left(s-4 M_Z^2\right){}^2 \left(s-M_Z^2\right)M_H^2\nn\\
&-&2 M_Z^2 \left(s-4 M_Z^2\right){}^3 \left(3 s-2 M_Z^2\right)\bigg)\, ,\\
{C_{(Z,H)}}_0^6  &=&  -\frac{i \kappa\, \alpha}{576 \pi  s c_w^2 \left(s-4 M_Z^2\right){}^2 s_w^2}
\bigg(96 M_Z^6-172 s M_Z^4+80 s^2 M_Z^2-7 s^3 +12 M_H^4 \left(3 s-2 M_Z^2\right)\nn\\
&+&24 M_H^2\left(10 M_Z^4-9 s M_Z^2+s^2\right)\bigg)\, ,\nn\\
{C_{(Z,H)}}_1^6  &=&     -\frac{i \kappa\, \alpha}{96 \pi  s c_w^2 M_Z^2 \left(s-4 M_Z^2\right){}^2 s_w^2}
\left(-16 M_Z^6+6 s M_Z^4-3 s^2 M_Z^2+2 M_H^2 \left(4 M_Z^4+s^2\right)\right)\, , \nn\\
{C_{(Z,H)}}_2^6  &=&      \frac{i \kappa\, \alpha}{96 \pi  s c_w^2 M_Z^2 \left(s-4 M_Z^2\right){}^2 s_w^2}
\left(2 M_H^2 \left(4 M_Z^4+s^2\right)-s M_Z^2 \left(6 M_Z^2+s \right)\right) \, ,\nn\\
{C_{(Z,H)}}_3^6  &=&   \frac{i \kappa\, \alpha}{192 \pi  s c_w^2 \left(s-4 M_Z^2\right){}^3 s_w^2}
\bigg(\left(24 M_Z^2-36 s \right) M_H^6-6 \left(7 s-6 M_Z^2\right) \left(s-4 M_Z^2\right) M_H^4 \nn \\
&-& 2 \left(s-4 M_Z^2\right){}^2 \left(3 s-4 M_Z^2\right) M_H^2+\left(s-4 M_Z^2\right){}^2 \left(8 M_Z^4-30 s
M_Z^2+s^2\right)\bigg)\, ,\nn\\
{C_{(Z,H)}}_4^6  &=&    \frac{i \kappa\, \alpha}{192 \pi  s c_w^2 \left(s-4 M_Z^2\right){}^3 s_w^2}
\bigg(4 \left(40 M_H^4+2 s M_H^2-3 s^2\right) M_Z^4-4(6 M_H^6+49 s M_H^4 \nn \\
&-& 26 s^2 M_H^2+4 s^3) M_Z^2+s \left(36 M_H^6-6 s M_H^4-4 s^2 M_H^2+s^3\right)\bigg)\, ,\nn\\
{C_{(Z,H)}}_5^6  &=&   \frac{i \kappa\, \alpha}{48 \pi  s c_w^2 M_Z^2 \left(s-4 M_Z^2\right){}^3 s_w^2}
\bigg(12 \left(7 s-4 M_H^2\right)M_Z^8+6 \left(2 M_H^4+5 s M_H^2-5 s^2\right) M_Z^6 \nn \\
&+& s \left(-6 M_H^4-43 s M_H^2+6 s^2\right) M_Z^4+4 s^2 M_H^2 \left(4 M_H^2+s \right) M_Z^2-s^3 M_H^4\bigg)\, , \nn\\
{C_{(Z,H)}}_6^6  &=&  \frac{i \kappa\, \alpha}{16 \pi  s c_w^2 \left(s-4 M_Z^2\right){}^3 s_w^2}
\left(M_H^4-4 M_Z^2 M_H^2+s M_Z^2\right) \bigg(\left(2 M_Z^2-3 s \right) M_H^4\nn\\
&+&\left(-8 M_Z^4+8 s M_Z^2+s^2\right) M_H^2+s M_Z^2 \left(2 M_Z^2-3 s \right)\bigg)\, ,\nn \\
{C_{(Z,H)}}_7^6  &=&     -\frac{i \kappa\, \alpha}{16 \pi  s c_w^2 \left(s-4 M_Z^2\right){}^3 s_w^2}
\left(M_H^2-4 M_Z^2+s \right) \bigg(\left(3 s-2 M_Z^2\right) M_H^6\nn\\
&+&2 \left(4 M_Z^4-5 s M_Z^2+s^2\right) M_H^4+2 s M_Z^2 \left(s-4 M_Z^2\right){}^2\bigg)\, ,
\eea
\bea
{C_{(Z,H)}}_0^7  &=&    \frac{i \kappa\, \alpha}{96 \pi  s c_w^2 \left(s-4 M_Z^2\right){}^2 s_w^2}
\bigg(-16 M_Z^6+10 \left(s-4 M_H^2\right) M_Z^4 +\left(4 M_H^4+s^2\right) M_Z^2+4 s M_H^4\bigg)\, ,\nn\\
{C_{(Z,H)}}_1^7  &=&   \frac{i \kappa\, \alpha}{48 \pi  s c_w^2 \left(s-4 M_Z^2\right){}^2 s_w^2}
\left(8 M_Z^4-7 s M_Z^2+M_H^2 \left(6 s-4 M_Z^2\right)\right)\, ,\nn\\
{C_{(Z,H)}}_2^7  &=&     \frac{i \kappa\, \alpha}{48 \pi  s c_w^2 \left(s-4 M_Z^2\right){}^2 s_w^2}
\left(\left(4 M_Z^2-6 s \right) M_H^2+5 s M_Z^2\right)\, ,\nn\\
{C_{(Z,H)}}_3^7  &=&    \frac{i \kappa\, \alpha}{96 \pi  s c_w^2 \left(s-4 M_Z^2\right){}^3 s_w^2}
\bigg(12 \left(M_Z^2+s \right) M_H^6+6 \left(s-4 M_Z^2\right) \left(3 M_Z^2+2 s \right) M_H^4 \nn \\
&+&  \left(s-4 M_Z^2\right){}^2 \left(4 M_Z^2+s \right) M_H^2+M_Z^2 \left(s-4 M_Z^2\right){}^2 \left(4 M_Z^2+11 s \right)\bigg)
\, , \nn\\
{C_{(Z,H)}}_4^7  &=&   \frac{i \kappa\, \alpha}{96 \pi  s c_w^2 \left(s-4 M_Z^2\right){}^3 s_w^2}   \bigg(-12 \left(M_Z^2+s \right)
M_H^6+\left(80 M_Z^4+54 s M_Z^2+4 s^2\right) M_H^4 \nn \\
&-&  s \left(124 M_Z^4+10 s M_Z^2+s^2\right) M_H^2+s^2 M_Z^2 \left(26 M_Z^2+s \right)\bigg)\, ,\nn\\
{C_{(Z,H)}}_5^7  &=&  -\frac{i \kappa\, \alpha}{24 \pi  s c_w^2 \left(s-4 M_Z^2\right){}^3 s_w^2}
\bigg(6 \left(4 M_H^2+9 s \right) M_Z^6-\left(6 M_H^4+59 s M_H^2+18 s^2\right) M_Z^4\nn\\
&+&s \left(20 M_H^4+2 s M_H^2+3 s^2\right) M_Z^2+s^2 M_H^4\bigg)\, ,\nn\\
{C_{(Z,H)}}_6^7  &=&   \frac{i \kappa\, \alpha}{16 \pi  s c_w^2 \left(s-4 M_Z^2\right){}^3 s_w^2}
\left(M_H^4-4 M_Z^2 M_H^2+s M_Z^2\right) \bigg(2 \left(M_Z^2+s \right) M_H^4\nn\\
&-&\left(8 M_Z^4+4 s M_Z^2+s^2\right) M_H^2+2 s M_Z^2 \left(M_Z^2+s \right)\bigg) \, ,\nn \\
{C_{(Z,H)}}_7^7  &=&    \frac{i \kappa\, \alpha  M_H^2}{16 \pi  s c_w^2 \left(s-4 M_Z^2\right){}^3 s_w^2} \bigg(2 \left(M_Z^2+s
\right) M_H^6+\left(s-4 M_Z^2\right) \left(4 M_Z^2+3 s \right) M_H^4 \nn\\
&+& \left(s-4 M_Z^2\right){}^2 \left(2 M_Z^2+s \right) M_H^2+2 s M_Z^2 \left(s-4 M_Z^2\right){}^2\bigg)\, ,\\
{C_{(Z,H)}}_0^8  &=&   -\frac{i \kappa\, \alpha}{288 \pi  s^2 c_w^2 \left(s-4 M_Z^2\right){}^3 s_w^2}  \bigg(96 M_Z^8-100 s
M_Z^6+48 s^2 M_Z^4+15 s^3 M_Z^2+s^4 \nn \\
&+& 12 M_H^4 \left(-6 M_Z^4+3 s M_Z^2+4 s^2\right) +12 M_H^2 \left(44 M_Z^6-22 s M_Z^4-10 s^2 M_Z^2+s^3\right)\bigg)\, ,\nn\\
{C_{(Z,H)}}_1^8  &=&   \frac{i \kappa\, \alpha}{48 \pi  s^2 c_w^2 M_Z^2 \left(s-4 M_Z^2\right){}^3 s_w^2}
\bigg(16 M_Z^8+2 \left(4 M_H^2-9 s \right) M_Z^6\nn\\
&+& s \left(8 M_H^2+13 s \right) M_Z^4+2 s^2 \left(s-8 M_H^2\right) M_Z^2-s^3 M_H^2\bigg) \, ,\nn\\
{C_{(Z,H)}}_2^8  &=&   \frac{i \kappa\, \alpha}{48 \pi  s^2 c_w^2 M_Z^2 \left(s-4 M_Z^2\right){}^3 s_w^2}
\bigg(32 M_Z^8-2 s M_Z^6-19 s^2 M_Z^4\nn\\
&+&M_H^2 \left(-8 M_Z^6-8 s M_Z^4+16 s^2 M_Z^2+s^3\right)\bigg)\, , \nn\\
{C_{(Z,H)}}_3^8  &=&  -\frac{i \kappa\, \alpha} {96 \pi  s^2 c_w^2 \left(s-4 M_Z^2\right){}^4 s_w^2}
\bigg(12 \left(-6 M_Z^4+3 s M_Z^2+4 s^2\right) M_H^6+6 \left(s-4 M_Z^2\right)(-18 M_Z^4 \nn \\
&+&  5 s M_Z^2+8 s^2) M_H^4+4 \left(s-4 M_Z^2\right){}^2 \left(-8 M_Z^4+s M_Z^2+2 s^2\right) M_H^2+M_Z^2 \left(s-4
M_Z^2\right){}^3 \left(2 M_Z^2+7 s \right)\bigg) \, ,\nn
\eea
\bea
{C_{(Z,H)}}_4^8  &=&   -\frac{i \kappa\, \alpha}{96 \pi  s^2 c_w^2 \left(s-4 M_Z^2\right){}^4 s_w^2}
\bigg(4 \left(-88 M_H^4+42 s M_H^2+s^2\right) M_Z^6+4 \left(M_H^2-s \right)(18 M_H^4 +57 s M_H^2 \nn \\
&-& 17 s^2) M_Z^4+s \left(-36 M_H^6+218 s M_H^4-86 s^2 M_H^2+9 s^3\right) M_Z^2-2 s^2 M_H^2 \left(24 M_H^4-10 s
M_H^2+s^2\right)\bigg)\, ,\nn\\
{C_{(Z,H)}}_5^8  &=&    -\frac{i \kappa\, \alpha} {48 \pi  s^2 c_w^2 M_Z^2 \left(s-4 M_Z^2\right){}^4 s_w^2} \bigg(\left(-8 M_Z^8+60 s
M_Z^6-100 s^2 M_Z^4-22 s^3 M_Z^2+s^4\right) M_H^4 \nn \\
&+& \left(96 M_Z^{10}-228 s M_Z^8+262 s^2 M_Z^6+42 s^3 M_Z^4-4 s^4 M_Z^2\right) M_H^2
- 6 s M_Z^4 \left(-44 M_Z^6+38 s M_Z^4-2 s^2 M_Z^2+s^3\right)\bigg)\, , \nn\\
{C_{(Z,H)}}_6^8  &=&  -\frac{i \kappa\, \alpha} {48 \pi  s^2 c_w^2 \left(s-4 M_Z^2\right){}^4 s_w^2}
\bigg(-4 \left(80 M_H^4-24 s M_H^2+s^2\right) M_Z^8+2(112 M_H^6+44 s M_H^4 \nn \\
&-& 48 s^2 M_H^2+9 s^3) M_Z^6-6 \left(6 M_H^8+16 s M_H^6-38 s^2 M_H^4+20 s^3 M_H^2-3 s^4\right) M_Z^4 \nn \\
&+& s \left(18 M_H^8-144 s M_H^6+92 s^2 M_H^4-18 s^3 M_H^2+s^4\right) M_Z^2+2 s^2 M_H^4 \left(12 M_H^4-7 s
M_H^2+s^2\right)\bigg)\, ,\nn\\
{C_{(Z,H)}}_7^8  &=&  -\frac{i \kappa\, \alpha  M_H^2 }{48 \pi  s^2 c_w^2 \left(s-4 M_Z^2\right){}^4 s_w^2}
\bigg(6 \left(-6 M_Z^4+3 s M_Z^2+4 s^2\right) M_H^6+12 \left(s-4 M_Z^2\right)(-6 M_Z^4 \nn \\
&+& 2 s M_Z^2+3 s^2) M_H^4+2 \left(7 s-10 M_Z^2\right) \left(s-4 M_Z^2\right){}^2 \left(2 M_Z^2+s \right) M_H^2 \nn \\
&+& \left(s-4 M_Z^2\right){}^3 \left(-4 M_Z^4+4 s M_Z^2+s^2\right)\bigg)\, ,\\
{C_{(Z,H)}}_0^9  &=&   \frac{i \kappa\, \alpha}{144 \pi  s^2 c_w^2 \left(s-4 M_Z^2\right){}^3 s_w^2}
\bigg(-48 M_Z^8+26 s M_Z^6-3 s^2 M_Z^4+9 s^3 M_Z^2+s^4 \nn \\
&+& 12 M_H^2 \left(s-11 M_Z^2\right) \left(2 M_Z^4-2 s M_Z^2+s^2\right)+12 M_H^4 \left(3 M_Z^4-4 s M_Z^2+3 s^2\right)\bigg)
\, ,\nn\\
{C_{(Z,H)}}_1^9  &=&   \frac{i \kappa\, \alpha}{24 \pi  s^2 c_w^2 M_Z^2 \left(s-4 M_Z^2\right){}^3 s_w^2}
\left(8 M_Z^8-5 s M_Z^6-2 s^3 M_Z^2+M_H^2 \left(4 M_Z^6-10 s M_Z^4+7 s^2 M_Z^2+s^3\right)\right)\, , \nn\\
{C_{(Z,H)}}_2^9  &=&  -\frac{i \kappa\, \alpha}{24 \pi  s^2 c_w^2 M_Z^2 \left(s-4 M_Z^2\right){}^3 s_w^2}
\bigg(-16 M_Z^8+25 s M_Z^6-14 s^2 M_Z^4\nn\\
&+&M_H^2 \left(4 M_Z^6-10 s M_Z^4+7 s^2 M_Z^2+s^3\right)\bigg)\, ,\nn\\
{C_{(Z,H)}}_3^9  &=&   \frac{i \kappa\, \alpha}{48 \pi  s^2 c_w^2 \left(s-4 M_Z^2\right){}^4 s_w^2}  \bigg(12 \left(3 M_Z^4-4 s
M_Z^2+3 s^2\right) M_H^6+6 \left(s-4 M_Z^2\right)(9 M_Z^4 \nn \\
&-& 10 s M_Z^2+6 s^2) M_H^4+\left(s-4 M_Z^2\right){}^2 \left(16 M_Z^4-18 s M_Z^2+7 s^2\right) M_H^2+M_Z^2
\left(M_Z^2+4 s \right) \left(4 M_Z^2-s \right){}^3\bigg) \, ,\nn \\
{C_{(Z,H)}}_4^9  &=&    -\frac{i \kappa\, \alpha}{48 \pi  s^2 c_w^2 \left(s-4 M_Z^2\right){}^4 s_w^2}
\bigg(12 \left(3 M_Z^4-4 s M_Z^2+3 s^2\right) M_H^6-2(88 M_Z^6-91 s M_Z^4 \nn \\
&+& 64 s^2 M_Z^2+8 s^3) M_H^4+s \left(84 M_Z^6-26 s M_Z^4+76 s^2 M_Z^2+s^3\right) M_H^2+s^2 M_Z^2
\left(2 M_Z^4-21 s M_Z^2-8 s^2\right)\bigg)\, , \nn \\
{C_{(Z,H)}}_5^9  &=& \frac{i \kappa\, \alpha }{24 \pi  s^2 c_w^2 M_Z^2 \left(s-4 M_Z^2\right){}^4 s_w^2}  \bigg(-6 s \left(22
M_Z^4-16 s M_Z^2+7 s^2\right) M_Z^6+M_H^4(4 M_Z^8-12 s M_Z^6 \nn \\
&-& 23 s^3 M_Z^2+s^4)+M_H^2 \left(-48 M_Z^{10}+90 s M_Z^8-30 s^2 M_Z^6+58 s^3 M_Z^4-4 s^4 M_Z^2\right)\bigg)\, , \nn
\eea
\bea
{C_{(Z,H)}}_6^9  &=&    \frac{i \kappa\, \alpha}{24 \pi  s^2 c_w^2 \left(s-4 M_Z^2\right){}^4 s_w^2}  \bigg(6 \left(3 M_Z^4-4 s
M_Z^2+3 s^2\right) M_H^8-(112 M_Z^6-120 s M_Z^4 \nn \\
&+&  84 s^2 M_Z^2+11 s^3) M_H^6+2 \left(80 M_Z^8-46 s M_Z^6+24 s^2 M_Z^4+31 s^3 M_Z^2+s^4\right) M_H^4 \nn \\
&-&  3 s M_Z^2 \left(16 M_Z^6-8 s M_Z^4+16 s^2 M_Z^2+5 s^3\right) M_H^2+s^2 M_Z^2 \left(2 M_Z^6+9 s^2 M_Z^2+s^3\right)\bigg)
\, ,\nn\\
{C_{(Z,H)}}_7^9  &=&    \frac{i \kappa\, \alpha  M_H^2} {24 \pi  s^2 c_w^2 \left(s-4 M_Z^2\right){}^4 s_w^2}
\bigg(6 \left(3 M_Z^4-4 s M_Z^2+3 s^2\right) M_H^6+3 \left(s-4 M_Z^2\right)(12 M_Z^4 \nn \\
&-& 14 s M_Z^2+9 s^2) M_H^4+\left(s-4 M_Z^2\right){}^2 \left(20 M_Z^4-22 s M_Z^2+11 s^2\right) M_H^2 \nn \\
&+& \left(s-4 M_Z^2\right){}^3 \left(2 M_Z^4-5 s M_Z^2+s^2\right)\bigg)\, .
\eea

\subsection{The improvement contribution}
\label{imprformfactors}
The two form factors with the improvement contribution are given by
\bea
 \Phi^{(I)}_1 (s,M_Z^2,M_Z^2,M_W^2,M_Z^2,M_H^2) &=& - i \frac{\kappa}{2} \frac{\alpha}{12 \pi \, s_w^2 \, c_w^2 \, s (s-4M_Z^2)^2} \bigg\{
 (c_w^2 - s_w^2)^2 \bigg[ s^2 - 6 M_Z^2 s + 8 M_Z^4 \nn \\
 && \hspace{-5cm} + 2 M_Z^2 (s + 2 M_Z^2) \, \mathcal D_0 \left( s, M_Z^2 , M_W^2, M_W^2 \right)   +  2 \left( c_w^2 M_Z^2 (8 M_Z^4 - 6 M_Z^2 s + s^2) - 2 M_Z^6 + 2 M_Z^4 s \right) \times \nn \\
 && \hspace{-5cm} \times \, \mathcal C_0 \left( s, M_Z^2, M_Z^2, M_W^2,M_W^2,M_W^2 \right) \bigg]  +  s^2 - 6 M_Z^2 s + 8 M_Z^4  + 2 M_Z^2 (s + 2 M_Z^2) \big[
 \mathcal B_0 \left(s, M_Z^2, M_Z^2 \right)  \nn \\
 && \hspace{-5cm} -  \mathcal B_0 \left(M_Z^2, M_Z^2, M_H^2 \right)  \big] + \left( 3 M_Z^2 s - 2 M_H^2 (s - M_Z^2) \right) \big[ \mathcal B_0 \left(s, M_H^2, M_H^2 \right) -  \mathcal B_0 \left(s, M_Z^2, M_Z^2 \right)\big] \nn \\
 && \hspace{-5cm} + M_H^2 \left( 2 M_H^2 (s-M_Z^2) + 8 M_Z^4 - 6 M_Z^2 s + s^2 \right) \mathcal C_0\left( s, M_Z^2,M_Z^2, M_H^2,M_Z^2,M_Z^2\right) \nn \\
 && \hspace{-5cm} + \left( 2 M_H^2 (M_H^2 - 4 M_Z^2)(s-M_Z^2) + s M_Z^2 (s+ 2 M_Z^2)\right) \mathcal C_0\left( s, M_Z^2,M_Z^2, M_Z^2,M_H^2,M_H^2\right)
 \bigg\} \, , \\
 \Phi^{(I)}_2 (s,M_Z^2,M_Z^2,M_W^2,M_Z^2,M_H^2) &=& i \frac{\kappa}{2} \frac{\alpha}{48 \pi \, s_w^2 \, c_w^2 (s-4 M_Z^2)^2} \bigg\{
 (c_w^2 -s_w^2)^2 \bigg[ 4 M_Z^4 (s - 4 M_Z^2) \nn \\
 && \hspace{-5cm} + 8 M_Z^4 (s-M_Z^2) \, \mathcal D_0 \left( s, M_Z^2, M_W^2, M_W^2\right) + 2 M_Z^4 \big( s^2 - 2 M_Z^2 s + 4 M_Z^4 + 4 c_w^2 M_Z^2 (s - 4 M_Z^2) \big) \times \nn \\
 && \hspace{-5cm}  \times \, \mathcal C_0 \left( s, M_Z^2, M_Z^2, M_W^2,M_W^2,M_W^2 \right) \bigg] + 4 M_Z^2 s_w^4 c_w^2 s (s - 4 M_Z^2)^2 \mathcal C_0 \left( s, M_Z^2, M_Z^2, M_W^2,M_W^2,M_W^2 \right)  \nn \\
 && \hspace{-5cm}  + 4 M_Z^2 (s-4 M_Z^2) + \big( M_Z^2 s (s + 2 M_Z^2) - M_H^2 (s^2 - 2 M_Z^2 s + 4 M_Z^4) \big) \big[ \mathcal B_0 \left(s, M_H^2, M_H^2 \right) -  \mathcal B_0 \left(s, M_Z^2, M_Z^2 \right) \big] \nn \\
 && \hspace{-5cm}  + 8 M_Z^4 (s - M_Z^2) \big[ \mathcal B_0 \left(s, M_Z^2, M_Z^2 \right) -  \mathcal B_0 \left(M_Z^2, M_Z^2, M_H^2 \right)  \big] +
 M_H^2 \big( 4 M_Z^4 (s-4 M_Z^2) \nn \\
 && \hspace{-5cm}  + M_H^2 (s^2 - 2 MZ^2 s + 4 M_Z^4)\big) \mathcal C_0\left( s, M_Z^2,M_Z^2, M_H^2,M_Z^2,M_Z^2\right) + \big( M_H^2(M_H^2 - 4 M_Z^2) (s^2 - 2 M_Z^2 s + 4 M_Z^4) \nn \\
 && \hspace{-5cm}  + 2 M_Z^2 s (s^2 - 6 M_Z^2 s + 14 M_Z^4)\big) \mathcal C_0\left( s, M_Z^2,M_Z^2, M_Z^2,M_H^2,M_H^2\right)
 \bigg\}\, .
\eea
\subsection{Coefficients of the external leg corrections}
\label{externalleg}
\bea
{C^{(I)}_{(F)}}^1_0 &=&  \frac{i \, \kappa \, \alpha \, m_f^2 (C^{f \, 2}_v + C^{f \, 2}_a) (s- 2 M_Z^2)}{6 \pi s_w^2 c_w^2 s (s-M_H^2)(s-4M_Z^2)} \,, \nn\\
{C^{(I)}_{(F)}}^1_1 &=& {C^{(I)}_{(F)}}^1_2 = 0 \,, \nn\\
{C^{(I)}_{(F)}}^1_3 &=& \frac{i \, \kappa \, \alpha \, m_f^2}{3 \pi s_w^2 c_w^2 s (s-M_H^2)(s-4M_Z^2)^2 } \bigg[ M_Z^2 (C^{f \, 2}_v + C^{f \, 2}_a) (s+ 2 M_Z^2) + C^{f \, 2}_a (s-4M_Z^2) s \bigg] \,, \nn\\
{C^{(I)}_{(F)}}^1_4 &=& \frac{i \, \kappa \, \alpha \, m_f^2 (s-2 M_Z^2)}{12 \pi s_w^2 c_w^2 s (s-M_H^2)(s-4M_Z^2)^2} \bigg[ (C^{f \, 2}_v + C^{f \, 2}_a)
\left( 4 m_f^2 (s-4M_Z^2) + 4 M_Z^4 + 6 MZ^2 s - s^2 \right)\nn\\
&+& 2  C^{f \, 2}_a s (s-4 M_Z^2) \bigg]\, ,
\eea
\bea
{C^{(I)}_{(F)}}^2_0 &=&  - \frac{i \, \kappa \, \alpha \, m_f^2 \, M_Z^2  (C^{f \, 2}_v + C^{f \, 2}_a) }{3 \pi s_w^2 c_w^2 s
(s-M_H^2)(s-4M_Z^2)} \,, \nn\\
{C^{(I)}_{(F)}}^2_1 &=& 0 \,, \nn\\
{C^{(I)}_{(F)}}^2_2 &=& - \frac{i \, \kappa \, \alpha \, m_f^2 \, C^{f \, 2}_a}{6 \pi s_w^2 c_w^2  (s-M_H^2)} \,,\nn \\
{C^{(I)}_{(F)}}^2_3 &=&  - \frac{i \, \kappa \, \alpha \, m_f^2 }{3 \pi s_w^2 c_w^2 s (s-M_H^2)(s-4M_Z^2)^2} \bigg[ 2 M_Z^4
(C^{f \, 2}_v + C^{f \, 2}_a) (s - M_Z^2) + C^{f \, 2}_a M_Z^2 s (s-4M_Z^2) \bigg]\,, \nn\\
{C^{(I)}_{(F)}}^2_4 &=& - \frac{i \, \kappa \, \alpha \, m_f^2}{6 \pi s_w^2 c_w^2 s (s-M_H^2)(s-4M_Z^2)^2} \bigg[
(C^{f \, 2}_v + C^{f \, 2}_a) M_Z^4 \left( 4 m_f^2 (s-4 M_Z^2) + 4 M_Z^4 - 2M_Z^2 s + s^2\right)\nn\\
&+& 2 C^{f \, 2}_a s (s-4M_Z^2)\left( m_f^2 (s-4M_Z^2)+M_Z^4\right)\bigg]\, ,
\eea
\bea
{C^{(I)}_{(W)}}^1_0 &=& - \frac{i \, \kappa \, \alpha (s-2M_Z^2)}{24 \pi s_w^2 c_w^2 s (s-M_H^2)(s-4M_Z^2)}\bigg[ M_H^2 (1-2
s_w^2)^2 + 2 M_Z^2 (-12 s_w^6 + 32 s_w^4 - 29 s_w^2 +9)\bigg] \,, \nn\\
{C^{(I)}_{(W)}}^1_1 &=& {C^{(I)}_{(W)}}^1_2 = 0 \,, \nn\\
{C^{(I)}_{(W)}}^1_3 &=& - \frac{i \, \kappa \, \alpha \, M_Z^2}{12 \pi s_w^2 c_w^2 s (s-M_H^2)(s-4M_Z^2)^2} \bigg[
M_H^2 (1-2s_w^2)^2 (s+ 2 M_Z^2) - 2 (s_w^2-1)\bigg( 2 M_Z^4 (12 s_w^4 - 20 s_w^2 + 9) \nn\\
&+& s M_Z^2 (12 s_w^4 - 20 s_w^2 + 1) + 2 s^2 \bigg) \bigg] \,, \nn\\
{C^{(I)}_{(W)}}^1_4 &=& - \frac{i \, \kappa \, \alpha \, M_Z^2}{12 \pi s_w^2 c_w^2 s (s-M_H^2)(s-4M_Z^2)^2} \bigg[
2 (s_w^2-1) \bigg( 2 M_Z^6 (4 s_w^2 -3) (12 s_w^4 - 20 s_w^2 + 9) \nn\\
&+& 2 M_Z^4 s (-36 s_w^6 + 148 s_w^4 - 163 s_w^2 + 54) + M_Z^2 s^2 (12 s_w^6 - 96 s_w^4 + 125 s_w^2 - 43) + 4 s^3 (2 s_w^4 - 3
s_w^2 + 1)\bigg)\nn\\
&-& M_H^2 (1-2 s_w^2)^2 \left( M_Z^4 (8 s_w^2 - 6) + 2 M_Z^2 s (2-3 s_w^2) + s^2 (s_w^2 -1)\right)\bigg]\, ,
\eea
\bea
{C^{(I)}_{(W)}}^2_0 &=& \frac{i \, \kappa \, \alpha \, M_Z^4}{ 12 \pi s_w^2 c_w^2 s (s-M_H^2)(s-4M_Z^2)}
\bigg[ M_H^2 (1-2 s_w^2)^2 + 2 M_Z^2 (-12 s_w^6 + 32 s_w^4 - 29 s_w^2 +9)\bigg] \,, \nn\\
{C^{(I)}_{(W)}}^2_1 &=& 0 \,, \nn\\
{C^{(I)}_{(W)}}^2_2 &=& - \frac{i \, \kappa \, \alpha \, M_Z^2}{24 \pi s_w^2 c_w^2 (s-M_H^2)} \bigg[ 8 s_w^4 - 13 s_w^2 + 5\bigg] \,, \nn\\
{C^{(I)}_{(W)}}^2_3 &=&  - \frac{i \, \kappa \, \alpha M_Z^2}{6 \pi s_w^2 c_w^2 s (s-M_H^2)(s-4M_Z^2)} \bigg[
M_H^2 M_Z^2 (1-2 s_w^2)^2 (M_Z^2 -s) -2 (s_w^2-1)\bigg( M_Z^6 (12 s_w^4 - 20 s_w^2 + 9)\nn\\
&-& 3 M_Z^4 s \left( 4 (s_w^2 - 3)s_w^2 + 7 \right) + M_Z^2 s^2 (7 - 8 s_w^2) + s^3(s_w^2-1)\bigg)  \bigg]\,, \nn\\
{C^{(I)}_{(W)}}^2_4 &=& -\frac{i \, \kappa \, \alpha \, M_Z^4}{ 24 \pi s_w^2 c_w^2 s (s-M_H^2)(s-4M_Z^2)}\bigg[
M_H^2 \bigg( - 4 M_Z^6 (1-2 s_w^2)^2 (4 s_w^2 -3) + 2 M_Z^4 s (24 s_w^6 - 28 s_w^4 + 6 s_w^2 -1) \nn\\
&+& M_Z^2 s^2 (-16 s^6 + 12 s_w^4 + 4 s_w^2 -1) + 2 s^3 s_w^4 (s_w^2-1)\bigg)\nn\\
&+& 2 (s_w^2 -1) \bigg( 4 M_Z^8 (4 s_w^2 - 3)(12
s_w^4 - 20 s_w^2 + 9) - 2 M_Z^6 s (24 s_w^6 - 52 s_w^4 + 6 s_w^2 + 15)\nn\\
&+& M_Z^4 s^2 (45 - 4 s_w^2 (s_w^2 +13)) + 2M_Z^2 s^3 (4 s_w^4 + 2 s_w^2 - 5) - s^4 (s_w^4 - 1) \bigg) \bigg]\, ,
\eea
\bea
{C^{(I)}_{(Z,H)}}^1_0 &=& - \frac{i \, \kappa \, \alpha (2 M_H^2 + M_Z^2) (s - 2 M_Z^2)}{24 \pi s_w^2 c_w^2 s (s-M_H^2)(s-4M_Z^2)} \,, \nn\\
{C^{(I)}_{(Z,H)}}^1_1 &=& - {C^{(I)}_{(Z,H)}}^1_2 = \frac{i \, \kappa \, \alpha (M_H^2 -M_Z^2)}{12 \pi s_w^2 c_w^2 s (s-M_H^2)(s-4 M_Z^2)} \,, \nn\\
{C^{(I)}_{(Z,H)}}^1_3 &=& - \frac{i \, \kappa \, \alpha}{24 \pi s_w^2 c_w^2 s (s-M_H^2)(s-4 M_Z^2)^2} \bigg[ 2 M_H^4 (s-M_Z^2) + 3 M_H^2 M_Z^2 s + 2 M_Z^2 (4 M_Z^4 - 9 M_Z^2 s + 2 s^2)\bigg] \,, \nn\\
{C^{(I)}_{(Z,H)}}^1_4 &=&  \frac{i \, \kappa \, \alpha}{8 \pi s_w^2 c_w^2 s (s-M_H^2)(s-4M_Z^2)^2} \bigg[ 2 M_H^4 (s-M_Z^2) - 3 M_H^2 M_Z^2 s \bigg] \,, \nn\\
{C^{(I)}_{(Z,H)}}^1_5 &=&  \frac{i \, \kappa \, \alpha}{12 \pi s_w^2 c_w^2 s (s-M_H^2)(s-4M_Z^2)^2} \bigg[ M_H^2 (s + 2 M_Z^2) (4 M_Z^2 - M_H^2 ) + 2 M_Z^2 s (s- 4 M_Z^2)\bigg] \,, \nn\\
{C^{(I)}_{(Z,H)}}^1_6 &=&  \frac{i \, \kappa \, \alpha \, M_H^2}{8 \pi s_w^2 c_w^2 s (s-M_H^2)(s-4M_Z^2)^2} \bigg[ 2 M_H^2 (s -M_Z^2) (4 M_Z^2 - M_H^2) - M_Z^2 s (s + 2 M_Z^2)\bigg] \,, \nn\\
{C^{(I)}_{(Z,H)}}^1_7 &=& - \frac{i \, \kappa \, \alpha \, M_H^2}{24 \pi s_w^2 c_w^2 s (s-M_H^2)(s-4M_Z^2)^2} \bigg[
2 M_H^4 (s - M_Z^2) + M_H^2 (4 M_Z^4 - 2 M_Z^2 s + s^2)\nn\\
&+& 2 M_Z^2 (8 M_Z^4 - 14 M_Z^2 s + 3 s^2 ) \bigg]\, ,
\eea
\bea
{C^{(I)}_{(Z,H)}}^2_0 &=& \frac{i \, \kappa \, \alpha \, M_Z^4 (2 M_H^2 + M_Z^2)}{12 \pi s_w^2 c_w^2 s (s - M_H^2)(s-4 M_Z^2)}
\, ,\nn \\
{C^{(I)}_{(Z,H)}}^2_1 &=& - {C^{(I)}_{(Z,H)}}^2_2 = \frac{i \, \kappa \, \alpha (M_Z^2 - M_H^2)(s-2M_Z^2)}{24 \pi s_w^2 c_w^2 s
(s - M_H^2)(s -4 M_Z^2)} \, , \nn\\
{C^{(I)}_{(Z,H)}}^2_3 &=& \frac{i \, \kappa \, \alpha}{48 \pi s_w^2 c_w^2 s (s-M_H^2)(s-4M_Z^2)^2} \bigg[
M_H^4 (4 M_Z^4 - 2 M_Z^2 s + s^2) + M_H^2 M_Z^2 s (s+ 2 M_Z^2) \nn\\
&-& M_Z^2 (16 M_Z^6 - 28 M_Z^4 s + 18 M_Z^2 s^2 - 3 s^3) \bigg] \,,\nn\\
{C^{(I)}_{(Z,H)}}^2_4 &=& - \frac{i \, \kappa \, \alpha \, M_H^2}{16 \pi s_w^2 c_w^2 s (s- M_H^2)(s-4M_Z^2)^2} \bigg[
M_H^2 (4 M_Z^4 - 2 M_Z^2 s + s^2) - M_Z^2 s (s+ 2 M_Z^2) \bigg] \,, \nn\\
{C^{(I)}_{(Z,H)}}^2_5 &=&  \frac{i \, \kappa \, \alpha }{6 \pi s_w^2 c_w^2 s (s-M_H^2)(s-4M_Z^2)^2} \bigg[
M_Z^2 s (M_H^2 - 2 M_Z^2)^2 - M_H^2 M_Z^4 (M_H^2 - 4 M_Z^2) - M_Z^4 s^2 \bigg] \,, \nn\\
{C^{(I)}_{(Z,H)}}^2_6 &=& \frac{i \, \kappa \, \alpha \, M_H^2}{16 \pi s_w^2 c_w^2 s (s-M_H^2)(s-4M_Z^2)^2} \bigg[
M_H^2 (4 M_Z^4 - 2 M_Z^2 s + s^2) (M_H^2 - 4 M_Z^2) + 2 MZ^2 s (16 M_Z^4 - 6 M_Z^2 s + s^2)  \bigg] \,,\nn\\
{C^{(I)}_{(Z,H)}}^2_7 &=& \frac{i \, \kappa \, \alpha}{ 48 \pi s_w^2 c_w^2 s (s-M_H^2)(s-4M_Z^2)^2} \bigg[
M_H^6 (4 M_Z^4 - 2 M_Z^2 s + s^2) + 2 M_H^4 M_Z^2 (s^2 - 4 M_Z^4)\nn\\
&-& 4 M_H^2 M_Z^2 (8 M_Z^6 - 10 M_Z^4 s + 6 M_Z^2 s^2 -s^3) + 4 M_Z^4 s (s-4M_Z^2)^2 \bigg]\, .
\eea
The one - loop graviton - Higgs mixing amplitude is given by
\bea
\Sigma^{\mu\nu}_{hH}(k) &=& \Sigma^{\mu\nu}_{Min, \, hH}(k) + \Sigma^{\mu\nu}_{I, \, hH}(k) =  i \frac{\kappa}{2} \frac{e}{288
\pi^2 \, s_w \, c_w \, M_Z \, s}
\bigg\{
2 m_f^2 \bigg[ 3 (s - 4 m_f^2) \mathcal B_0\left( s, m_f^2,m_f^2 \right) \nn \\
&+& 12 \mathcal A_0\left( m_f^2 \right)+ 2 s - 12 m_f^2 \bigg] +
6 ( 6 M_W^2 + M_H^2) \big( M_W^2 \mathcal \, B_0 \left( s, M_W^2 , M_W^2 \right) + M_W^2 -  \mathcal A_0\left( M_W^2 \right)\big)
\nn \\
&+& s\big( 18 M_W^2 \, \mathcal B_0 \left( s, M_W^2 , M_W^2 \right) + 18 M_W^2 + M_H^2 \big) + 3 M_Z^2 (M_H^2 + 6 M_Z^2 - 3 s)
\mathcal B_0 \left( s, M_Z^2, M_Z^2\right) \nn \\
&+& 9 M_H^4 \mathcal B_0 \left( s, M_H^2, M_H^2\right) - 3 (M_H^2 + 6 M_Z^2) \mathcal A_0\left( M_Z^2 \right) -9 M_H^2  \mathcal
A_0\left( M_H^2 \right) + 3 (3 M_H^4 + M_H^2 M_Z^2 \nn \\
&+& 6 M_Z^4) - s (2 M_H^2 + 9 M_Z^2)
\bigg\} \left( s \, \eta^{\mu\nu} - k^{\mu}k^{\nu} \right) + i \frac{\kappa}{2} \frac{e}{16 \pi^2 \, s_w \, c_w } \bigg\{ 2 c_w
M_W \mathcal A_0\left( M_W^2 \right) + M_Z \mathcal A_0\left( M_Z^2 \right) \bigg\} \eta^{\mu\nu}\, .\nn\\
\eea


\begin{thebibliography}{10}

\bibitem{Armillis:2009pq}
R.~Armillis, C.~Corian\`o, and L.~Delle~Rose,
\newblock (2009), arXiv:0910.3381.

\bibitem{Armillis:2010qk}
R.~Armillis, C.~Corian\`o, and L.~Delle~Rose,
\newblock Phys.Rev. {\bf D82}, 064023 (2010), arXiv:1005.4173.

\bibitem{Armillis:2010pa}
R.~Armillis, C.~Corian\`o, L.~Delle~Rose, and L.~Manni,
\newblock (2010), arXiv:1003.3930.

\bibitem{Giannotti:2008cv}
M.~Giannotti and E.~Mottola,
\newblock Phys. Rev. {\bf D79}, 045014 (2009), arXiv:0812.0351.

\bibitem{Coriano:2011ti}
C.~Corian\`o, L.~Delle Rose, A.~Quintavalle, and M.~Serino,
\newblock (2011), arXiv:1101.1624.

\bibitem{Duff:1977ay}
M.~J. Duff,
\newblock Nucl.Phys. {\bf B125}, 334 (1977).

\bibitem{Duff:1993wm}
M.~J. Duff,
\newblock Class. Quant. Grav. {\bf 11}, 1387 (1994), arXiv:hep-th/9308075.

\bibitem{Freedman:1974gs}
D.~Z. Freedman, I.~J. Muzinich, and E.~J. Weinberg,
\newblock Ann. Phys. {\bf 87}, 95 (1974).

\bibitem{Callan:1970ze}
C.~G. Callan, Jr., S.~R. Coleman, and R.~Jackiw,
\newblock Ann. Phys. {\bf 59}, 42 (1970).

\bibitem{Adler:1976zt}
S.~L. Adler, J.~C. Collins, and A.~Duncan,
\newblock Phys. Rev. {\bf D15}, 1712 (1977).

\bibitem{Collins:1976yq}
J.~C. Collins, A.~Duncan, and S.~D. Joglekar,
\newblock Phys. Rev. {\bf D16}, 438 (1977).

\bibitem{Armillis:2009sm}
R.~Armillis, C.~Corian\`o, L.~Delle~Rose, and M.~Guzzi,
\newblock JHEP {\bf 12}, 029 (2009), arXiv:0905.0865.

\bibitem{Armillis:2009im}
R.~Armillis, C.~Corian\`o, and L.~Delle Rose,
\newblock Phys. Lett. {\bf B682}, 322 (2009), arXiv:0909.4522.

\bibitem{Armillis:2008bg}
R.~Armillis, C.~Corian\`o, M.~Guzzi, and S.~Morelli,
\newblock JHEP {\bf 10}, 034 (2008), arXiv:0808.1882.

\bibitem{Coriano:2008pg}
C.~Corian\`o, M.~Guzzi, and S.~Morelli,
\newblock Eur. Phys. J. {\bf C55}, 629 (2008), arXiv:0801.2949.

\bibitem{Dolgov:1993vg}
A.~Dolgov,
\newblock Phys. Rev. {\bf D48}, 2499 (1993), arXiv:hep-ph/9301280.

\bibitem{Corradini:2007gd}
O.~Corradini and A.~Iglesias,
\newblock JCAP {\bf 0805}, 012 (2008), arXiv:0708.1052.

\bibitem{Starobinsky:1980te}
A.~A. Starobinsky,
\newblock Phys. Lett. {\bf B91}, 99 (1980).

\bibitem{Bastianelli:2002qw}
F.~Bastianelli, O.~Corradini, and A.~Zirotti,
\newblock Phys.Rev. {\bf D67}, 104009 (2003), arXiv:hep-th/0211134.

\bibitem{Riegert:1984kt}
R.~J. Riegert,
\newblock Phys. Lett. {\bf B134}, 56 (1984).

\bibitem{Mottola:2006ew}
E.~Mottola and R.~Vaulin,
\newblock Phys.Rev. {\bf D74}, 064004 (2006), arXiv:gr-qc/0604051.

\bibitem{Dolgov:1971ri}
A.~D. Dolgov and V.~I. Zakharov,
\newblock Nucl. Phys. {\bf B27}, 525 (1971).

\bibitem{Ross:1973fp}
D.~A. Ross and J.~C. Taylor,
\newblock Nucl. Phys. {\bf B51}, 125 (1973).

\bibitem{LopesCardoso:1991zt}
G.~Lopes~Cardoso and B.~A. Ovrut,
\newblock Nucl.Phys. {\bf B369}, 351 (1992).

\bibitem{Bagger:1999rd}
J.~A. Bagger, T.~Moroi, and E.~Poppitz,
\newblock JHEP {\bf 0004}, 009 (2000), arXiv:hep-th/9911029.

\bibitem{Caracciolo:1989pt}
S.~Caracciolo, G.~Curci, P.~Menotti, and A.~Pelissetto,
\newblock Annals Phys. {\bf 197}, 119 (1990).

\bibitem{Denner:1991kt}
A.~Denner,
\newblock Fortschr. Phys. {\bf 41}, 307 (1993), arXiv:0709.1075.

\end{thebibliography}

\end{document}